\documentclass[twocolumn,epjc3]{svjour3}          

\RequirePackage[numbers,sort&compress]{natbib}
\RequirePackage[colorlinks,citecolor=blue,urlcolor=blue,linkcolor=blue]{hyperref}

\usepackage{listings}
\pdfoutput=1
\usepackage[utf8]{inputenc}
\usepackage[english]{babel}

\usepackage{lmodern}
\usepackage[safe]{textcomp}
\usepackage{bbm}

\usepackage{amstext}  
\usepackage{amsfonts}
\usepackage{graphicx}
\usepackage{amsmath}
\usepackage{amsfonts}
\usepackage{mathtools}

\usepackage{array}                
\usepackage{xcolor} 
\usepackage{slashed}
\usepackage[absolute]{textpos}
\usepackage{multirow}

\usepackage{relsize}

\journalname{Eur. Phys. J. C}

\newcommand{\OLtwo}{\text{OL2}}
\newcommand{\onthefly}{\text{on-the-fly}}

\newcommand{\Collier}{\text{Collier}}

\renewcommand{\refeq}[1]{\mbox{\eqref{#1}}}

\newcommand{\reffi}[1]{\mbox{Fig.~\ref{#1}}}
\newcommand{\reffis}[2]{\mbox{Figs.~\ref{#1}--\ref{#2}}}
\newcommand{\refta}[1]{\mbox{Table~\ref{#1}}}

\newcommand{\refse}[1]{\mbox{Section~\ref{#1}}}
\newcommand{\refses}[2]{\mbox{Sections~\ref{#1}--\ref{#2}}}

\newcommand{\ie}{i.e.\ }
\newcommand{\eg}{e.g.\ }

\newcommand{\f}[2]{\frac{#1}{#2}}

\newcommand{\ssst}[1]{\scriptscriptstyle{\text{#1}}}
\newcommand{\nosss}[1]{#1}

\newcommand{\bit}{\begin{itemize}}
\newcommand{\eit}{\end{itemize}}
\newcommand{\bce}{\begin{center}}
\newcommand{\ece}{\end{center}}
\newcommand{\bea}{\begin{eqnarray}}
\newcommand{\eea}{\end{eqnarray}}
\newcommand{\be}{\begin{equation}}
\newcommand{\ee}{\end{equation}}
\newcommand{\ba}{\begin{align}}
\newcommand{\ea}{\end{align}}
\newcommand{\beas}{\begin{eqnarray*}}
\newcommand{\eeas}{\end{eqnarray*}}
\newcommand{\bes}{\begin{equation*}}
\newcommand{\ees}{\end{equation*}}
\newcommand{\bas}{\begin{align*}}
\newcommand{\eas}{\end{align*}}

\newcommand{\eps}{{\varepsilon}}

\newcommand{\lb}{\left(}
\newcommand{\rb}{\right)}

\newcommand{\idop}{1\!\!1}

\newcommand{\ceps}{C_{\eps}}
\newcommand{\Dbar}[1]{\bar{D}_{\nosss{#1}}}
\newcommand{\Dtilde}[1]{\widetilde{D}_{\nosss{#1}}}

\newcommand{\momk}[1]{k_{\nosss{#1}}}
\newcommand{\momp}[1]{p_{#1}}
\newcommand{\mass}[1]{m_{\nosss{#1}}}
\newcommand{\heli}{h}
\newcommand{\helibar}{\bar h}
\newcommand{\helicheck}{\check h}

\newcommand{\helihat}{\hat h}
\newcommand{\hset}{\mathcal{H}}
\newcommand{\hsetcheck}{\mathcal{\check H}}
\newcommand{\hsetbar}{\mathcal{\bar H}}

\newcommand{\hsethat}{\mathcal{\hat H}}

\newcommand{\np}{\nosss{-1}}
\newcommand{\momq}{\bar{q}}
\newcommand{\tilq}{\tilde{q}}
\newcommand{\npart}{N_{\mathrm{p}}}
\newcommand{\nmax}{N_{\mathrm{max}}}
\newcommand{\calA}{\mathcal{A}}
\newcommand{\calC}{\mathcal{C}}
\newcommand{\calI}{\mathcal{I}}

\newcommand{\calK}{\mathcal{K}}
\newcommand{\calM}{\mathcal{M}}
\newcommand{\calN}{\mathcal{N}}

\newcommand{\calE}{\mathcal{E}}
\newcommand{\calU}{\mathcal{U}}
\newcommand{\calS}{\mathcal{S}}
\newcommand{\calV}{\mathcal{V}}
\newcommand{\calVtilde}{\widetilde{\mathcal{V}}}
\newcommand{\calW}{\mathcal{W}}
\newcommand{\calWmin}{\calW_{\min}}
\newcommand{\Omegatilde}{\widetilde\Omega}
\newcommand{\seg}{S}
\newcommand{\segtilde}{\widetilde S}

\newcommand{\rmaxk}{R^{\max}_k}

\newcommand{\col}{\mathrm{col}}

\newcommand{\re}{\mathrm{Re}}
\newcommand{\Tr}{\mathrm{Tr}}

\newcommand{\unpinched}{\mathrm{unpinched}}

\newcommand{\rem}{\mathrm{rem}}

\newcommand{\ri}{\mathrm i}
\newcommand{\rd}{\mathrm d}
\newcommand{\ord}{\mathcal O}

\definecolor{bluemar}{rgb}{0,0,.5}
\definecolor{redmar}{rgb}{.8,0,0}
\definecolor{greenmar}{rgb}{0,.5,0}

\newcommand{\qperp}{q_\perp}
\newcommand{\qpar}{q_{\parallel}}

\newcommand{\bzero}{B_0}
\newcommand{\dbzero}{\Delta B_0}
\newcommand{\deltathr}{\delta_{\mathrm{thr}}}

\newcommand{\tree}{\mathrm{tree}}
\newcommand{\oneloop}{\mathrm{1-loop}}
\newcommand{\Rnext}{R_{\mathrm{next}}}
\newcommand{\Rlast}{R_{\mathrm{last}}}

\newcommand{\calVpinch}[2]{\calV_{#1}\big({\Omega_N^{#1}[#2]}\big)}

\newlength{\defheight}
\newcommand{\heightA}{\defheight}
\newcommand{\heightB}{1.3\defheight}
\newcommand{\heightC}{2.25\defheight}
\newcommand{\heightD}{1.08\defheight}
\newcommand{\heightI}{1.16\defheight}
\newcommand{\heightF}{2.05\defheight}

\newcommand{\showredpinchdiag}[3]{
\raisebox{-1mm}{\parbox[t]{4.1\defheight}{\includegraphics[height=#2]{#1}\\\mbox{#3}}}}

\allowdisplaybreaks

\usepackage{tocloft}
\makeatletter
\renewcommand{\numberline}[1]{%
  \@cftbsnum #1\@cftasnum~\@cftasnumb%
}
\newcommand{\eqnum}[1]{\leavevmode\hfill\refstepcounter{equation}\label{#1}\textup{\tagform@{\theequation}}}
\makeatother
 
\begin{document}
\title{On-the-fly reduction of open loops}

\author{Federico Buccioni\thanksref{e1,inst1} \and
Stefano Pozzorini\thanksref{e2,inst1}\and
and Max Zoller\thanksref{e3,inst1}}

\thankstext{e1}{e-mail: buccioni@physik.uzh.ch}
\thankstext{e2}{e-mail: pozzorin@physik.uzh.ch}
\thankstext{e3}{e-mail: zoller@physik.uzh.ch}

\institute{Physik-Institut, Universit\"at Z\"urich, CH-8057 Z\"urich, Switzerland \label{inst1}}
\date{Received: date / Revised version: date}
%

\maketitle

\begin{abstract}
Building on the open-loop algorithm we introduce a new method for the
automated construction of one-loop amplitudes and their reduction to scalar
integrals.
The key idea is that the factorisation of one-loop integrands
in a product of loop segments makes it possible to perform various operations
on-the-fly while constructing the integrand.
Reducing the integrand on-the-fly,
after each segment multiplication,
the construction of loop diagrams and their reduction are unified 
in a single numerical recursion.
In this way we entirely avoid objects with high tensor rank,
thereby reducing the complexity of the calculations in a drastic way.  
Thanks to the on-the-fly approach, which is applied also to helicity summation and 
for the  merging of different diagrams, the speed of the original open-loop algorithm
can be further augmented in a very significant way.  
Moreover, addressing spurious singularities of the employed reduction identities 
by means of simple expansions in rank-two Gram determinants, we achieve a
remarkably high level of numerical stability.
These features of the new algorithm, 
which will be made publicly available in a forthcoming release of the {\sc OpenLoops} program,
are particularly attractive for NLO multi-leg
and NNLO real--virtual calculations.

\end{abstract} 

\setcounter{tocdepth}{2}
\tableofcontents

\section{Introduction} \label{se:intro}

The continuous improvement of statistics and experimental systematics at the
Large Hadron Collider (LHC) permits to challenge the Standard Model of
particle physics at steadily increasing levels of energy and precision.  In
this context, the uncertainty of theoretical predictions starts playing a
critical role in many areas of the physics program of the LHC, providing
strong motivation for developing new techniques that make it possible to
push theoretical calculations towards more complex processes and higher
perturbative orders.

In the last decade, the advent of new powerful methods for the calculation of
one-loop scattering amplitudes~\cite{Britto:2004nc,
delAguila:2004nf,
Bern:2005cq,
Denner:2005nn,
Ossola:2006us,
Forde:2007mi,Giele:2008ve,
vanHameren:2009vq,
Cascioli:2011va} 
has opened the door to the
automation of next-to-leading order (NLO) calculations.  Nowadays, 
one-loop calculations are supported by a number of
highly automated tools~\cite{Ossola:2007ax,
Berger:2008sj,
vanHameren:2009dr,
Hirschi:2011pa,
Mastrolia:2010nb,
Badger:2012pg,
hepforge,
Cullen:2014yla,
Peraro:2014cba,
Denner:2016kdg,
Actis:2016mpe} 
that provide the key to achieve NLO precision
in the context of multi-purpose Monte Carlo generators~\cite{Gleisberg:2008ta,
Alioli:2010xd,
Bevilacqua:2011xh,
Alwall:2014hca,
Sjostrand:2014zea,
Weiss:2015npa,
Bellm:2015jjp}.
This recent progress has enabled NLO calculations for a huge 
number of processes and has extended their reach up to 
multi-particle final states of unprecedented complexity~\cite{Bern:2013gka,
Badger:2013yda,
Bevilacqua:2015qha,
Hoche:2016elu,
Denner:2016wet}.
Nevertheless, in various cases the technical limitations of one-loop
generators still represent a serious bottleneck or even a show stopper. 
These issues can be encountered in processes  with many final-state particles and 
for kinematic configurations  with two or more widely separated scales.
An important example is given by the real--virtual contributions to
next-to-next-to leading order (NNLO) calculations, which require very fast and highly stable
one-loop amplitudes in deeply infrared regions of phase space.

Motivated by these considerations, in this paper we introduce a new method
that leads to very significant efficiency and stability improvements in the
construction of one-loop amplitudes.  This new method builds on {\sc
OpenLoops}~\cite{Cascioli:2011va,hepforge}, a fully automated framework for the automated generation
of scattering amplitudes in the Standard Model.
The original implementation of the open-loop approach~\cite{Cascioli:2011va,hepforge}
supports NLO QCD~\cite{Cascioli:2013gfa,Cascioli:2013era,Cascioli:2013wga,Hoeche:2014qda,Hoeche:2014rya,Jezo:2016ujg,Hoche:2016elu,Nejad:2016bci} 
as well as  NLO EW~\cite{Kallweit:2014xda,Kallweit:2015dum,Kallweit:2017khh,Lindert:2017olm,Granata:2017iod}
calculations and is interfaced to  various multi-purpose Monte Carlo tools.
The {\sc OpenLoops} program is also part of {\sc Matrix}~\cite{Wiesemann:2016eiw}
and has already been applied to several NNLO calculations~\cite{Grazzini:2013bna,Cascioli:2014yka,Gehrmann:2014fva,Grazzini:2015nwa,
Grazzini:2015wpa,Grazzini:2015hta,
Grazzini:2016swo,Grazzini:2016ctr,deFlorian:2016uhr,Grazzini:2017ckn}.
The essence of the open-loop method~\cite{Cascioli:2011va} consists of a numerical recursion that
generates cut-open loop diagrams, called open loops, by multiplying, one after
the other, the various building blocks that are connected through loop propagators. More precisely,
the construction of $N$-point loop integrands is organised 
through the factorisation of
$N$ loop {\it segments}, which consist each of a loop propagator and 
a corresponding external subtree.
Segment multiplications are implemented through process-independent numerical
routines that correspond to the Feynman rules of the model at hand.
This type of recursion was first proposed  in the context of
off-shell recurrence relations for colour-ordered gluon-scattering amplitudes~\cite{vanHameren:2009vq}.
Thanks to a tensorial representation that retains the loop-momentum
dependence of all building blocks, this approach can be used in combination
with reduction techniques based on tensor integrals~\cite{Denner:2005nn} or with the OPP
reduction method~\cite{Ossola:2006us}, resulting in both cases in very fast computer
code~\cite{Cascioli:2011va}.

The new method presented in this paper exploits the factorised structure of
the open-loop representation in a completely new way.  The key idea is that
certain operations, which are usually done when all building blocks of
Feynman diagrams have been assembled, can be anticipated and performed
{\it on-the-fly} during the construction of the diagrams.
Exploiting the factorised structure of the integrands, this 
on-the-fly approach permits to perform various types of operations
at a much lower level of complexity, thereby boosting their efficiency.
As we will show, it can be exploited in order to factorise helicity
summations as well as the sums over different Feynman diagrams that share
the same one-loop topology.
Moreover, based on the integrand reduction method by del\,Aguila and
Pittau~\cite{delAguila:2004nf}, we will introduce an on-the-fly technique for the reduction
of open loops.  In this way, we will promote {\sc OpenLoops} to an
algorithm that combines the construction and the reduction of loop
amplitudes in a unified numerical recursion.  A notable feature of this
approach is that it permits to avoid high-rank objects at any stage of the calculations.  More
precisely, tensor integrals are always kept at rank two or lower, thereby
reducing the computational complexity in a dramatic way.

The on-the-fly technique leads to very significant improvements of CPU
efficiency.  For what concerns numerical stability, 
in order to avoid severe instabilities that result
from squared inverse Gram determinants in the reduction identities
of~\cite{delAguila:2004nf}, we present a method that isolates such instabilities in certain
triangle topologies and circumvents them via analytic expansions in the limit
of small Gram determinants.  In this way we obtain the first
integrand-reduction algorithm that is essentially free from Gram-determinant
instabilities.  The achieved level of stability in double precision is competitive
with the most sophisticated tools on the market~\cite{Denner:2016kdg} and 
with public implementations of OPP reduction in quadruple
precision.

The paper is organized as follows.
In \refse{se:olreview} we review the original open-loop method.
The on-the-fly approach is introduced in \refse{sec:OFhelmerging} 
for the case of helicity sums and for the merging of topologically equivalent 
open loops.
In \refse{sec:red} the on-the-fly approach is generalised to the 
reduction of open loops.
Details on the employed  integrand-reduction identities and 
our treatment of Gram-determinant instabilities are discussed
in \refse{se:red}.
The entire algorithm and its implementation are outlined in 
\refse{sec:algorithm}, where we also present technical studies on the
CPU performance and numerical stability.
Our conclusions are presented in \refse{se:conclusions}, and
\ref{app:intred} deals with 
low-rank integrals that remain to be solved at the end of 
the on-the-fly recursion.

\section{The open-loop method} \label{se:olreview}

In this section we review the original open-loop method~\cite{Cascioli:2011va},
which is implemented in the publicly available {\sc OpenLoops\,1} program~\cite{hepforge}.
At variance with the original publication~\cite{Cascioli:2011va}, here we
refine various aspects of the notation and we adopt a particular perspective
that sets the stage for the new methods introduced in
\refses{sec:OFhelmerging}{se:red}.  These new techniques are going
to become publicly available in the {\sc OpenLoops\,2} release.

\subsection{Helicity and colour bookkeeping}
\label{se:helcolbook}

The task carried out by the open-loop algorithm
is the calculation of the tree-level 
and one-loop contributions
to the scattering probability 
density, 
\bea
\label{eq:W01}
\calW_{\tree}
&=&
\sum_{\heli}\sum_{\col}|\calM_{0}(\heli)|^2, \nonumber\\
\calW_{\oneloop}
&=&
\sum_{\heli}\sum_{\col} 2\,\re \Big[\calM_{0}^*(\heli)\calM_{1}(\heli)\Big],
\eea
or the squared one-loop contribution
\bea
\calW_{\oneloop^2}=\sum_{\heli}\sum_{\col}|\calM_1(\heli)|^2,
\label{eq:W1}
\eea
for loop-induced processes.
The polarised matrix elements $\calM_0(\heli)$ and $\calM_1(\heli)$ should
be understood as generic tree and one-loop amplitudes, in the sense that the
techniques presented in this paper are applicable to any renormalisable
theory, including the QCD and electroweak sectors of the Standard Model, as
well as BSM theories.
The sums in~\refeq{eq:W01}--\refeq{eq:W1}
run over all helicity and colour degrees of
freedom of the scattering particles.  While colour indices are kept
implicit, the helicity dependence is characterised by a single index
$\heli$, which corresponds to the global helicity configuration of the
event, as described below.

Scattering amplitudes are computed as sums of Feynman diagrams,
\bea
\label{eq:amplitudes}
\calM_{0}(\heli) 
&=& 
\sum_{\calI\in\Omega_{\tree}} \calM_{0}(\calI,\heli),\nonumber\\
\calM_{1}(\heli) 
&=& 
\sum_{\calI\in\Omega_{\oneloop}} \calM_{1}(\calI,\heli),
\qquad
\eea
where $\Omega_{\tree}$ and $\Omega_{\oneloop}$ stand for the sets of 
tree and one-loop diagrams.
Each tree and one-loop diagram can be factored into a colour factor 
$\calC(\calI)$ and a colour-stripped diagram amplitude,%
\footnote{Quartic gluon couplings involve three independent colour structures. Thus each diagram
involving $k$ quartic couplings needs to be split into $3^k$ contributions of type
\eqref{eq:colfact} that are effectively handled as separate diagrams.}
\bea
\label{eq:colfact}
\calM_{L}(\calI,\heli)
=
\calC(\calI)\,\calA_L(\calI,\heli),
\eea
for $L=0,1$. The colour-stripped amplitudes $\calA_L(\calI,\heli)$ 
are the main source of complexity.
In the open-loop approach, their calculation is 
addressed with numerical recursions as 
described in \refses{se:treelevel}{se:parent-child}.
For what concerns colour factors, 
exploiting the factorisation properties~\refeq{eq:colfact}, all relevant operations
can be reduced to the calculation of colour-summed interference terms of the form
\bea
\calK(\calI_a,\calI_b)=
\sum_{\col}\,\mbox{$\calC(\calI_a)$}^{*}\,\calC(\calI_b),
\label{eq:colintA}
\eea
which appear in the calculation of the scattering probabilities~\refeq{eq:W01}-\refeq{eq:W1}.
This task must be addressed only once per process. It is handled by
algebraically reducing all $\calC(\calI)$ to a standard basis $\{\calC_i\}$
and relating the~terms \refeq{eq:colintA} to the interference
matrix~\cite{Bredenstein:2010rs,Cascioli:2011va} 
\bea 
\calK_{ij}=
\sum_{\col}\,\mbox{$\calC_i$}^*\, \calC_j.  
\label{eq:colintB} 
\eea
In {\sc OpenLoops} we use a basis where all colour factors are
expressed through products and traces of the SU(3) generators $T^a_{ij}$ in
the fundamental representation.

For the bookkeeping of external momenta and helicities
in a process with $\npart$
scattering particles we introduce the set of particle indices
\bea
\calE=\{1,2,\dots, \npart\}.
\label{eq:fullpartset}
\eea
To characterise the helicity configurations $s$ of individual 
particles we use labels
\bea
\lambda_i=\begin{cases}
\;1,3   & \mbox{for fermions with}\,s= -1/2, 1/2\\[2mm] 
\;1,2,3 & \mbox{for gauge bosons with}\,s=-1,0,1\\[2mm]
\;0     & \mbox{for scalars with}\,s=0 
\end{cases}
\label{eq:hellabellam}
\eea
$\forall\; i\in\calE$. The configuration 
$\lambda_i=0$ will also be used to characterise unpolarised
particles, \ie fermions or gauge bosons whose helicity is still
unassigned or has already been summed over.
Since a particle can have up to four different
helicity states, it is convenient to 
adopt a helicity numbering scheme based on the labels
\bea
\helibar_{i}=\lambda_i\,4^{i-1},
\label{eq:quatrep}
\eea
which correspond to a quaternary number with $\lambda_i\in \{0,1,2,3\}$ as
$i^{\mathrm{th}}$-last digit and all other digits equal to
zero.  In this way, the helicity configurations 
$(\lambda_1,\dots,\lambda_{\npart})$
of the full event can be uniquely identified with 
the label
\bea
\heli=\helibar_1+\dots +\helibar_{\npart},
\label{eq:globhellabel}
\eea
which corresponds to a quaternary number of $\npart$ digits, each 
of which describes the helicity of a particular external particle.
Let us also introduce the single-particle 
helicity spaces, $\hsetbar_{i}\ni\helibar_{i}$, defined as
\bea
\hsetbar_{i}=\begin{cases}
\;\{\lambda_i\,4^{i-1} | \lambda_i=1,3\} & \;\mbox{for fermions or mass-}\\
                                         & \;\mbox{less gauge bosons,}\\[2mm]
\;\{\lambda_i\,4^{i-1} | \lambda_i=1,2,3\}& \;\mbox{for massive}\\
                                          & \;\mbox{gauge bosons,}\\[2mm]
\;\{0\}& \;\mbox{for scalars,} 
\end{cases}
\label{eq:hsetbari}
\eea
where we do not include unpolarised states. 
Finally, the global helicity space for the full set of scattering
particles, $\hset\ni\heli$, is defined as
\bea
\hset=\hsetbar_1\otimes\dots\otimes\hsetbar_{\npart},
\label{eq:globhelset}
\eea
where the product is understood as 
$\mathcal{A}\otimes\mathcal{B}=\{a+b|a\in\mathcal{A},b\in\mathcal{B}\}$.

\subsection{Tree amplitudes}
\label{se:treelevel}

At tree level, each colour-stripped Feynman diagram is constructed by
contracting two so-called subtrees, which arise by cutting the diagram in
two pieces in correspondence of an internal propagator,
\bea
\calA_0(\calI,h)
\;=\;\;
\vcenter{\hbox{\scalebox{1.}{\includegraphics[width=40mm]{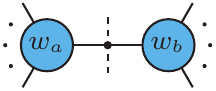}}}}
\label{eq:cuttreeA}
\eea
A generic subtree $w_a$ corresponds to the part of 
a certain Feynman diagram that 
connects an internal off-shell line with outgoing momentum $\momk{a}$
to
a subset of external particles, 
\bea
\calE_a=\{\alpha_{a1},\dots,\alpha_{an_a}\}\subset\calE.
\label{eq:partsubset}
\eea
In the numbering scheme~\refeq{eq:quatrep}--\refeq{eq:globhellabel},
the helicity configurations of a subtree are labeled
\bea
\heli_{a}= \helibar_{\alpha_{a1}}+\dots+\helibar_{\alpha_{an_a}},
\label{eq:quatrepsum}
\eea
and the corresponding helicity space, $\hset_a\ni\heli_a$, is defined as
\bea
\hset_{a}=\hsetbar_{\alpha_{a1}}\otimes\dots\otimes \hsetbar_{\alpha_{an_a}}.
\label{eq:hseti}
\eea
Subtrees are represented as complex $n$-tuples,
$w^{\sigma_a}_a(\momk{a},h_a)$, where $\sigma_a$ is the spinor or Lorentz
index of the cut line. With this notation, the contraction~\refeq{eq:cuttreeA} takes 
the form
\bea
\calA_0(\calI,h)
\;=\;\;
w^{\sigma_a}_a(k_a,h_a)
\,\delta_{\sigma_a\sigma_b}
\widetilde{w}^{\sigma_b}_b(k_b,h_b),
\label{eq:cuttreeB}
\eea
where $k_b=-k_a$, $h=h_a+h_b$, and summation over repeated indices is implicitly understood.
The propagator associated with the cut line is included
only in the subtree $w_a$ and not in  $\widetilde{w}_b$.

\begin{figure}[t!]
\bea
w^{\sigma_a}_a(k_a,h_a)
\;=\;\;
\vcenter{\hbox{\scalebox{1.}{\includegraphics[width=60mm]{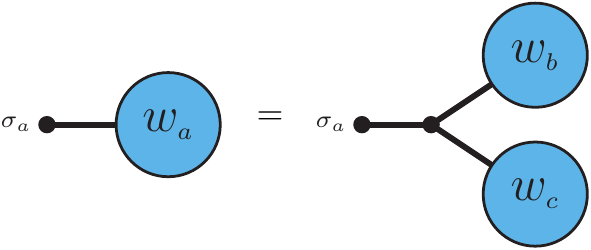}}}}
\nonumber
\eea
\caption{Diagrammatic representation of a subtree and its
numerical construction through the recurrence relation~\refeq{eq:treerecursion}.
The outgoing momentum $k_a$ and the spin or Lorentz index $\sigma_a$ are associated with the
off-shell internal line, which is shown explicitly, while the on-shell external 
particles with helicity $h_a$ are implicitly understood.
}
\label{fig:Tree2}
\end{figure}

Subtrees are constructed by means of a numerical recursion that starts from
the external wave functions and recursively merges subtrees by attaching
their off-shell lines to the vertices that occur in the various tree
diagrams. A recursion step for the case of a generic three-particle vertex
is depicted\footnote{To draw Feynman diagrams we use {\sc Axodraw}~\cite{Vermaseren:1994je}.}
in~\reffi{fig:Tree2}.  Its algebraic form reads
\bea
\label{eq:treerecursion}
w^{\sigma_a}_a(\momk{a},\heli_{a}) =
\f{X_{\sigma_b\sigma_c}^{\sigma_a}(\momk{b},\momk{c})}{\momk{a}^2-\mass{a}^2}\; 
w^{\sigma_b}_b(\momk{b},\heli_{b})\;
w^{\sigma_c}_c(\momk{c},\heli_{c}){},\nonumber\\
\eea
where the tensor $X_{\sigma_b\sigma_c}^{\sigma_a}$ describes the vertex that
connects $w_b$ and $w_c$ to $w_a$, as well as the numerator of the
propagator that connects to $w_a$. The related denominator,
$(k_a^2-m_a^2)$, appears explicitly in~\refeq{eq:treerecursion}.
The momentum of the resulting subtree is 
$\momk{a}=\momk{b}+\momk{c}$  and 
its helicity is $\heli_{a}=\heli_{b}+\heli_{c}$.
Each recursion step must be repeated for all independent helicity
configuration $\heli_a\in\hset_a=\hset_b\otimes\hset_c$.
The corresponding recursion step for quartic vertices reads
\bea
\label{eq:treerecursionB}
w^{\sigma_a}_a(\momk{a},\heli_{a}) &=&
\f{X_{\sigma_b\sigma_c\sigma_d}^{\sigma_a}(\momk{b},\momk{c},\momk{d})}{\momk{a}^2-\mass{a}^2}\;
w^{\sigma_b}_b(\momk{b},\heli_{b})\;\nonumber\\ 
&\times& w^{\sigma_c}_c(\momk{c},\heli_{c}){}\;
w^{\sigma_d}_d(\momk{d},\heli_{d}).
\eea

The recursion ends when all off-shell propagators that have been cut in the
beginning can be reconnected, such as to obtain the colour-stripped
amplitudes \refeq{eq:cuttreeB} for the full set of tree diagrams.

Note that~\refeq{eq:treerecursion}--\refeq{eq:treerecursionB} are analogous to 
Berends--Giele recurrence relations
for off-shell currents~\cite{Berends:1987me}. However, 
while each subtree corresponds to a single topology, 
off-shell currents incorporate all 
possible subtrees associated with a certain internal line.
The inefficiency due to the usage of individual subtrees is compensated,
especially at one-loop level, 
by the optimisation opportunities that result from the
colour-factorisation identities~\refeq{eq:colfact} and 
from the fact that each subtree can occur in multiple Feynman diagrams at tree and loop level.

\subsection{One-loop amplitudes} 
\label{sec:oneloopamp}

The amplitude of a colour-stripped $N$-point 
one-loop diagram, $\calI_N$, has the general form
\bea \bar{\mathcal{A}}_{1}(\calI_N,\heli) 
&=&
\quad\vcenter{\hbox{\scalebox{1.}{\includegraphics[height=\heightF]{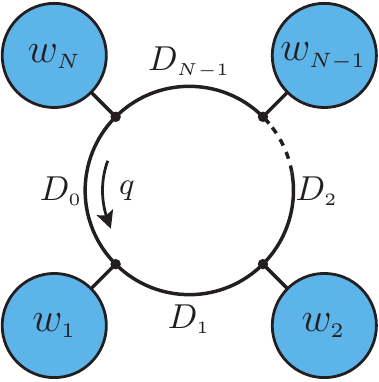}}}}\;=
\nonumber\\
&=& 
\int\!\rd^D\!\momq\, 
\f{\Tr\Big[\bar{\calN}(\mathcal{I}_{N},\momq,\heli)\Big]}
{\Dbar{0} \Dbar{1}\cdots \Dbar{N-1}}\,.
\label{eq:OLfulldia} 
\eea
Symbols carrying a bar denote quantities in $D=4-2\varepsilon$ dimensions,
and for the loop momentum $\momq$ we adopt the decomposition
\bea
\momq=q+\tilde{q},
\eea
where $q$ and $\tilde{q}$ denote its four-dimensional and $(D-4)$-dimensional parts, respectively. 
The denominators read 
\bea \Dbar{i}=(\momq + \momp{i})^2-\mass{i}^2\,,
\;\,\;\,\; \momp{i}=\sum_{j=1}^i \momk{j}{}, \label{eq:propdef}
\eea
where $\momk{j}$ is the external momentum flowing into the loop at the
$j^{\mathrm{th}}$ loop vertex.  Internal momenta are chosen such that
$\momp{0}=\momp{N}=0$, i.e.~the momentum flowing through the
$\Dbar{0}=\Dbar{N}$ propagator is $\momq$.
The one-loop diagram $\calI_N$ in~\refeq{eq:OLfulldia} can be regarded as a sequence 
of loop segments, 
\bea 
\calI_{N}=\{\calS_1,\calS_2, \ldots,\calS_N\},
\label{eq:calIN}
\eea
where the segment $\calS_i$
consists of a subtree $w_i$ 
that involves a certain set 
$\calE_i$ of external particles
and is connected to the $i^{\mathrm{th}}$ loop vertex, $v_i$, 
and to the adjacent loop propagator associated with 
$\Dbar{i}$. Segments associated to a quartic vertex involve two subtrees,
$w_{i_1}$ and $w_{i_2}$.
The helicity configurations of the whole diagram
are related to the ones
of individual segments,
$\heli_i\in \hset_i$, via
\bea
\heli=\heli_1+\dots+\heli_{N}.
\eea

The trace in~\refeq{eq:OLfulldia} 
stands for the full contraction of the spinor and Lorentz indices
of propagators and vertices along the loop.
In general, the numerator $\bar{\calN}(\momq)$ consists of a 4-dimensional part $\calN(q)$ and an
$\eps$-dependent remnant $\widetilde{{\calN}}(\momq)$,
\bea
\Tr\Big[\bar{\calN}(\momq)\Big] = \Tr\Big[\calN(q)\Big]+\Tr\Big[\widetilde{{\calN}}(\momq)\Big].
\eea
The terms that result from 
$\widetilde{\calN}(\momq)$ are known as rational terms of type $R_2$ 
and can be reconstructed separately as counterterms using appropriate Feynman rules
\cite{Ossola:2008xq,Draggiotis:2009yb,Garzelli:2009is,Pittau:2011qp}. Thus, the full 
amplitude can be decomposed as 
\bea
\bar{\calA}_{1}(h)=\mathcal{A}_{1}(h)+\mathcal{A}_{1,R_2}{(h)}, \label{eq:diaA1dbar}
\eea
and in the following we focus on the nontrivial part
\bea 
\mathcal{A}_{1}(\calI_N,h)=\int\!\rd^D\!\momq\, 
\f{\Tr\Big[{\calN}(\mathcal{I}_{N},q,\heli)\Big]}{\Dbar{0} \Dbar{1}\cdots \Dbar{N-1}},
\label{eq:A1d}
\eea
which stems from the four-dimensional part of the numerator
but involves the $D$-dimensional denominators~\eqref{eq:propdef}.

In the open-loop approach, loop diagrams 
are cut-open in correspondence of the $\Dbar{0}$ propagator,
in the sense that the loop numerator is constructed as a tensor, 
\bea 
\Big[\mathcal{N}(\calI_{N},q,\heli)\Big]_{\beta_0}^{\beta_{N}}
\quad=\quad 
\vcenter{\hbox{\includegraphics[height=\heightF]{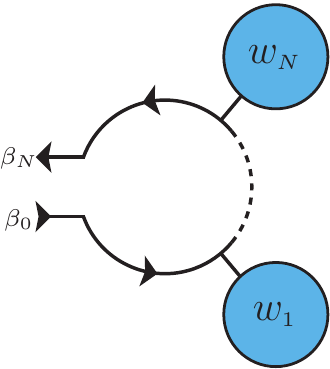}}}\quad, 
\label{eq:Ncoeff}
\eea 
where $\beta_0$ and $\beta_N$ are the spinor or Lorentz indices associated with the cut propagator.
We use the Feynman gauge, which means that the numerator of the gluon propagator is
simply $-\ri g^{\beta}_{\alpha}$.
Once $\big[\mathcal{N}(\calI_{N},q,\heli)\big]_{\beta_0}^{\beta_N}$ is determined, we take
its trace, 
\bea
\Tr\big[\calN(\mathcal{I}_{N},q,\heli)\big] = \delta^{\beta_0}_{\beta_N}\,
\big[\calN(\mathcal{I}_{N},q,\heli)\big]^{\beta_N}_{\beta_0},
\label{eq:cutopenloop}
\eea
where summation over repeated indices is implicitly understood.

\begin{figure*}[t!] \begin{center} 
\includegraphics[height=\heightB]{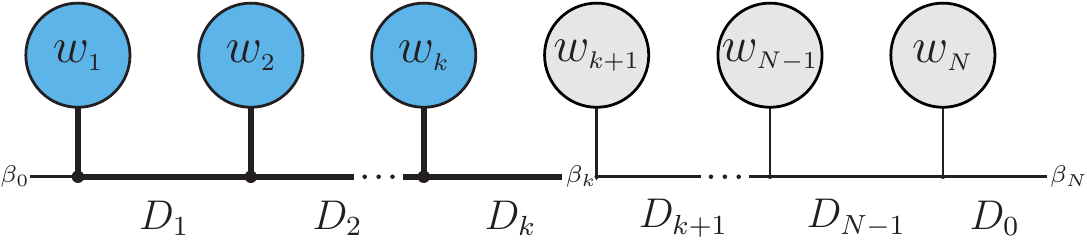}
\end{center} \caption{
Diagrammatic representation of an $N$-point open loop with $k$ dressed and $N-k$ undressed segments.
The segment containing the last subtree,  $w_N\equiv w_0$,
is associated with the propagator 
$\Dbar{N}\equiv\Dbar{0}=\momq^2-m_0^2$.
\label{fig:OL1} }
\end{figure*}

A key feature of the open-loop approach is that,
similarly to the product of 
loop denominators $\Dbar{0}\cdots\Dbar{N-1}$ in~\refeq{eq:A1d},
the loop numerator~\eqref{eq:Ncoeff} is factored
into a product of \textit{segments}, 
\bea
\calN(\mathcal{I}_{N},q,\heli)=\seg_1(q,\heli_1)\cdots\seg_{N}(q,\heli_{N}).
\label{eq:numfactorizationN}
\eea
Here and in the following, the matrix structure 
is implicitly understood, \ie \refeq{eq:numfactorizationN}
should be interpreted as 
\bea
\big[\calN(\mathcal{I}_{N},q,\heli)\big]_{\beta_0}^{\beta_N}&=&
\big[\seg_1(q,\heli_1)\big]_{\beta_0}^{\beta_1}\,
\big[\seg_2(q,\heli_2)\big]_{\beta_1}^{\beta_2} \cdots\nonumber\\ &\cdots&
\big[\seg_{N}(q,\heli_{N})\big]_{\beta_{N-1}}^{\beta_N}.
\label{eq:numfactorizationNbeta}
\eea
Segments involving a triple vertex have the generic form
\bea
\big[\seg_i(q,\heli_i)\big]^{\beta_{i}}_{\beta_{i-1}}
= X^{\beta_{i}}_{\beta_{i-1}\sigma_i}(q+p_{i-1},k_{i})
\, w_{i}^{\sigma_i}(\momk{i},\heli_i),
\eea
where $w_{i}^{\sigma_i}(\momk{i},\heli_i)$ is the corresponding external subtree.
The tensor $X^{\beta_{i}}_{\beta_{i-1}\sigma_{i}}(q+p_{i-1},k_{i})$ 
corresponds to the interaction term in 
\eqref{eq:treerecursion}
and embodies the $q$-dependent 
contributions of the
loop vertex $v_i$ and of the numerator of the adjacent
$D_i$ propagator. In renormalisable theories, each segment $\seg_i(q,\heli_i)$
is a $q$-polynomial of rank $R\le 1$.
In the SM, the structure of three-point vertices is
\bea
& &\Big[\seg_{i}(q,\heli_i)\Big]_{\beta_{i-1}}^{\beta_{i}}
=\quad
\raisebox{3mm}{\parbox{1.45\defheight}{
\includegraphics[height=\heightB]{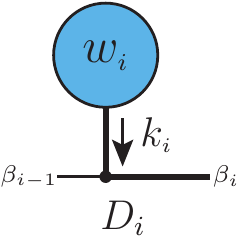}}}= \label{eq:seg3point}\\
& &=\,\Bigg\{\Big[Y_{\sigma_i}^{i}\Big]_{\beta_{i-1}}^{\beta_{i}}
+ \Big[Z_{\nu;\sigma_i}^{i}\Big]_{\beta_{i-1}}^{\beta_{i}}\,q^\nu 
\Bigg\}\, w^{\sigma_i}_i(\momk{i},\heli_i), \nonumber
\eea
while four-point vertices have rank zero,
\bea
& &\Big[\seg_{i}(q,\heli_i)\Big]_{\beta_{i-1}}^{\beta_{i}}
=\quad
\raisebox{3mm}{\parbox{2.0\defheight}{
\includegraphics[height=22mm]{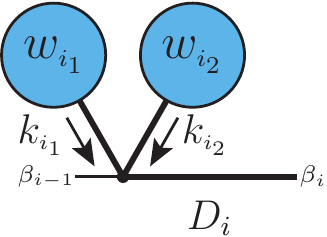}}}=\label{eq:seg4point}\\
& &=\, \Big[Y_{\sigma_1\sigma_2}^i\Big]_{\beta_{i-1}}^{\beta_{i}}
\, w^{\sigma_{1}}_{i_1}(\momk{i_1},\heli_{i_1})
\, w^{\sigma_{2}}_{i_2}(\momk{i_2},\heli_{i_2}),\nonumber
\eea 
with $\heli_i=\heli_{i_1}+\heli_{i_2}$ and  $k_i=k_{i_1}+k_{i_2}$. 

The numerator \eqref{eq:numfactorizationN} is built as a sequence of $N$ segment 
multiplications, and we refer to such a  multiplication as the dressing of a segment.
In the following, we will represent the state of the numerator after 
$k$ dressing steps as,
\bea
& &\calN(\mathcal{I}_{N},q,\heli)
=
\calN_N(\mathcal{I}_{N},q,\heli)=\label{eq:numfactorization}\\
& &\;=
\calN_k(\mathcal{I}_{N},q,\helihat_k)
\seg_{k+1}(q,\heli_{k+1})\cdots\seg_{N}(q,\heli_{N}),\nonumber
\eea
where $\seg_{k+1},\dots,\seg_N$ are the
still undressed segments, and
\bea
\calN_k(\mathcal{I}_{N},q,\helihat_k)
=
\seg_{1}(q,\heli_{1})\cdots\seg_{k}(q,\heli_{k})
\label{eq:numfactorizationB}
\eea
is a $q$-polynomial of rank $R\le k$ that 
incorporates the $k$ dressed segments.
The symbol $\helihat_k$ and its counterpart
$\helicheck_k=h-\helihat_k$
denote, respectively,
the helicity configurations of the dressed and undressed parts
of a diagram with $k$ dressed segments and 
$N-k$ undressed ones.
They are defined through
\bea
\underbrace{\heli_1+\ldots+\heli_k}_{\mathlarger{\helihat_k}}
\;+\;
\underbrace{\heli_{k+1}+\ldots+\heli_{N}}_{\mathlarger{\helicheck_k}}
\;=\;h,
\label{eq:helidress}
\eea
where $h$ is the global helicity state.

The corresponding helicity spaces, $\hsethat_k$ and $\hsetcheck_k$, are defined by
\bea
\underbrace{\hset_{1}\otimes\dots\otimes \hset_{k}}_{\mathlarger{\hsethat_k}} 
\;\otimes\;
\underbrace{\hset_{k+1}\otimes\dots\otimes \hset_{N}}_{\mathlarger{\hsetcheck_k}}
\;=\;\hset.
\label{eq:hsetdresspart}
\eea

The $q$-dependent polynomials~\refeq{eq:numfactorizationB} are denoted 
{\it open loops}, and this notion implicitly includes also the
corresponding undressed segments $\seg_{k+1},\dots,\seg_N$ and loop
denominators $\Dbar{0},\dots,\Dbar{N}$.
A graphical representation of a generic open loop with $k$ dressed segments 
is depicted in \reffi{fig:OL1}. 

The dressing of open loops is implemented through a numerical recursion
\bea
\calN_k(\mathcal{I}_{N},q,\helihat_{k})\,=\,\calN_{k-1}(\mathcal{I}_{N},q,\helihat_{k-1})\seg_{k}(q,\heli_{k}),
\label{eq:OLrecfullpolya}
\eea
where $\helihat_{k}=\helihat_{k-1}+\heli_{k}$.  This operation needs to be
performed separately for all relevant helicity configurations
$\helihat_{k}\in\hsethat_k=\hsethat_{k-1}\otimes\hset_k$ and iterated for $k=1,\dots,N$.  The initial
condition is
\bea
\calN_0(\calI_{N},q,\helihat_{0}) =\idop,
\eea
where $\helihat_{0}\in\hsethat_{0}=\{0\}$, and the identity operator is
understood as $\left[\idop\right]_\beta^{\beta'} = \delta_\beta^{\beta'}$.

In order to capture the full $q$-dependence of open-loop polynomials we use
the tensorial representation
\bea
\calN_k(\mathcal{I}_{N},q,\helihat_{k})
\;=\;
\sum_{r=0}^R
\calN_{k;\,\mu_1\dots\mu_r}(\mathcal{I}_{N},\helihat_{k})
\,q^{\mu_1}\cdots q^{\mu_r},
\label{eq:OLcoeff}
\eea
and numerical operations are always performed at the level of the
tensor coefficients $\calN_{k;\,\mu_1\dots\mu_r}(\mathcal{I}_{N},\helihat_{k})$.
In particular, the explicit form of a step of the dressing
recursion~\eqref{eq:OLrecfullpolya} is
\bea
& &\Big[\calN_{k;\,\mu_1\dots\mu_r}(\mathcal{I}_{N},\helihat_{k})\Big]_{\beta_{0}}^{\beta_{k}} = \nonumber\\
&&=
\Bigg\{
\Big[\calN_{k-1;\,\mu_1\dots\mu_r}(\mathcal{I}_{N},\helihat_{k-1})\Big]_{\beta_{0}}^{\beta_{k-1}}
\Big[Y^k_{\sigma_k}\Big]_{\beta_{k-1}}^{\beta_{k}}
\nonumber\\
&&\phantom{=}+\,
\Big[\calN_{k;\,\mu_2\dots\mu_r}(\mathcal{I}_{N},\helihat_{k-1})\Big]_{\beta_{0}}^{\beta_{k-1}}
\Big[Z^k_{\mu_1;\sigma_k}\Big]_{\beta_{k-1}}^{\beta_{k}}
\Bigg\}\nonumber\\
&&\phantom{=}\times\, w_{k}^{\sigma_k}(\momk{k},\heli_k)
\label{eq:OLrecfullcoeffa}
\eea
for a three-point vertex as defined in \eqref{eq:seg3point}.
For an efficient implementation the $\mu_1\dots\mu_r$ indices shall be symmetrised throughout.

\subsection{Parent--child relations and cutting rule}
\label{se:parent-child}

The original open-loop algorithm can be boosted  
by using parts of pre-computed $(N-1)$-point diagrams
as a starting point for the construction of more involved $N$-point
diagrams.
This approach is based on so-called parent--child relations,
which connect open loops of type
\bea
&&\calN(\mathcal{I}_{N},q,\heli)=\nonumber\\ &&= 
\raisebox{7mm}{\parbox{3.85\defheight}{\includegraphics[height=\heightA]{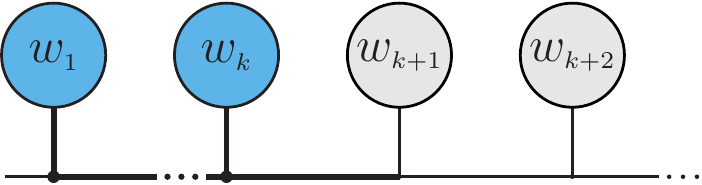}}}
\label{eq:olmerginidA}
\eea
and\vspace{-3mm}
\bea
&&\calN(\tilde{\mathcal{I}}_{N-1},q,\heli)=\nonumber\\ &&= 
\raisebox{7mm}{\parbox{3.85\defheight}{\includegraphics[height=\heightA]{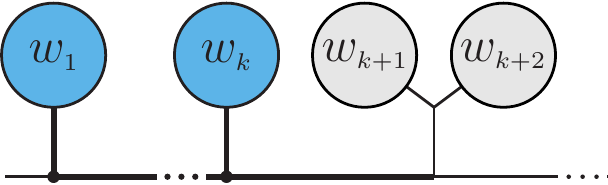}}}
\label{eq:olmerginidB}
\eea
that start with identical segments $\calS_1,\dots,\calS_k$.
Since open loops are colour-stripped, \ie they do not depend on the different
colour factors of the loop diagrams 
\bea
\label{eq:parent-childA}
\calI_{N}&=&\{\calS_1,\dots\calS_k,\calS_{k+1},\dots,\calS_N\}
\quad\mbox{and}\nonumber\\
\tilde{\calI}_{N-1}&=&\{\calS_1,\dots\calS_k,\tilde\calS_{k+1},\dots,\tilde\calS_{N-1}\},
\eea
it is clear that the
dressed parts of~\refeq{eq:olmerginidA} and~\refeq{eq:olmerginidB} remain
identical up to step $k$ of the recursion, \ie
\bea
\calN_k(\mathcal{I}_{N},q,\helihat_k)= 
\calN_k(\tilde{\mathcal{I}}_{N-1},q,\helihat_k). 
\label{eq:parent-childB}
\eea
This allows one to construct the more involved $N$-point parent
diagram~\refeq{eq:olmerginidA} using a building block of the simpler
$(N-1)$-point child diagram~\refeq{eq:olmerginidB}.  In general, such relations can be
applied for any $k$ with $2\le k\le N-2$, and the maximum gain in efficiency is
obtained when $k=N-2$, so that only the last two segments of the parent
diagram remain to be dressed.

The availability of child diagrams of type~\refeq{eq:olmerginidB} is an
obvious prerequisite for the applicability of the parent--child approach,
and in QCD most one-loop diagrams turn out to be the parent of a corresponding child.
Moreover, the correspondence between the first $k$ dressed segments 
in~\refeq{eq:olmerginidA} and \refeq{eq:olmerginidB}
requires an appropriate \textit{cutting rule},
\ie a prescription that determines the cut propagator and the dressing
direction in a similar way as in~\refeq{eq:olmerginidA}--\refeq{eq:olmerginidB}.

To this end, for each segment $\calS_i$ with external particles
$\calE_i=\{\alpha_{i1},\dots,\alpha_{in_i}\}$
we introduce a binary weight defined as the sum of the weights 
$2^{\alpha-1}{}$ for each particle $\alpha$, \ie
\bea
F(\calS_i)=\sum_{\alpha\in\calE_i} 2^{\alpha-1}{}. 
\label{eq:cuttingruleS}
\eea
For example, $F(\calS_i)=2^0+2^1+2^3=11$ for a segment connected to the
external legs $\calE_i=\{1,2,4\}$.  For the merging of subtrees $\calS_i$
and $\calS_j$ into a single segment $\calS_i\oplus\calS_j$ with external
legs $\calE_i\cup\calE_j$, the weight function obeys the useful distributive
property
\bea
F(\calS_i\oplus\calS_j)=
F(\calS_i)+F(\calS_j).
\label{eq:cuttingruleSadd}
\eea
This implies that merged segments always outweigh 
the original segments. Based on this feature, for $N$-point diagrams 
we adopt the cutting rule~\cite{Cascioli:2011va}
\begin{align}
F(\calS_k) &> F(\calS_1) 
\quad\forall\quad k>1,\quad &\text{(selection rule)}\label{eq:cuttingrulesel}\\
F(\calS_N) &> F(\calS_2). &\text{(direction rule)}\label{eq:cuttingruledir}
\end{align}
The fact that the first segment is identified as the one with lowest weight
guarantees its stability with respect to the merging of 
$\calS_{k+1}\oplus\calS_{k+2}$ in~\refeq{eq:olmerginidA}--\refeq{eq:olmerginidB},
while~\refeq{eq:cuttingruledir} guarantees the stability of the
dressing direction for all configurations with $k\ge 2$.
In this way, the parent--child approach permits to 
recycle the longest possible open loops.

Note that relations of type~\refeq{eq:parent-childB} can be exploited also
for diagrams that involve the same number $N$ of loop propagators 
and identical dressed segments, but different undressed ones.

\begin{figure*}[t!] 
$$
\sum_{\heli_1,\dots,\heli_k}\quad
\left(\;\;\parbox{2\defheight}{\includegraphics[height=\heightF]{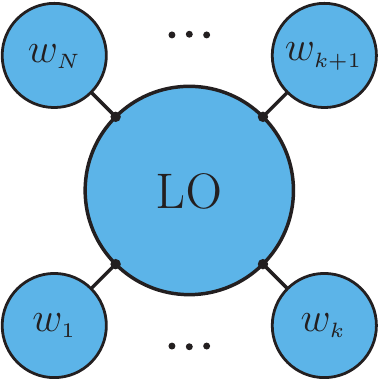}}\;\;\right)^{\hspace{-.5ex}\mathlarger{*}}
\;\,\times\;\;
\parbox{2\defheight}{\includegraphics[height=\heightF]{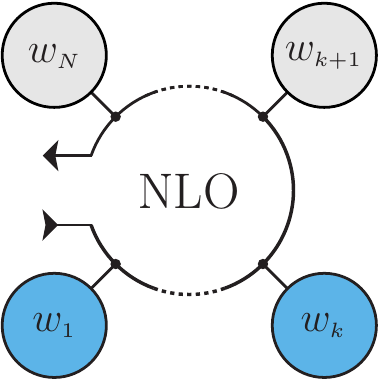}}
$$
 \caption{Schematic representation of the on-the-fly helicity
sums in~\refeq{eq:vOLdef}.  Taking the interference with the Born amplitude
makes it possible to sum over the helicities $\heli_1,\dots,\heli_k$ of the
first $k$ dressed segments of an open loop, while the remaining segments are still undressed.}
\label{fig:helbook} \end{figure*}

\subsection{Helicity treatment and reduction to scalar integrals}
\label{se:helicity+reduction}

In the following we discuss the operations that are required in order to
determine the contribution of a loop diagram $\calI_N$ to the scattering probability
density~\eqref{eq:W01}, starting form the output of the open-loop
recursion, \ie from an open-loop numerator~\refeq{eq:numfactorization} with
$k=N$ dressed segments.

Instead of proceeding via a direct construction 
of the one-loop amplitude $\calA_1(\calI_N,h)$ defined in~\refeq{eq:A1d}, 
we start with the associated colour structure 
$\calC(\calI_N)$ 
defined in~\refeq{eq:colfact},
and we proceed by building the colour-summed interference with 
the Born amplitude,
\bea
\calU_0(\calI_{N},\heli)
=2\left(\sum_{\col}\calM^*_0(h)\,\calC(\calI_N)\right)
\qquad
\forall\;h\in \hset,
\label{eq:initOL2}
\eea
combining it with the trace of the colour-stripped loop numerator,
\bea
\calU(\calI_N,q,\heli)=
\calU_0(\calI_{N},\heli)
\Tr\Big[\calN(\calI_{N},q,\heli)\Big],
\label{eq:colsumint}
\eea
and performing helicity sums,
\bea
\calU(\calI_N,q,0)
=\sum_{\heli}\calU(\calI_N,q,h).
\label{eq:unpolnum}
\eea
Here we use $h=0$ for the configuration
where all particles are unpolarised, in the sense that 
their helicities have been summed over.
The above operations are performed at the level of $q$-coefficients
in the tensorial representation~\refeq{eq:OLcoeff}, i.e.~in practice 
we compute
\bea
&&\calU_{\mu_1\dots\mu_r}(\calI_N,0)=
\sum_{\heli}
\calU_{\mu_1\dots\mu_r}(\calI_N,\heli)=\nonumber\\ &&=
\sum_{\heli}
\calU_0(\calI_N,\heli)\;
\Tr\Big[\calN_{\mu_1\dots\mu_r}(\calI_{N},\heli)\Big].
\label{eq:unpolnumcoeff}
\eea

After the summation over colours and helicities
it is possible to combine all diagrams with the same one-loop topology, \ie diagrams
of type
\bea
\calI_N^{\alpha_1\dots\alpha_N}=\{\calS_1^{\alpha_1},\dots,\calS_N^{\alpha_N}\}
\label{eq:topeqdiag}
\eea
with all possible combinations of segments,
\bea
\calS_i\equiv \big\{\calS_i^{\alpha_i}\big|\,\alpha_i=1,\dots,m_i\big\},
\label{eq:segcomb}
\eea
that have the same external legs $\calE_i$ and loop propagators $\Dbar{i}$
but different 
external subtrees $w_i^{\alpha_i}$ and/or loop vertices $v_i^{\alpha_i}$.
To filter out combinations of segments 
that are not allowed by the Feynman rules we introduce the 
tensor
\bea
\delta_{\alpha_1\ldots\alpha_N}=\begin{cases}
                                  1 & \quad\text{if $\calS_1^{\alpha_1},\ldots,\calS_N^{\alpha_N}$ form a}\\
                                    & \quad\text{valid one-loop diagram}\\[2mm]
                                  0 & \quad\text{else}\,.
                                 \end{cases}
\label{eq:deltatensor}
\eea
In this way, the full set of topologically equivalent one-loop diagrams can be defined 
as
\bea
\Omega_N =
\Big\{\calI_N^{\alpha_1\dots\alpha_N}\!\Big|
\alpha_i=1,\dots,m_i
\,\,\mbox{with}\,\,
\delta_{\alpha_1\dots\alpha_N}\neq 0
\Big\},
\label{eq:omegaN}
\eea
and their sum yields
\bea
&&\calV_{\mu_1\dots\mu_r}(\Omega_N,0)= \nonumber\\ &&
=\!\!\!\!\sum\limits_{\alpha_1\dots\alpha_N}\!\!
\sum\limits_{\heli}
\delta_{\alpha_1\dots \alpha_N}\,
\calU_{\mu_1\dots\mu_r}(\calI_N^{\alpha_1\dots\alpha_N},h).
\label{eq:fulldiagsum}
\eea

The contribution of the diagrams~\refeq{eq:omegaN} to the scattering 
probability~\eqref{eq:W01} reads 
\bea
&&\calW_{\oneloop}=\re\left\{\int\!\rd^D\!\momq \,\f{\calV(\Omega_N,q,0)}{\Dbar{0} \cdots\Dbar{N-1}}\right\}= \nonumber\\
&&=
\re\left\{\sum_{r=0}^R \calV_{\mu_1\dots\mu_r}(\Omega_N,0)\!\!\int\!\!\rd^D\!\momq \f{q^{\mu_1}\cdots q^{\mu_r}}{\Dbar{0} \cdots\Dbar{N-1}}\right\}.\qquad
\label{eq:W01b}
\eea
In {\sc OpenLoops\,1}, the
calculation of the coefficients~\refeq{eq:fulldiagsum} 
is entirely based on the open-loop approach,
but  the 
reduction of the loop integrals~\eqref{eq:W01b} to scalar integrals, as well as the numerical
evaluation of the latter, are performed by means of 
external libraries.

By default, {\sc OpenLoops\,1} adopts the tensor representation on
the rhs of~\eqref{eq:W01b} and computes the relevant tensor integrals with
the {\sc Collier} library~\cite{Denner:2016kdg}, 
which implements the reduction
techniques of~\cite{Denner:2002ii,Denner:2005nn} and the scalar integrals
of~\cite{Denner:2010tr}.
One of the powerful features of {\sc Collier} lies in sophisticated analytic
expansions~\cite{Denner:2005nn} that avoid dangerous numerical
instabilities in phase space regions with small Gram
determinants.

Alternatively, the reduction to scalar integrals is performed with {\sc
Cuttools}~\cite{Ossola:2007ax}, and scalar integrals are computed with {\sc
OneLoop}~\cite{vanHameren:2010cp}.  The {\sc Cuttools} program implements
the OPP reduction method~\cite{Ossola:2006us}, which is based on double,
triple and quadruple cuts of the integrand on the lhs of~\eqref{eq:W01}.
This requires a large number of evaluations of
$\calV(\Omega_N,q,0)$, and the high efficiency of the open-loop representation,
$
\calV(\Omega_N,q,0)=
\sum_{r=0}^R
\calV_{\mu_1\dots\mu_r}(\Omega_N,0)
\,q^{\mu_1}\cdots q^{\mu_r},
$
results in a dramatic boost of the OPP method.

Another key feature behind the high speed of the open-loop method is the fact
that, irrespectively of the reduction method, the time-consuming reduction
to scalar integrals is performed after summing over colour and helicity
degrees of freedom.

\section{Summing helicities and diagrams on-the-fly} \label{sec:OFhelmerging}

In this section we introduce a new technique that makes it possible to sum
helicities and to merge different one-loop diagrams on-the-fly, \ie after each
step of the open-loop dressing recursion.
Besides boosting the open-loop algorithm in a significant way, this approach
is also a key aspect of the on-the-fly reduction technique introduced in
the~\refse{sec:red}.

\subsection{On-the-fly helicity summation} 
\label{sec:heltreat}

In the original formulation of the open-loop method, helicity
sums~\refeq{eq:unpolnum} are performed at the end of the dressing recursion. 
This implies that the $k^\mathrm{th}$ dressing
step~\refeq{eq:OLrecfullpolya} needs to be performed for all helicity
configurations of the dressed segments $\calS_1,\dots,\calS_k$.
This feature, combined with the fact that the number of relevant helicity
states and the cost of a single dressing step scale exponentially with
$k$, result in a very rapid growth of the CPU cost of dressing operations
in the course of the open-loop recursion.
To avoid this negative trend, in this section we introduce a method 
that exploits the factorisation properties of the open-loop representation~\refeq{eq:numfactorizationN}
in a way that makes it possible to perform helicity sums 
{\it on-the-fly}, after the dressing of each new segment.

The idea, sketched in~\reffi{fig:helbook}, is that, upon taking the
interference of open loops with the Born amplitude, it is possible to sum over the helicities of all dressed segments,
irrespectively of the presence of still undressed segments.
To introduce the technical aspects of this approach, let us rewrite the
interference~\eqref{eq:unpolnum} between the colour-summed Born term
$\calU_0(\calI_N,\heli)$
and the one-loop numerator as
\bea
\calU(\calI_N,q,0)&=&\sum_\heli
\calU_0(\calI_N,\heli)
\Tr\big[\calN(\calI_N,q,\heli)\big]=\nonumber\\
&=&
\sum_\heli
\calU_0(\calI_N,\heli)
\Tr\big[\calN_k(\calI_N,q,\helihat_k) \nonumber\\&&\times\,\seg_{k+1}(q,\heli_{k+1})\cdots
\seg_{N}(q,\heli_N)\big].\quad
\label{eq:tracefact}
\eea
To take advantage of the factorisation of loop segments on the rhs
of~\refeq{eq:tracefact}, we then postpone the trace operation 
and generalise~\refeq{eq:tracefact} by defining
\bea
&&\calU_k(\calI_N,q,\helicheck_{k})
=
\sum_{\helihat_k}\;
\calU_0(\calI_{N},\heli)\,
\calN_k(\calI_{N},q,\helihat_{k})=
\nonumber\\
&&=
\hspace{-2mm}
\sum_{\heli_1,\dots,\heli_k}
\calU_0(\calI_{N},\heli)
\seg_1(q,\heli_1)
\cdots
\seg_k(q,\heli_k),
\label{eq:vOLdef}
\eea
where the interference with $\calU_0(\calI_N,\heli)$ is restricted to the
first $k$ dressed segments of the open loop, and the corresponding 
helicities, $\helihat_k=\heli_1+\dots+\heli_k$, are summed over.
As a result, the first $k$ segments become effectively unpolarised, 
and~\refeq{eq:vOLdef} depends only on the helicities of the
remaining $N-k$ undressed segments,
$\helicheck_k=\heli_{k+1}+\dots+\heli_N$.
Due to this dependence, which is induced by the fact that the 
Born term~\refeq{eq:initOL2} depends on the helicities 
$h=\helihat_k+\helicheck_k$ of all external particles~\refeq{eq:helidress}, 
the sums over $\heli_1,\dots,\heli_k$ in~\refeq{eq:vOLdef} 
cannot be entirely factorised. However, they can be cast in the
nested form
\bea
\calU_k(\mathcal{I}_{N},q,\helicheck_k) &=&
\sum_{\heli_k}\Bigg[\dots
\sum_{\heli_2}
\Bigg[\sum_{\heli_1}\;
\calU_0(\calI_{N},\heli)\seg_1(q,\heli_1)\Bigg]\nonumber\\ && \times\, 
\seg_2(q,\heli_2)
\cdots
\Bigg]\seg_{k}(q,\heli_{k}),
\label{eq:numfactorizationv}
\eea
which highlights the fact that each segment 
becomes effectively unpolarised after its dressing.

\begin{figure*}[t!]\hspace{0.08\textwidth}
\begin{minipage}{0.8\textwidth} 
\bea
&&\hspace{-7mm}\parbox[t]{4.8\defheight}{\includegraphics[height=\heightA]{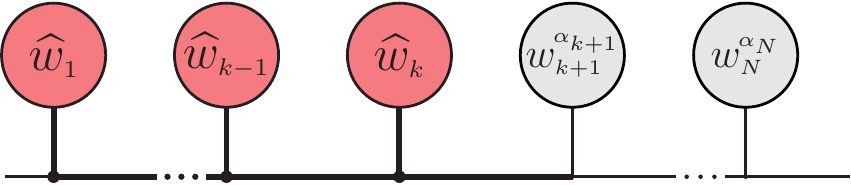}}= 
\nonumber\\[10mm]
&&=\sum\limits_{\alpha_1,\ldots,\alpha_k}
\sum_{\heli_1,\ldots,\heli_k}\;
\left(\;\;\parbox{2\defheight}{\includegraphics[height=\heightF]{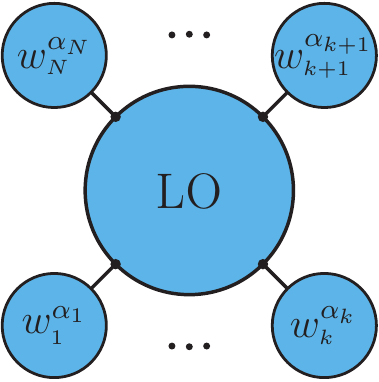}}\;\;\right)^{\hspace{-.5ex}\mathlarger{*}}
\;\,\times\;\;
\raisebox{0mm}{\parbox[t]{4.8\defheight}{\includegraphics[height=\heightA]{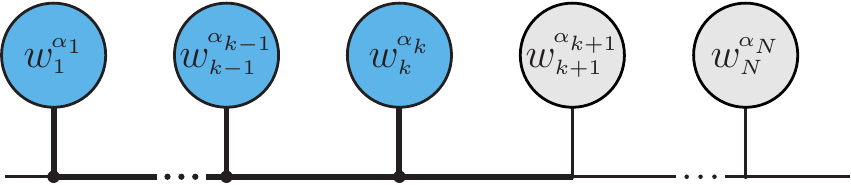}}}
\nonumber
\eea
\end{minipage} 
\caption{Diagrammatic representation of helicity sums and diagram merging.
Upon taking the interference with the 
Born amplitude, the helicities of the $k$ dressed segments
are summed, and the full set of topologically equivalent
diagrams with the same 
undressed segments $\{\calS_{k+1}^{\alpha_{k+1}},\dots, \calS_N^{\alpha_N}\}$
is merged in a single
open loop.
See \refeq{eq:mergingA} and~\refeq{eq:mergingB}.
}
\label{fig:merginggen}
\end{figure*}

In practice, in analogy with the standard open-loop
recursion~\refeq{eq:OLrecfullpolya}, the helicity-summed open
loops~\eqref{eq:vOLdef} are constructed with the recurrence relation
\bea
\calU_k(\calI_N,q,\helicheck_k) 
&=& 
\sum_{\heli_k}\;\calU_{k-1}(\calI_{N},q,\helicheck_{k-1}) \seg_k(q,\heli_{k}),
\label{eq:helsumrecursion} 
\eea
where the helicities $\heli_k\in\hset_k$ of the dressed segment are summed
on-the-fly.  To this end, the dressing operation needs to be performed for
all $\helicheck_{k-1}=\helicheck_{k}+\heli_{k}\in\hsetcheck_{k-1}$.
The initial condition reads
\bea
\calU_0(\calI_{N},q,\heli)
&=&
\calU_0(\calI_{N},\heli)\cdot\idop= \nonumber\\
&=&
2\,\sum_{\col}\calM^*_0(h)\,\calC(\calI_N)\cdot\idop,
\label{eq:initOL2b}
\eea
\ie a fully undressed open loop
is given by the interference of 
its colour structure with the Born amplitude, whose helicity states
$\heli$ live in the 
global helicity space $\hset$. 
At each dressing step,
helicity degrees of freedom are reduced by a factor equal to the number of
helicity states of the dressed segment, \ie by factor two for each external
fermion or massless vector boson and a factor three for each external massive
vector boson in the segment.

At the end of the recursion, when all $N$ segments are dressed,
no helicity dependence is left over ($\helicheck_{N}\equiv 0$),
and the
unpolarised loop numerator \eqref{eq:unpolnum}
is obtained by taking the trace
\bea
\calU(\calI_N,q,0)=\Tr\Big[\calU_N(\calI_{N},q,\helicheck_{N})\Big].
\label{eq:OLhelsumtrace}
\eea

The recursion~\refeq{eq:helsumrecursion} is understood as matrix multiplication, 
\bea
&&\big[\calU_k(\mathcal{I}_{N},q,\helicheck_k)\big]_{\beta_0}^{\beta_k}=\nonumber\\ &&=
\hspace{-1mm}
\sum_{\heli_k}
\big[\calU_{k-1}(\mathcal{I}_{N},q,\helicheck_{k-1})\big]_{\beta_0}^{\beta_{k-1}}\,
\big[\seg_{k}(q,\heli_{k})\big]_{\beta_{k-1}}^{\beta_k},\qquad
\label{eq:numfactorizationvbeta}
\eea
in the tensor representation
\bea
\calU_k(\mathcal{I}_{N},q,\helicheck_k)
&=&
\sum_{r=0}^R\;
\calU_{k;\,\mu_1\dots\mu_r}(\mathcal{I}_{N},\helicheck_k)\,
\,q^{\mu_1}\cdots q^{\mu_r},\qquad
\eea
which leads to the same tensorial recursion as in~\refeq{eq:OLrecfullcoeffa}.

In summary, performing helicity sums on-the-fly leads to a decreasing number
of helicity degrees of freedom when the number $k$ of dressed segments
increases.  In this way, the effect of the growing CPU cost of dressing
operations at large $k$ can be strongly attenuated.
The price to pay is that the parent--child
approach~\refeq{eq:olmerginidA}--\refeq{eq:parent-childB} is not
applicable anymore, due to the fact that~\refeq{eq:initOL2b} incorporates the colour
structure $\calC(\calI_N)$ of the whole one-loop diagram.  
However, as we will see in \refse{sec:merging}, the parent--child relations
can be replaced by a
similarly efficient method based on the merging of topologically equivalent
one-loop diagrams.
Finally, let us note that the
recursion~\refeq{eq:helsumrecursion}--\refeq{eq:initOL2b} is not applicable
to squared one-loop amplitudes.  For this case we still rely on the original
open-loop algorithm.

\subsection{On-the-fly merging of topologically equivalent open loops}
\label{sec:merging}

The key idea behind the recursion
\refeq{eq:helsumrecursion}--\refeq{eq:initOL2b} is that, taking 
the interference between the Born amplitude and the one-loop colour
structure $\calC(\calI_N)$ as initial condition makes it possible to anticipate operations that
are usually performed after completion of the construction of a one-loop diagram.
In particular, such operations become applicable {\it on-the-fly} after the dressing of
individual loop segments.  This technique will be denoted as on-the-fly
approach, and its applicability goes well beyond helicity sums.

As sketched in \reffi{fig:merginggen}, the on-the-fly technique can be
extended to the double sums over helicity states and
topologically equivalent loop diagrams in~\refeq{eq:fulldiagsum}.
The idea is that, rather than being constructed one by one, the
topologically equivalent diagrams
\bea
\calI_N^{\alpha_1\dots\alpha_N}=\{\calS_1^{\alpha_1},\dots,\calS_N^{\alpha_N}\}\in \Omega_N,
\eea
defined in~\refeq{eq:topeqdiag}--\refeq{eq:omegaN},
can be merged in a
recursive way by summing over the various subtrees $\calS_i^{\alpha_i}$ as
soon as they get dressed. 

\begin{figure*}[t!] \hspace{0.08\textwidth}
\begin{minipage}{0.8\textwidth}
\bea
\parbox[t]{3.8\defheight}{\includegraphics[height=\heightA]{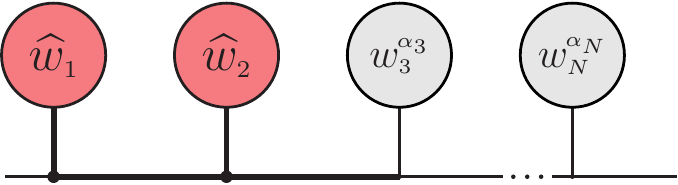}} 
&=&
\sum_{\heli_2}\;
\parbox[t]{4.2\defheight}{\includegraphics[height=\heightA]{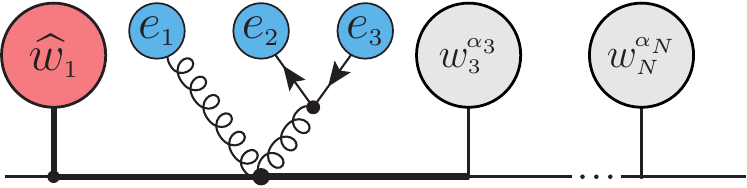}}
\nonumber\\[4mm]
&+&
\sum_{\heli_2}\;
\parbox[t]{4.2\defheight}{\includegraphics[height=\heightA]{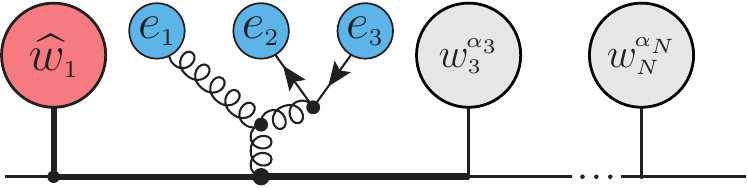}} 
+\dots
\nonumber
\eea
\end{minipage} 
\caption{
Example of a step 
of the diagram merging recursion~\refeq{eq:mergingD}.
After dressing the second segment
and summing over its helicity
configurations $\heli_2$,
all diagrams with equivalent one-loop topology
and identical undressed segments
$\calS_3^{\alpha_3},\dots, \calS_N^{\alpha_N}$  
are merged into a single open loop. 
}
\label{fig:mergingex}
\end{figure*}

To this end, let us define subsets of diagrams, $\Omega_N^k\subset\Omega_N$,
that share the same undressed segments, 
$\{\calS_{k+1}^{\alpha_{k+1}},\dots, \calS_N^{\alpha_N}\}$, after $k$ dressing steps,
\bea
\Omega^k_N &\equiv& \Omega^k_N(\alpha_{k+1},\dots,\alpha_N) = \nonumber\\
&=&\Big\{\calI_N^{\rho_1\dots\rho_k\alpha_{k+1}\dots\alpha_N}\;\Big|\;
1\le \rho_j\le m_j \nonumber\\ &&
\;\;\mbox{with}\;\;
\delta_{\rho_1\dots\rho_k\alpha_{k+1}\dots\alpha_N}\neq 0
\Big\},
\label{eq:mergingA}
\eea
where the tensor $\delta$, defined in \refeq{eq:deltatensor}, filters out one-loop
diagrams that are not allowed by the Feynman rules.  By construction, all
diagrams in the set~\refeq{eq:mergingA} must undergo identical future
dressing steps, which
can be performed only once after merging the first $k$ segments.  This operation can be organised in a very similar way
as helicity summations in \refse{sec:heltreat}.
Technically, taking  as a starting point the nested helicity
sums in~\refeq{eq:numfactorizationv}, it is sufficient to generalise the
loop segments and the Born term~\refeq{eq:initOL2b} by replacing
\bea
S_i(q,\heli_i)
&\to& 
S^{\alpha_i}_i(q,h_i),\nonumber\\
\calU_0(\calI_{N},\heli) 
&\to&
\delta_{\alpha_1\dots \alpha_N}\,
\calU_0(\calI_{N}^{\alpha_1\dots\alpha_N},\heli),
\label{eq:initOL5}
\eea
and to extend the summation over the helicities $\heli_1,\dots,\heli_k$ of
the dressed segments to the corresponding ``diagrammatic'' degrees of freedom
$\alpha_1,\dots,\alpha_k$.  This leads to the identity
\bea
&&\calV^{\alpha_{k+1}\dots\alpha_N}_k(\Omega^k_N ,q,\helicheck_{k})=
\nonumber\\
&&=\hspace{-2mm}
\sum_{\alpha_k,\heli_k}\Bigg[
\dots
\sum_{\alpha_2,\heli_2}\Bigg[
\sum_{\alpha_1,\heli_1}
\delta_{\alpha_1\dots
 \alpha_N}\,
\calU_0(\calI_{N}^{\alpha_1\dots
\alpha_N},\heli) \nonumber\\ &&\phantom{=}\times\,
\seg_1^{\alpha_1}(q,\heli_1)\Bigg]
\seg_2^{\alpha_2}(q,\heli_2)\Bigg]
\cdots\Bigg]
\seg_k^{\alpha_{k}}(q,\heli_k),
\label{eq:mergingB}
\eea
which defines an open-loop object with
fixed undressed segments $\{\calS_{k+1}^{\alpha_{k+1}},\dots
\calS_N^{\alpha_N}\}$ and helicities
$\helicheck_k=\heli_{k+1}+\dots+\heli_N$ that 
incorporates  all possible chains of dressed segments
$\{\calS_{1}^{\alpha_1},\dots, \calS_k^{\alpha_k}\}$
forming a valid Feynman
diagram, summed over the corresponding 
helicities
$\heli_1,\dots,\heli_k$.

Note that the dependence of~\refeq{eq:mergingB} on the helicities
$\heli_{k+1},\dots,\heli_N$ and 
diagrammatic indices $\alpha_{k+1},\dots,\alpha_N$ of the undressed segments
is due to the fact that the Born term defined in~\refeq{eq:initOL2b} and~\refeq{eq:initOL5}
retains the full helicity dependence of the Born amplitude as well as the 
tensor~\refeq{eq:deltatensor} 
and the colour structure of the whole one-loop diagram.

In analogy with~\refeq{eq:OLrecfullpolya} and~\refeq{eq:helsumrecursion},
the open-loop objects~\refeq{eq:mergingB} can be constructed with the 
recurrence relation
\bea
&&\calV^{\alpha_{k+1}\dots\alpha_N}_k(\Omega_N^k,q,\helicheck_{k})=\nonumber\\
&&=
\sum_{\alpha_{k}}
\sum_{\heli_{k}}
\calV^{\alpha_{k}\dots\alpha_N}_{k-1}(\Omega_N^{k-1},q,\helicheck_{k-1})
\seg_{k}^{\alpha_{k}}(q,\heli_{k}),\qquad
\label{eq:mergingD}
\eea
where helicity sums and diagram merging are performed on-the-fly.  An
explicit example of an on-the-fly merging step is illustrated in
\reffi{fig:mergingex}.  Similarly as for~\refeq{eq:helsumrecursion}, the
recursion~\refeq{eq:mergingD} is implemented in the form of tensorial
relations~\refeq{eq:OLrecfullcoeffa}.
The relevant initial conditions at $k=0$ are 
\bea
&&\calV^{\alpha_{1}\dots\alpha_N}_0(\Omega_N^0,h)
=
\delta_{\alpha_1\dots \alpha_N}\,
\calU_0(\calI_{N}^{\alpha_1\dots\alpha_N},\heli)
\cdot\idop      =       \nonumber\\
&&=
2\, \delta_{\alpha_1\dots \alpha_N}\,
\sum_{\col}\calM_0(h)^*\calC(\calI^{\alpha_{1}\dots\alpha_N})
\cdot\idop,
\label{eq:mergingE}
\eea
where each fully undressed contribution corresponds to an individual diagram,
\bea
\Omega^0_N \equiv \Omega^0_N(\alpha_{1},\dots,\alpha_N) 
=\big\{\calI_N^{\alpha_1\dots\alpha_N}\big\},
\eea
with helicities $\heli$ that live in the 
full helicity space $\hset$. 
Let us point out that, thanks to the absorption of the colour factors
$\calC(\calI_N^{\alpha_1\dots\alpha_N})$ in the Born interference
term~\refeq{eq:mergingE}, in~\refeq{eq:mergingD} it is possible to merge
parts of diagrams that carry different colour structures in a single object,
while respecting the exact colour dependence.

After $N$ recursion steps one obtains a single open-loop object
$\calV_N(\Omega^N_N,q,\helicheck_N)$, which merges the full set of
topologically equivalent diagrams ($\Omega^N_N\equiv\Omega_N$) and is
entirely unpolarised ($\helicheck_N\equiv 0$). At this stage,
taking the trace that closes the loop one arrives at
\bea
\calV(\Omega_N,q,0)
&=&
\Tr\left[\calV_N(\Omega^N_N,q,\helicheck_N)\right],
\label{eq:mergingtrace}
\eea
which is equivalent to~\refeq{eq:fulldiagsum}.

\begin{figure*}[t!]\begin{center} \begin{tabular}{ccc}
\includegraphics[height=66mm]{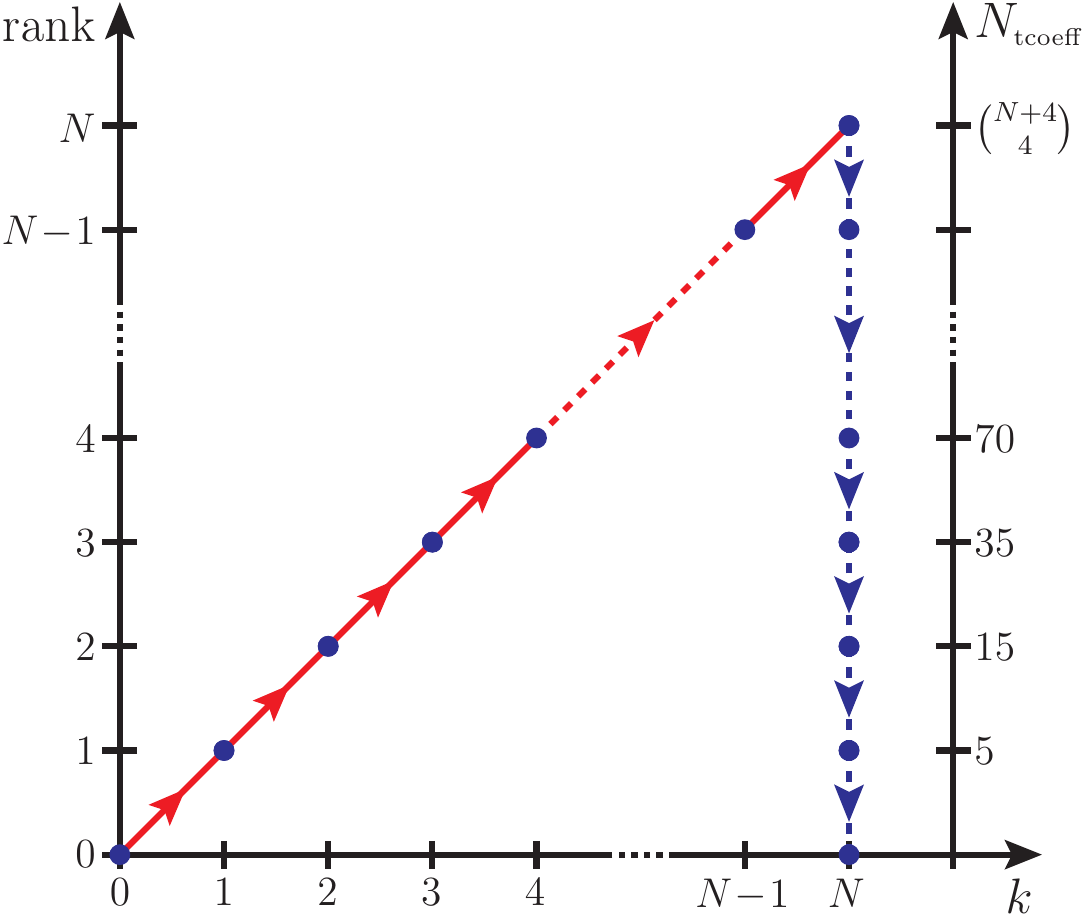} & \quad\; &
\includegraphics[height=66mm]{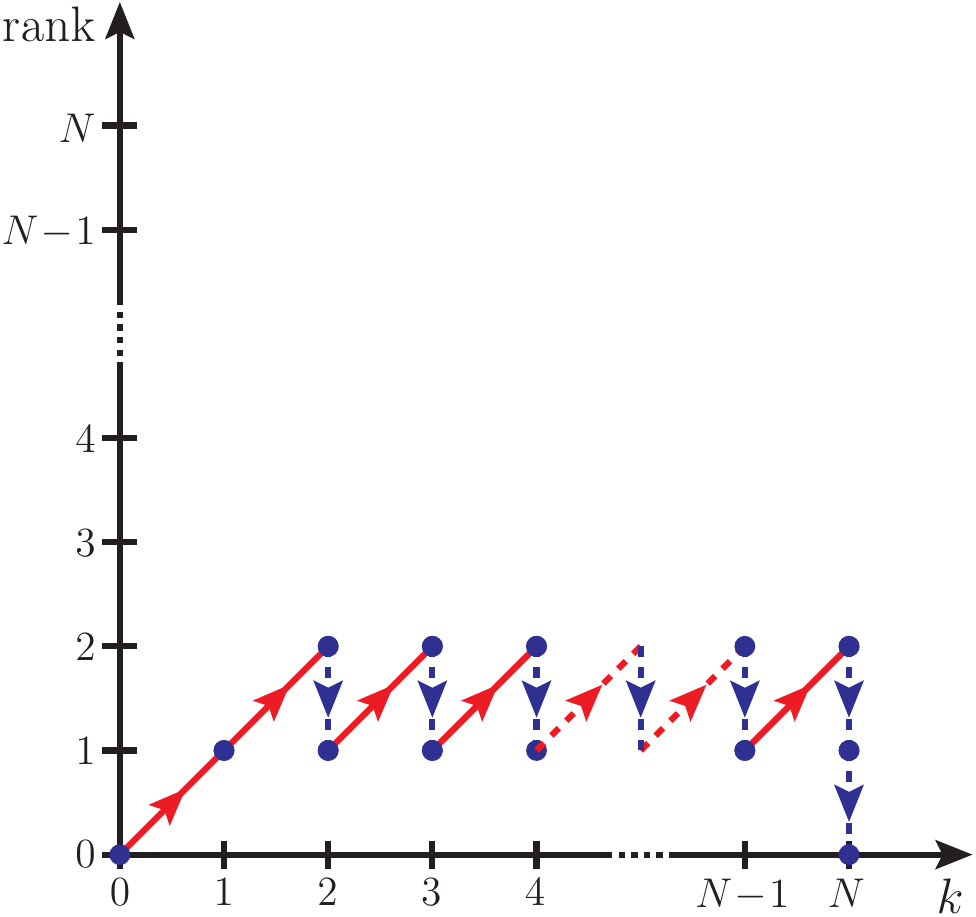}\\[3ex] (a) A posteriori reduction & & (b) On-the-fly reduction
 \end{tabular} \end{center}
\caption{Evolution of the tensor rank $R$ and the number
$N_{\mathrm{tcoeff}}\,(R)=\binom{R+4}{4}$ of open-loop tensor coefficients
(right vertical axis) as a function of the number $k$ of dressed segments
during the open-loop recursion.  Each dressing step is assumed to increase
the rank by one.  The original open-loop algorithm, where tensor reduction
is applied a posteriori (left), is compared to the on-the-fly reduction
approach (right).  The red diagonal lines illustrate the dressing steps
 and the blue vertical lines the reduction steps.  
\label{fig:OL1OL2_r_vs_n} } 
\end{figure*}

As demonstrated in \refse{sec:stab}, performing helicity sums and merging
diagrams on-the-fly yields a very significant efficiency improvement with respect
to the original open-loop algorithm.
More precisely, if helicity sums are performed at the end of the recursion
as in \refeq{eq:unpolnum}, the merging approach and the parent--child
relations~\refeq{eq:parent-childB} permit to achieve a similar speed-up
factor of the order of two.
However, contrary to the parent--child technique, the on-the-the-fly
approach is applicable both to diagram merging and helicity sums.  This
leads to a further speed-up factor that can vary from two to three,
depending on the process.

As we will see in \refse{sec:red}, the on-the-fly approach will be a crucial
ingredient in order to arrive at a new efficient algorithm that combines the
operations of open-loop dressing and tensor reduction at the level of each
individual step of the open-loop recursion.

\section{On-the-fly reduction of open loops} \label{sec:red}

In the original version of the {\sc OpenLoops} program, the construction of
integrand numerators and the reduction to scalar integrals are performed
independently of one another using different tools.  Open-loop numerators of
$N$-point diagrams are constructed by recursively dressing $N$ segments as
described in \refse{sec:oneloopamp}.
Each step of the recursion can increase the tensor rank by one, and, upon
symmetrisation of all $q^{\mu_1}\ldots q^{\mu_r}$ monomials with $r\le R$,
open-loop polynomials of rank $R$ involve $\binom{R+4}{4}$ independent
tensor coefficients.
Thus their complexity grows exponentially with the number of recursion
steps.  For instance, open loops with $R=6$ and $R=7$ involve, respectively,
$210$ and $330$ components, while only $5$ components are present for $R=1$.
As illustrated in the left plot of \reffi{fig:OL1OL2_r_vs_n}, in the
original open-loop algorithm tensorial complexity keeps growing until the
maximum rank $R\le N$ is reached at the end of the dressing recursion.
At this stage, upon summation of helicity configurations and loop diagrams
with equivalent one-loop topology, tensor integrals are reduced to scalar
integrals using external libraries, such as {\sc Collier}~\cite{Denner:2014gla}
or {\sc Cuttools}~\cite{Ossola:2007ax}, as described in
\refse{se:helicity+reduction}.

Dealing with intermediate results with a large number of tensor components
requires a considerable amount of computing power, both for the reduction of
high-rank objects and at the level of the tensorial structure of the
open-loop recursion~\refeq{eq:OLrecfullcoeffa}, which needs to be performed
for each relevant helicity configuration and each 
$[\dots]_{\beta_0}^{\beta_k}$
component.  These operations can be significantly
accelerated by means of the techniques introduced
in~\refse{sec:OFhelmerging}. Nevertheless, they remain the most CPU intensive
aspect of multi-particle one-loop calculations in {\sc OpenLoops}.

Motivated by the above considerations, in this section we introduce a new
approach that avoids the appearance of high-rank objects at any stage of the
calculation.  This is achieved by extending the on-the-fly approach
introduced in \refse{sec:OFhelmerging} to the reduction of open loops.
In this way, interleaving the operations of open-loop dressing and tensor
reduction, we build a single recursive algorithm, where each increase of
tensorial rank caused by a dressing step is compensated by an
integrand-reduction step.

\begin{figure*}[t!]
\hspace{0.09\textwidth}
\begin{minipage}{0.8\textwidth}
\begin{eqnarray} 
\showredpinchdiag{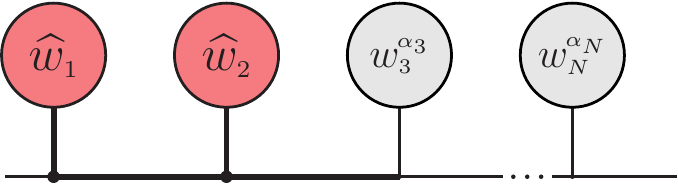}{\heightD}{\hspace{14mm}$\calV_2(\Omega_N^2)$}
&=&
\showredpinchdiag{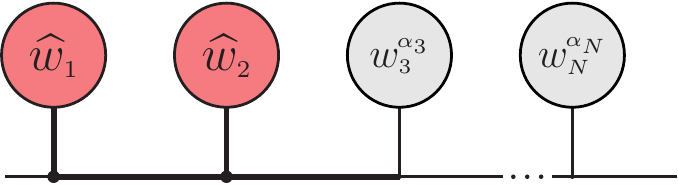}{\heightD}{\hspace{10mm}$\calVpinch{2}{-1}$}
\nonumber\\[4mm]
+\;\;
\showredpinchdiag{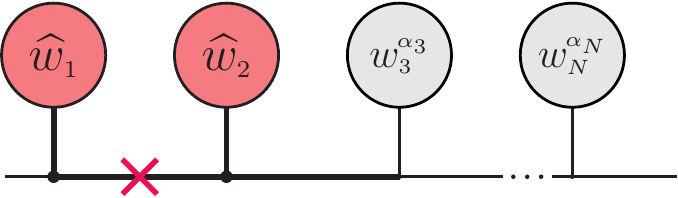}{\heightI}{\hspace{10mm}$\calVpinch{2}{1}$}
&+&
\showredpinchdiag{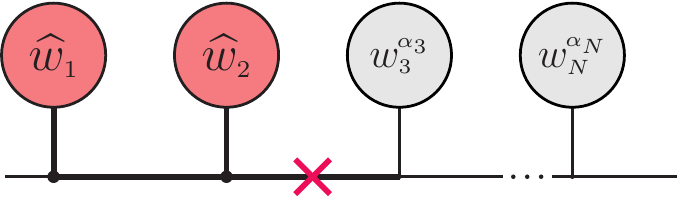}{\heightI}{\hspace{10mm}$\calVpinch{2}{2}$}
\nonumber\\[4mm]
+\;\;\showredpinchdiag{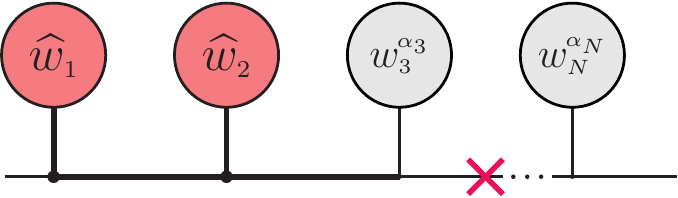}{\heightI}{\hspace{10mm}$\calVpinch{2}{3}$}
&+&\showredpinchdiag{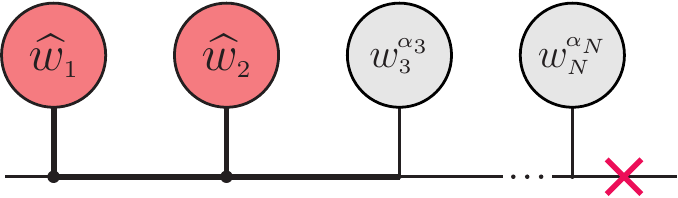}{\heightI}{\hspace{10mm}$\calVpinch{2}{0}$}
\nonumber
\end{eqnarray}\end{minipage}
\caption{Diagrammatic representation of the on-the-fly reduction
step~\refeq{eq:OFRmasterformula} for an $N$-point open loop at step $k=2$ of
the dressing recursion.  The symbols $\calV_k(\Omega^k_N)$ and
$\calV_k(\Omega^k_N[j])$ correspond, respectively, to the
rank-two polynomial on the lhs of~\refeq{eq:OFRmasterformula} and its
reduced rank-one counterparts on the rhs.  The red crosses indicate the
pinching of the $\Dbar{j}$ propagators in the $\calV_k(\Omega^k_N[j])$ terms
with $j=0,\dots,3$.  Since $\Dbar{0}=\Dbar{N}$, in our graphical
representation the $\Dbar{0}$ denominator is located on 
the $N^\mathrm{th}$ segment.
\label{fig:OLred}}
\end{figure*}

As illustrated in the right plot of \reffi{fig:OL1OL2_r_vs_n}, the
on-the-fly reduction approach avoids the appearance of any
intermediate object with rank higher than two.  Besides the CPU cost
needed for the processing of high-rank objects, this alleviates also possible memory
issues due to their storage.

\subsection{On-the-fly integrand reduction}
\label{se:OFR}

For the on-the-fly reduction of open-loop polynomials we are going to use the
method of~\cite{delAguila:2004nf},
which permits to reduce rank-two monomials of the loop momentum 
through identities of the form
\bea q^\mu q^\nu
 &=& \left[ A^{\mu\nu}_{\np} + A^{\mu\nu}_{0} D_0  \right] 
+ \left[ B^{\mu\nu}_{-1,\lambda}+
 \sum\limits_{j=0}^{3} B^{\mu\nu}_{j,\lambda} D_j \right]q^{\lambda}{}.
\nonumber\\
\label{eq:qmuqnuredfinal}
\eea
The rank-one polynomial on the rhs is a linear combination of four loop
denominators, $D_0,\dots,D_3$, and the corresponding tensor coefficients,
$A_j^{\mu\nu}$ and $B_{j,\lambda}^{\mu\nu}$, depend only on the three
external momenta $p_1,p_2,p_3$.
The coefficients of loop denominators are labeled with indices
$j=0,\dots,3$, while $j=-1$ is used for the constant parts.
Their explicit expressions are presented in \refse{se:redform4}.

The identity~\refeq{eq:qmuqnuredfinal} provides an exact reconstruction of
$q^\mu q^\nu$ in terms of four-dimensional loop
denominators, but can be easily generalised to $D$-dimensional denominators
by replacing
\bea
D_j\to \Dbar{j}-\tilq^2\quad\mbox{for}\quad j=0,1,2,3\,.
\label{eq:DiDibar}
\eea
Note that $\tilde{q}^2$ contributions resulting from 
the terms $B^{\mu\nu}_{j,\lambda} D_j$ with $j=0,1,2,3$ must cancel among each other 
in~\refeq{eq:qmuqnuredfinal} since they generate rank-three
terms of type $q^\lambda\,\tilde{q}^2$ that are
not consistent with the rank-two structure on the lhs.
Thus the substitutions~\refeq{eq:DiDibar} generate only an extra term 
$-\tilq^2 A_0^{\mu\nu}$ on the rhs of~\refeq{eq:qmuqnuredfinal}.

The integrand reduction~\refeq{eq:qmuqnuredfinal} holds at the
integrand level, irrespectively of the presence of extra loop denominators
$D_4,\dots,D_{N-1}$ or additional $q$-dependent factors that may multiply
the $q^\mu q^\nu$ monomial.  These properties, in combination with the
factorisation of open loops into segments,
make it possible to apply the reduction~\refeq{eq:qmuqnuredfinal}
at any intermediate stage of the recursion~\refeq{eq:mergingD}. 
This on-the-fly reduction approach is illustrated in \reffi{fig:OLred}, and
the corresponding reduction identities for $N$-point integrands at step $k$
of the dressing recursion have the form
\bea
&&\frac{\calV_{k}(\Omega^k_N,\bar q)\;\seg_{k+1}(q)\cdots\seg_{N}(q)}{\Dbar{0}\cdots \Dbar{3} \cdots \Dbar{N-1}}=\nonumber\\
&&=
\sum_{j=-1}^3
\frac{\calV_{k}(\Omega^k_N[j],\bar q)\;\seg_{k+1}(q)\cdots\seg_{N}(q)}{\Dbar{0}\cdots 
\slashed{\Dbar{j}}\cdots
\Dbar{3} \cdots \Dbar{N-1}},
\label{eq:OFRmasterformula}
\eea
where $\Omega^k_N[j]$ for $0\le j\le 3$ denote the $(N-1)$-point
subtopologies that arise from $\Omega^k_N$ by pinching the $\Dbar{j}$
propagator, while terms with $j=-1$ on the rhs correspond to the original
topology, $\Omega^k_N[-1]=\Omega^k_N$.  Note that the denominators
$\Dbar{j}$ can be pinched irrespectively of whether the related
$\seg_{j}(q)$ segments are already dressed or not.
In~\refeq{eq:OFRmasterformula} we adopt the approach of~\refse{sec:merging},
where open-loop polynomials incorporate the colour-summed interference with
the Born amplitude as well as helicity sums and merging of all dressed
segments.  However, for simplicity, the bookkeeping of helicities, merged
diagrams, and $[\dots]_{\beta_0}^{\beta_k}$ indices is kept implicit.

The partially dressed open loops on the lhs and rhs of
\refeq{eq:OFRmasterformula} have the general form
\bea
\calV_k(\Omega,\bar q)=
\sum_{s=0}^{S}
\sum_{r=0}^{R}
\calV^s_{k;\mu_1\dots\mu_r}(\Omega)\,
q^{\mu_1}\cdots q^{\mu_r}\,
\tilde{q}^{2s},\quad
\label{eq:qtildepolynomials}
\eea
where four-dimensional loop-momentum components are accompanied by
$\tilde{q}^2$ terms that arise from
\refeq{eq:qmuqnuredfinal}-\refeq{eq:DiDibar}.
As discussed in \refse{sec:rationalterms}, only a small fraction of the
$\tilq^2$-dependent terms can lead to non-vanishing contributions at the
end of the recursion.  Thus, in order to avoid the proliferation of tensor
coefficients, all $\tilq^2$ terms that are expected to vanish are identified
and discarded in advance at each dressing and reduction step.

In general, the relation \refeq{eq:OFRmasterformula} allows one to reduce
any polynomial $\calV_{k}(\Omega^k_N,\bar q)$ of rank $R\ge 2$ to rank $R-1$
polynomials $\calV_{k}(\Omega^k_N[j],\bar q)$.
But, in practice, the reduction~\refeq{eq:OFRmasterformula} can be interleaved
with the open-loop dressing recursion in a way that the tensor rank 
never exceeds two.  For $R=2$ the coefficients of the rank-one open-loop
polynomials that arise from the reduction~\refeq{eq:OFRmasterformula} read
\bea
\calV^s_{k;}(\Omega^k_N[j])&=&
\calV^s_{k;\nu_1\nu_2}(\Omega^k_N)
A_{j}^{\nu_1\nu_2}
+\delta_{-1j}
\left[
\calV^s_{k;}(\Omega^k_N)\right.
\nonumber\\
&&- \left. 
\calV^{s-1}_{k;\nu_1\nu_2}(\Omega^k_N)
A_{0}^{\nu_1\nu_2}
\right],
\nonumber\\
\calV^s_{k;\mu_1}(\Omega^k_N[j]) &=&
\calV^s_{k;\nu_1\nu_2}(\Omega^k_N) B_{j,\mu_1}^{\nu_1\nu_2}
+\delta_{-1j}\calV^s_{k;\mu_1}(\Omega^k_N).\nonumber\\
\label{eq:OFRmasterformulaB}
\eea
The transformations~\refeq{eq:OFRmasterformula}--\refeq{eq:OFRmasterformulaB}
can be used to reduce any rank-two open loop with $N\ge 4$ propagators to a
rank-one $N$-point object and four $(N-1)$-point pinched objects of rank
one.

Rank-two open loops with only $N=3$ loop propagators can be reduced to rank
one in a very similar way~\cite{delAguila:2004nf}.  The relevant identities
(see \refse{se:redform3}) have the same structure as
\refeq{eq:qmuqnuredfinal} but involve only three reconstructed propagators,
$\Dbar{0}$, $\Dbar{1}$ and $\Dbar{2}$. 
Moreover, they do not hold at the integrand level, but only upon integration over the
loop momentum.
The tensors $A_j^{\mu\nu}$ and $B_{j,\lambda}^{\mu\nu}$ for 
the case $N=3$ depend only on $p_1$ and $p_2$. They are obtained 
from the ones for $N=4$
by simply setting to zero the terms involving $\momp{3}$
(see \refse{se:redform3}).

The on-the-fly reduction formula
for $N=3$ has the form
\bea
&&\int\!\mathrm{d}^D\!\momq\,
\frac{\calV_{k}(\Omega^k_3,\bar q)\;\seg_{\rem}(q)}{\Dbar{0}\cdots \Dbar{2}}=\nonumber\\
&&=
\sum_{j=-1}^2
\int\!\mathrm{d}^D\!\momq\,
\frac{\calV_{k}(\Omega^k_3[j],\bar q)\;\seg_{\rem}(q)}{\Dbar{0}\cdots 
\slashed{\Dbar{j}}\cdots
\Dbar{2}},
\label{eq:qmuqnuredfinalB}
\eea
where $\calV_{k}(\Omega^k_3,\bar q)$ is an open loop of rank $R=2$ that
results from a certain number $k\ge 2$ of dressing steps and $k-2$ or less
reduction steps.  Possible undressed segments are denoted as
$\seg_{\rem}(q)$, and \refeq{eq:qmuqnuredfinalB} is valid only for terms
$\seg_{\rem}(q)$ of rank zero or one.  In general this allows for only
$N_{\rem}\le 1$ extra segments, \ie
\bea
\seg_\rem(q)= 
\begin{cases}
\;\seg_{k+1}(q) & \mbox{for}\; N_{\rem}=1 \\[2mm]
\;1 &\mbox{for}\; N_{\rem}=0
\end{cases}
\;\;,
\label{eq:remsegment}
\eea
and these two cases are sufficient to cover all relevant $N=3$ topologies
and pinched subtopologies in renormalisable theories.\footnote{The fact that
$N_{\rem}\le 1$ follows from the inequality
$
N_{\rem} + R - N_{\unpinched}\le 0,
$
where $R=2$ is the rank of $\calV_{k}(\Omega^k_3,\bar q)$, and
$N_{\unpinched}=3$ is the number of unpinched propagators in
\refeq{eq:qmuqnuredfinalB}.  The 
above inequality
is obviously fulfilled at the beginning of the open-loop recursion, where
$k=R=N_{\rem}-N_{\unpinched}=0$, and its validity is guaranteed by the fact
that,  in renormalisable theories, dressing and reduction steps cannot increase $N_{\rem} + R-
N_{\unpinched}$.}
The relations between the rank-two polynomial $\calV_{k}(\Omega^k_3,\bar q)$
and its reduced counterparts $\calV_{k}(\Omega^k_3[j],\bar q)$ have the same
form as in~\refeq{eq:OFRmasterformulaB}.

Although the above three-point reduction holds only at the integral level,
the fact that the term $\seg_{\rem}(q)$ can be factorised makes it possible
to apply~\refeq{eq:qmuqnuredfinalB} as soon as the dressed open loop
$\calV_{k}(\Omega^k_3,\bar q)$ has reached rank two, independently of the
remaining part of the numerator.

In summary, exploiting the fact that open loops are factorised into segments,
the identities~\refeq{eq:qmuqnuredfinal} and~\refeq{eq:qmuqnuredfinalB}
can be applied on-the-fly during the dressing recursion, while
an arbitrary number of segments is still undressed.\footnote{%
Note that the fact that the technique of~\cite{delAguila:2004nf} can be
applied on-the-fly is not a general feature of integrand
reduction methods.  For instance, the OPP method~\cite{Ossola:2006us} is not applicable
on-the-fly since it is based on the multiple poles of the entire
integrand.}
Thus, dressing and reduction steps can be interleaved in a way that the
increase of tensorial rank resulting from dressing is promptly compensated
through an on-the-fly reduction.
More precisely, on-the-fly reduction steps are applied to diagrams and
pinched sub-diagrams with $N\ge 3$ at those stages of the 
recursion where the next dressing step would generate a rank-three
object.\footnote{%
This means that, before performing an on-the-fly reduction,
rank-two open loops are first dressed with all possible adjacent segments of
rank zero.  Delaying the reduction step in this way reduces the additional 
CPU cost that results from the appearance of new pinched objects.}
The reduced rank-one objects with $N$ and $N-1$ loop propagators are further
dressed and possibly reduced until one arrives at fully dressed open loops of rank $R\le 2$
for all two- and higher-point contributions. 
An this stage, the open-loop matrix structure can be eliminated by taking
the trace~\refeq{eq:mergingtrace}, and all rank-two objects with $N\ge 3$ can be
reduced to rank one with a final reduction step of
type~\refeq{eq:qmuqnuredfinal} or~\refeq{eq:qmuqnuredfinalB}.

After the above dressing and on-the-fly reduction steps, the
following types of integrals remain to be reduced:
\bit 
\item[(i)] integrals with $N\ge 5$ loop propagators\\ and rank $R=1,0$; 

\item[(ii)] integrals with $N=4,3$ loop propagators\\ and rank $R=1$;\eqnum{list:TIs}

\item[(iii)] integrals with $N=2$ loop propagators\\ and rank $R=2,1$.
\eit 
For their reduction to scalar integrals with $N\le 4$ 
we use a combination of integral reduction and OPP reduction 
identities as described in \ref{app:intred}.

\subsection{Merging pinched topologies}
\label{sec:red_merging}

Since it allows one to keep the tensor rank low at any stage of the
calculations, the on-the-fly reduction approach has the potential to
accelerate one-loop calculations in a significant way.  However, some
aspects of the on-the-fly reduction approach could lead to a dramatic
increase of the computational cost.
First, the fact that the reduction is performed when the loop is
still open, \ie before taking the trace~\refeq{eq:mergingtrace}, implies
that the entries of the 
$[\dots]_{\beta_0}^{\beta_k}$
matrix have to be processed as 16 independent objects.
Second, the reduction has to be performed for all not yet summed helicity
configurations of the undressed segments.  
Third, each reduction step~\refeq{eq:OFRmasterformula} generates four
pinched topologies that need to be processed as independent contributions in
subsequent dressing and reduction steps.

Due to the proliferation of pinched subtopologies, the
naive iteration of on-the-fly reduction steps 
would lead to a dramatic increase of the CPU cost.
Fortunately, this can be avoided by means of
the merging technique introduced in \refse{sec:merging}, 
which makes it possible to absorb
pinched $N$-point open loops
into unpinched $(N-1)$-point open loops, in such a way that
they do not need to be processed as separate objects.
As explained in the following, the merging of pinched subtopologies requires
a different implementation depending on whether the pinch belongs
to the dressed part of the open loop or not, as well as for 
the special case of a $\Dbar{0}$ pinch. 

\begin{figure*}[t!]\hspace{0.09\textwidth}
\begin{minipage}{0.8\textwidth}
\begin{eqnarray} 
&&\raisebox{-5mm}{\parbox[t]{3.15\defheight}{\includegraphics[height=\heightB]{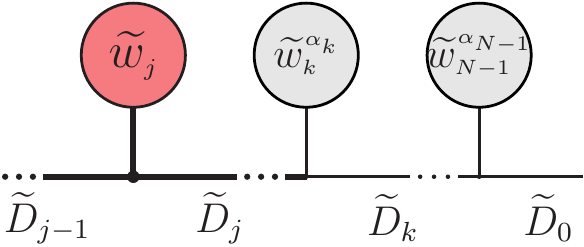}}}\quad=\quad
{\sum_{\scriptstyle{\alpha_j,\alpha_{j+1}}}}  \; 
\raisebox{-5mm}{\parbox[t]{3.5\defheight}{\includegraphics[height=\heightB]{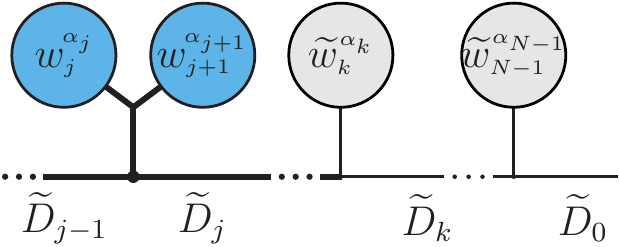}}}
\nonumber\\[6mm]
&&+
{\sum_{\scriptstyle{\alpha_j,\alpha_{j+1}}}}  \; 
\raisebox{-5mm}{\parbox[t]{3.5\defheight}{\includegraphics[height=\heightB]{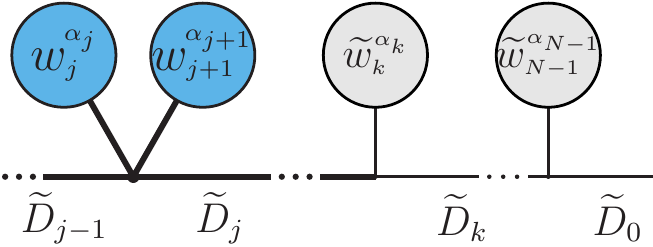}}}
\quad+\quad
\raisebox{-5mm}{\parbox[t]{3.7\defheight}{\includegraphics[height=\heightB]{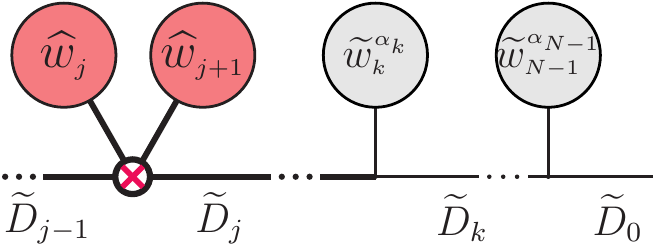}}}
\nonumber
\end{eqnarray}\end{minipage}
\caption{Diagrammatic representation of formula~\refeq{eq:OFRmergingA}
for the merging of pinched open loops.
The first two unpinched diagrams on the rhs,
where the generic subtrees
$w_j^{\alpha_j}$ and $w_{j+1}^{\alpha_{j+1}}$ are directly connected to the
loop propagators $\Dtilde{j-1}$ and $\Dtilde{j}$,
correspond to the first term on the rhs of~\refeq{eq:OFRmergingA}.  
The corresponding triple
or quartic vertices of the Feynman rules play an analogous role as the
crossed vertex in the pinched open loop (last diagram on the rhs).  Besides merging all
relevant unpinched combinations $\alpha_j,\alpha_{j+1}$, also possible
unpinched topologies where some of the external legs in $w_j$ and $w_{j+1}$
are interchanged (not shown) should be included.  }
\label{fig:OLredC}
\end{figure*}

\vspace{-4mm}
\subsubsection{Pinching a dressed propagator}
\label{se:dressedpinch}
\vspace{-4mm}
Let us consider the on-the-fly reduction of an $N$-point open loop with $k$
dressed segments, focusing on contributions that result from the
pinching of a $\Dbar{j}$ denominator with $j<k$, 
\vspace{2.5mm}
\bea
\label{eq:OFRmasterformulaC}
&&\raisebox{-5mm}{\parbox[t]{4.2\defheight}{\includegraphics[height=\heightB]{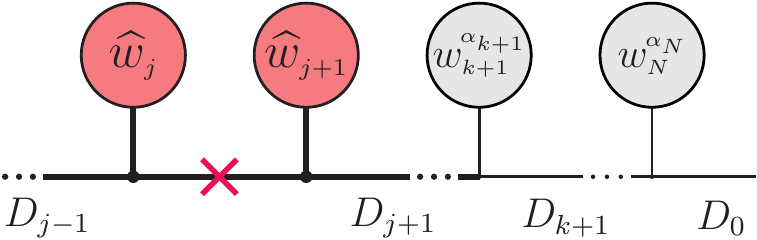}}}=\nonumber\\[3mm]
&&=
\frac{\calV^{\alpha_{k+1}\dots\alpha_{N}}_{k}(\Omega^k_N[j],\bar q)\;
\seg_{k+1}^{\alpha_{k+1}}(q)\cdots\seg_{N}^{\alpha_N}(q)}{\Dbar{0}\cdots 
\slashed{\Dbar{j}}\cdots
\Dbar{3} \cdots \Dbar{N-1}}.\qquad
\eea
Here, consistently with the notation of \refse{sec:merging}, we have
restored the indices $\alpha_i$ of the various undressed segments, while
helicity indices are kept implicit.  The pinched propagator
in~\refeq{eq:OFRmasterformulaC} is entirely dressed and merged, in the sense
that both adjacent segments, $\seg_{j}$ and $\seg_{j+1}$, are dressed
and merged.
Thus, for what concerns all future dressing and reduction steps, apart from
the disappearance of the $\Dbar{j}$ denominator, the pinch has no effect. 
Consequently, the above contribution can be {\it absorbed} into 
unpinched
$(N-1)$-point open loops that involve the same loop denominators,
$\Dbar{0}\dots\Dbar{j-1}\Dbar{j+1}\dots \Dbar{N-1}$, and the same undressed
segments, $\seg_{k+1}^{\alpha_{k+1}},\dots,\seg_{N-1}^{\alpha_{N-1}}$.
To this end, it is sufficient to bring the pinched open loop
\refeq{eq:OFRmasterformulaC} in the standard form 
\vspace{2.5mm}
\begin{eqnarray} 
\label{eq:pinchtra}
&&\;\raisebox{-5mm}{\parbox[t]{3.4\defheight}{\includegraphics[height=\heightB]{newdiaredpinchF}}}=\nonumber\\
&&=\;
\raisebox{-5mm}{\parbox[t]{4.2\defheight}{\includegraphics[height=\heightB]{newdiaredpinchE}}}\;,\qquad
\end{eqnarray}
which corresponds to an
unpinched $(N-1)$-point open loop with $k-1$ dressed segments. 
The crossed vertex on the lhs of~\refeq{eq:pinchtra} indicates that
the two original segments that are connected by the pinched propagator
should be regarded as a single effective segment.
This symbolic contraction of segments does not change anything in the numerics of
the open-loop numerator, and the transformation~\refeq{eq:pinchtra} 
is nothing but a trivial relabeling of the 
denominators and of the undressed segments that lie on the right side
of the pinch, 
\bea
\begin{array}{rclcrclcc}
\Dtilde{i}(q)&=&\Dbar{i}(q)  &     &&       &
 &\;\mbox{for}\;& i< j, \\[4mm]
\Dtilde{i}(q)&=&\Dbar{i+1}(q) &
\,\text{and}\,& 
\segtilde^{\alpha_i}_i(q) &=&\seg_{i+1}^{\alpha_{i+1}}(q) & 
\;\mbox{for}\;& i\ge j. \\[1mm]
\end{array}
\label{eq:pinchtraA}
\eea

The pinched open loop~\refeq{eq:pinchtra} can be merged with
corresponding unpinched open loops to form a single $(N-1)$-point object
with $k-1$ dressed segments. The corresponding formula reads
\vspace{1mm}
\bea
&&\frac{\calVtilde^{\alpha_k\dots\alpha_{N-1}}_{k-1}(\Omegatilde^{k-1}_{N-1},\bar q)\;
\segtilde^{\alpha_k}_{k}(q)\cdots\segtilde_{N-1}^{\alpha_{N-1}}(q)}{\Dtilde{0}\cdots \Dtilde{3} \cdots \Dtilde{N-2}}=\nonumber\\
&&=
\frac{\calV^{\alpha_k\dots\alpha_{N-1}}_{k-1}(\Omega_{N-1}^{k-1},\bar q)\;
\segtilde^{\alpha_k}_{k}(q)\cdots\segtilde_{N-1}^{\alpha_{N-1}}(q)}{\Dtilde{0}\cdots \Dtilde{3} \cdots \Dtilde{N-2}}
\nonumber\\[2mm]
&&+\!
\sum_{\Omega^k_N[j]}\!
\frac{\calVtilde_{k}^{\alpha_k\dots\alpha_{N-1}}(\Omega^k_N[j],\bar q)\;
\seg_{k+1}^{\alpha_k}(q)\cdots\seg_{N}^{\alpha_{N-1}}(q)}{\Dbar{0}\cdots 
\slashed{\Dbar{j}}\cdots
\Dbar{3} \cdots \Dbar{N-1}}.\qquad
\label{eq:OFRmergingA}
\eea

\noindent A diagrammatic representation of this identity is given in \reffi{fig:OLredC}.
The resulting object, for which we introduce the symbol
$\calVtilde^{\alpha_k\dots\alpha_{N-1}}_{k-1}$, is a combination of
unpinched and pinched open loops, which enter, respectively, through the
first and second term on the rhs of~\refeq{eq:OFRmergingA}.
By construction, the corresponding set of diagrams
($\Omegatilde_{N-1}^{k-1}$) includes all unpinched ($\Omega_{N-1}^{k-1}$)
and pinched ($\Omega^k_N[j]$) diagrams with loop propagators
$\Dtilde{0}\dots\Dtilde{N-2}$ and undressed segments
$\seg_{k+1}^{\alpha_{k+1}},\dots,\seg_{N-1}^{\alpha_{N-1}}$.  
The set of pinched diagrams $\Omega^k_N[j]$ corresponds to an $N$-point
topology that results from $\Omega_{k-1}^{N-1}$ by undoing a $\Dbar{j}$
pinch, and~\refeq{eq:OFRmergingA} involves all 
possible $\Omega^k_N[j]$ contributions with $1\le j \le k$.

As a necessary condition for the merging
operation~\refeq{eq:OFRmergingA} to be applicable, the different open loops on the rhs
of~\refeq{eq:OFRmergingA} have to feature the same undressed segments.  This
implies that they must be at the same stage of the dressing recursion, and,
most importantly, that the starting position and the directions of the
respective  dressing recursions should be equivalent to each other.  
With other words, $\Dtilde{0}\dots\Dtilde{N-2}$ and
$\Dbar{0}\dots\slashed{\Dbar{j}}\dots\Dbar{N-1}$ should be two identical
{\it ordered} sets of propagators.  In particular they should start from the
same cut propagator, $\Dtilde{0}=\Dbar{0}$.  As discussed in
\refse{se:cuttingrule} this can be guaranteed, to some extent, by means of an appropriate
cutting rule.

Another obvious prerequisite for the absorption of pinched $N$-point open loops
is the existence of corresponding unpinched $(N-1)$-point Feynman diagrams.
With other words, the crossed vertex
in~\refeq{eq:pinchtra} should have a physical counterpart
consisting of a 
triple or quartic vertex, which can directly connect the $\Dbar{j-1}$ and $\Dbar{j+1}$
propagators to subtrees involving the external legs
attached to $\hat w_j$ and $\hat w_{j+1}$ (see \reffi{fig:OLredC}).
In QCD, this turns out to be the case for most pinched
configurations.

Moreover, pinched $N$-point configurations of the form~\refeq{eq:pinchtra}
can also be merged with other pinched higher-point diagrams that get the
relevant pinches in past or future reduction steps.  Thus, 
pinched objects of type~\refeq{eq:pinchtra} will always be denoted as absorbable,
irrespective of whether corresponding unpinched $(N-1)$-point Feynman
diagrams exist or not.

The merging procedure~\refeq{eq:OFRmergingA} needs to be applied after any
on-the-fly reduction step that creates new pinched objects.  Thus merging
operations have to be interleaved with iterated dressing and reduction
steps.
Since pinched $N$-point objects are absorbed into 
lower-point objects, the algorithm should start with the dressing and
on-the-fly reduction of open loops with $N=N_{\max}$, and continue towards
lower $N$.  
The merging operation~\refeq{eq:OFRmergingA} starts at stage $N-1 =
N_{\max}-1$ and is applied after every dressing step, together with the
merging of unpinched open loops (see \refse{sec:merging}).  
Note that, due to the iterative nature of the algorithm, the term
$\calVtilde^{\alpha_k\dots\alpha_{N-1}}_{k}$ on the rhs
of~\refeq{eq:OFRmergingA} can be the result of multiple pinching and merging
steps.

\begin{figure*}[t!]
\hspace{0.09\textwidth}
\begin{minipage}{0.8\textwidth}
\begin{eqnarray} 
\raisebox{-5mm}{\parbox[t]{3.55\defheight}{\includegraphics[height=\heightB]{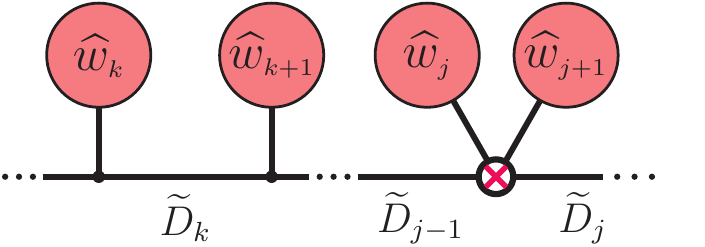}}}
&=&\hspace{-2mm}
\sum_{\alpha_{k+1},\dots,\alpha_{j+1}}\hspace{-2mm} 
\raisebox{-5mm}{\parbox[t]{4.2\defheight}{
\includegraphics[height=\heightB]{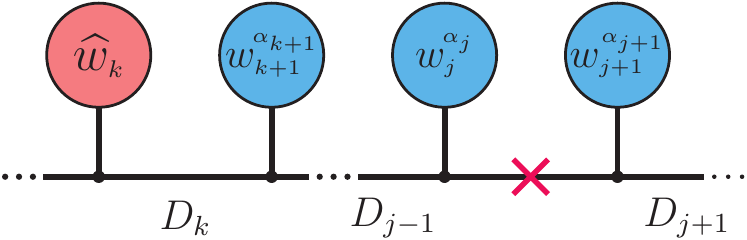}}}
\nonumber\\[0mm]
&&
\hspace{15mm}\parbox[t]{4.2\defheight}{\includegraphics[height=\heightC]{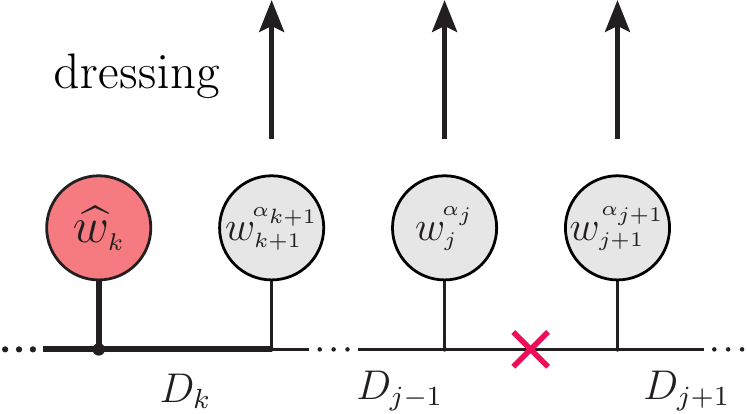}}
\nonumber
\end{eqnarray}\end{minipage}
\caption{Required dressing~\refeq{eq:OFRmergingC} and relabeling~\refeq{eq:pinchtraA} 
operations before merging open loops 
with an undressed pinched propagator $\Dbar{j}$.}
\label{fig:OLredE}
\end{figure*}

\subsubsection{Pinching an undressed propagator}
\label{se:undressedpinch}

The possibility to absorb pinched open loops as in \reffi{fig:OLredC} is
based on the fact that all future dressing and reduction operations can be
performed only once at the level of a merged object.  Thus, 
the undressed segments of pinched and unpinched open loops should be
identical.

However, the segments connected to the $\Dtilde{j-1}$ and
$\Dtilde{j}$ propagators are {\it different} for pinched and unpinched
terms.  In the pinched case there are two separate segments, which involve
$w_{j}$ and $w_{j+1}$ and require two subsequent dressing steps.  In
contrast, unpinched open loops require a single dressing step, since $w_{j}$
and $w_{j+1}$ are combined in a single segment.

It is thus clear that pinched open loops can be absorbed only after the
dressing of the segments that lie on the two sides of the pinch.
If this is not the case, \ie when a $\Dbar{j}$ pinch is applied to 
an open loop with $k\le j$ dressed segments, its absorption becomes possible only 
after $(k-j+1)$ additional dressing steps, which result in
\bea
\label{eq:OFRmergingC}
&&\calVtilde_{j+1}^{\alpha_{j+1}\dots\alpha_{N-1}}(\Omega^{j+1}_N[j],\bar q)=\nonumber\\
&&=\sum_{\alpha'_{k+1}\dots\alpha'_{j+1}}
\calVtilde_{k}^{\alpha'_{k+1}\dots\alpha'_{j+1}\alpha_{j+1}\dots\alpha_{N-1}}(\Omega^k_N[j],\bar q)\nonumber\\
&&\phantom{=\sum_{\alpha'_{k+1}\dots\alpha'_{j+1}}}\times\,
\seg_{k+1}^{\alpha'_{k+1}}(q)\cdots\seg_{j+1}^{\alpha'_{j+1}}(q),
\eea
where $\seg_{k+1}\dots\seg_{j+1}$ are dressed.
Note that in~\refeq{eq:OFRmergingC} we also sum over all possible $\alpha'_{i}$, and
use indices $\alpha_{j+1},\dots,\alpha_{N-1}$ with shifted
labels for the undressed segments $\seg_{j+2},\dots,\seg_{N}$ on the right
side of the pinch.

As illustrated in \reffi{fig:OLredE}, the dressing
operation~\refeq{eq:OFRmergingC} combined with the
relabeling~\refeq{eq:pinchtraA} brings the pinched open loops in a
configuration that can be absorbed with the merging
formula~\refeq{eq:OFRmergingA}.  However, the absorption of undressed pinched
propagators is more involved than the simplified picture
of~\reffi{fig:OLredE}.
Pinched open loops can require more than one dressing step to become
absorbable, in which case, in general, also new reduction steps are needed in
order to keep the tensor rank below three.  Such new reductions generate
additional pinches, and their iteration can lead to a proliferation of
multi-pinched configurations.

Thus, to avoid dramatic inefficiencies,  it is crucial to minimise the
number of required dressing steps by keeping pinchable propagators as close
as possible to the dressed part of the numerator. This is why we choose to perform the 
on-the-fly reduction using the  denominators $\Dbar{0},\dots,\Dbar{3}$.\footnote{%
The on-the-fly reduction~\refeq{eq:OFRmasterformula}
can be performed using any set of four propagators, 
$\Dbar{i_0},\ldots,\Dbar{i_3}$.}
In this way, when the first two segments are dressed and the first reduction
step is applied (see \reffi{fig:OLred}), the various pinched
propagators are located at most one left-dressing step ($\Dbar{0}$) and two
right-dressing steps ($\Dbar{3}$) away from the dressed part.

In order to identify pinches that cannot be directly absorbed and to
anticipate how they propagate through the recursion, let us consider
generic open-loop configurations before the creation of a new pinch through
a reduction step.  At this stage the rank $R$ must be equal to two.  We
first consider the very first reduction step, which can occur after $k\ge
R=2$ dressing steps, and we focus on a $\Dbar{2}$ pinch in the case where
only $k=2$ segments are dressed.  In this case, an on-the-fly
reduction step and a subsequent dressing step yield 
\bea
&&\raisebox{-4mm}{\parbox[t]{3.9\defheight}{\includegraphics[height=\heightB]{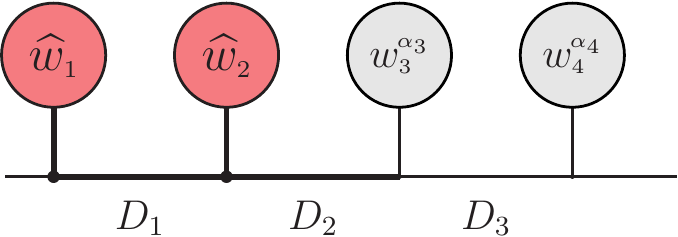}}}\label{eq:pinchevolA}\\
&&{\longrightarrow\atop \mbox{\parbox{15mm}{$\Dbar{2}$-pinch\\+dressing}}}
\raisebox{-4mm}{\parbox[t]{3.2\defheight}{\includegraphics[height=\heightB]{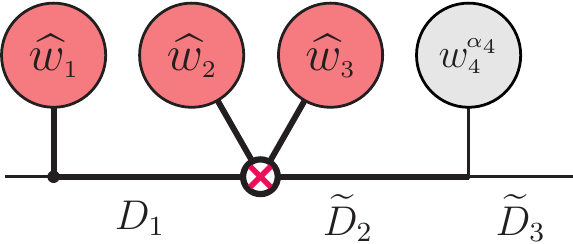}}}\nonumber
\eea
\ie the $\Dbar{2}$ pinch can be brought in the standard form~\refeq{eq:pinchtra}
and can thus be absorbed into unpinched contributions.
The same considerations apply also to $\Dbar{1}$ and $\Dbar{0}$ pinches (see
\refse{se:dzeropinch}) and can be generalised to any step of the recursion,
since the structure of the dressed parts 
on the lhs and
rhs of~\refeq{eq:pinchevolA} is the same.

Also $\Dbar{3}$  pinches can be absorbed in a similar way in case there
are at least three dressed segments before the reduction step.  Otherwise,
when only two segments are dressed, the combination of an on-the-fly
reduction step with a subsequent dressing step leads to
\vspace{1mm}
\bea
&&\raisebox{-4mm}{\parbox[t]{3.9\defheight}{\includegraphics[height=\heightB]{flowdiaredmain}}}\label{eq:pinchevolb}\\
&&{\longrightarrow\atop \mbox{\parbox{15mm}{$\Dbar{3}$-pinch\\+dressing}}}
\raisebox{-4mm}{\parbox[t]{3.9\defheight}{\includegraphics[height=\heightB]{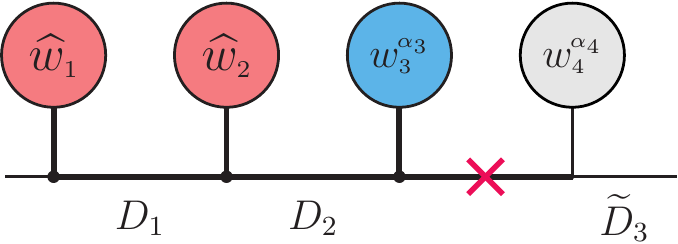}}}\nonumber
\eea
where the pinch is applied on the last dressed segment.  Unless another
dressing step can be applied before reaching rank three, this kind of pinch cannot
be absorbed without a further reduction and dressing step.
Dressing one more segment
allows one to absorb the original $D_3$ pinch as well as new
$\Dbar{0}$, $\Dbar{1}$, and $\Dbar{2}$ pinches that arise from the new reduction step.
However, the new reduction 
leads again to a configuration 
with a $\Dtilde{3}$ pinch 
on the last dressed segment,
\bea
&&\raisebox{-4mm}{\parbox[t]{3.9\defheight}{\includegraphics[height=\heightB]{flowdiaredpinchC}}}\label{eq:pinchevolc} \\
&&{\longrightarrow\atop \mbox{\parbox{15mm}{$\Dtilde{3}$-pinch\\+dressing}}}
\raisebox{-4mm}{\parbox[t]{3.9\defheight}{\includegraphics[height=\heightB]{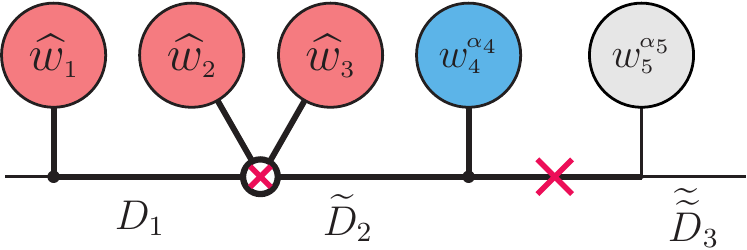}}}\nonumber
\eea 
Again, the dressed parts on the 
lhs and rhs of~\refeq{eq:pinchevolc} have the same structure,
which implies that such $\Dbar{3}$-pinched configurations are stable with respect to further
reduction steps. Thus, open loops with multiple non-absorbable pinches do
not occur, and the only type of configuration with a single non-absorbable
pinch is the one in~\refeq{eq:pinchevolc}.

\begin{figure*}[t!]
\hspace{0.09\textwidth}
\begin{minipage}{0.8\textwidth}
\begin{eqnarray} 
\raisebox{-5mm}{\parbox[t]{3.5\defheight}{\includegraphics[height=\heightB]{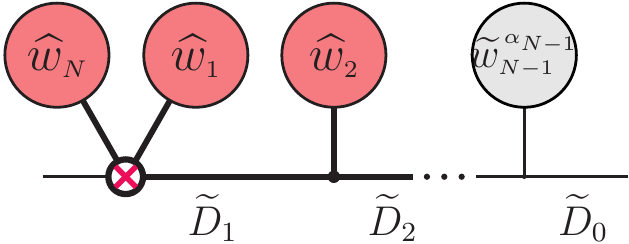}}}
&=&\hspace{0mm}
{\sum_{\alpha_{N}}}\hspace{2mm} 
\raisebox{-5mm}{\parbox[t]{4.2\defheight}{\includegraphics[height=\heightB]{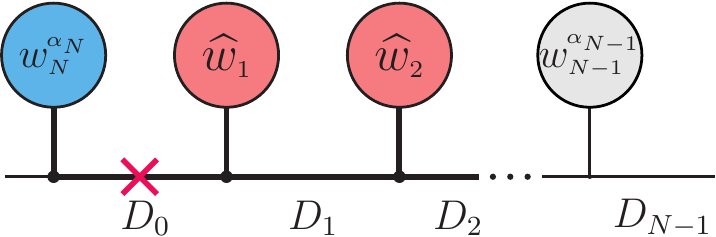}}}
\nonumber\\[0mm]
&&
\hspace{8mm}\parbox[t]{4.2\defheight}{
\includegraphics[height=\heightC]{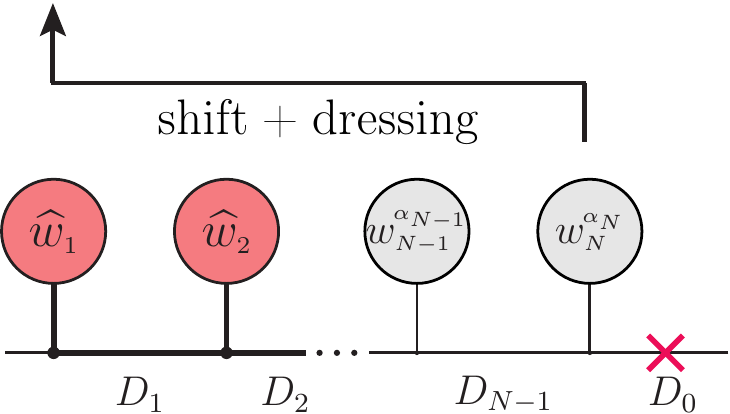}
}
\nonumber
\end{eqnarray}\end{minipage}
\caption{Required dressing, loop-momentum shift and 
relabeling operations~~\refeq{eq:OFRmergingE}--\refeq{eq:pinchtraB}
before merging an open loop 
with a pinched $\Dbar{0}$ propagator.}
\label{fig:OLredF}
\end{figure*}

\subsubsection{Pinching  the $\Dbar{0}$ propagator}
\label{se:dzeropinch}

Finally, we consider open loops with $k$ dressed segments
and a pinched $\Dbar{0}$ denominator,
\vspace{2.5mm}
\bea
\label{eq:OFRmasterformulaD}
&&\raisebox{-5mm}{\parbox[t]{4.2\defheight}{\includegraphics[height=\heightB]{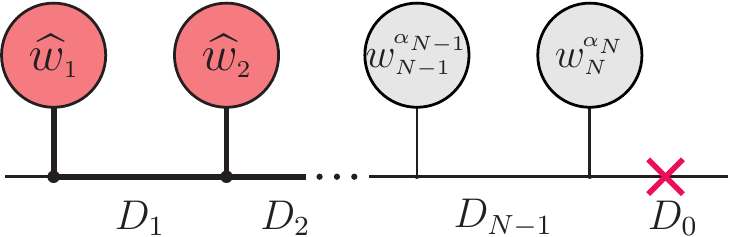}}}=\nonumber\\[3mm]
&&=
\frac{\calV^{\alpha_{k+1}\dots\alpha_{N}}_{k}(\Omega^k_N[0],\bar q)\;
\seg_{k+1}^{\alpha_{k+1}}(q)\cdots\seg_{N}^{\alpha_N}(q)}{\Dbar{1}\cdots 
\Dbar{3} \cdots \Dbar{N-1}\slashed{\Dbar{0}}}.\qquad
\eea
For convenience, here and in the following we put $\Dbar{0}$ at the end of
the chain of denominators.  Similarly as for the cases discussed in
\refse{se:undressedpinch}, the above pinched configuration becomes
absorbable only when the segments connected to the $\Dbar{0}$ propagator,
\ie $\seg_N(q)$ and $\seg_1(q)$, are dressed.  However, this happens only at
the end of the standard dressing recursion.

In order to anticipate the absorption of the most advanced pinch one could
replace the denominators $\Dbar{0},\dots, \Dbar{3}$ used for the on-the-fly
reduction by $\Dbar{1},\dots, \Dbar{4}$, which lie all directly on the right
side of the cut.  However, the absorption of each $\Dbar{4}$ pinch would
require up to three extra dressing steps and two related reduction
steps, resulting in the creation of multiple new pinches.
This problem can be circumvented by observing that the
$\Dbar{0}$ propagator 
lies only one step away 
from the dressed part of the open loop, 
if one reverts the dressing direction.
Therefore, as illustrated in \reffi{fig:OLredF}, the pinched $\Dbar{0}$
propagator can be entirely dressed by means of
a single left-dressing step.
This operation results in
\bea
&&\calVtilde_{k+1}^{\alpha_{k+1}\dots\alpha_{N-1}}(\Omega^{k+1}_N[0],\bar q)=\nonumber\\
&&= 
\sum_{\alpha_{N}}
\seg_{N}^{\alpha_{N}}(q)\;
\calVtilde_{k}^{\alpha_{k+1}\dots\alpha_{N-1}\alpha_{N}}(\Omega^{k}_N[0],\bar q)
,
\label{eq:OFRmergingE}
\eea
where the left multiplication of the $N^{\mathrm{th}}$ segment should be understood as
\bea
&&\Big[
\seg_N(q)\;
\calVtilde_{k}(\Omega,\bar q)
\Big]^{\beta_{k+1}}_{\beta_{0}}
=
\sum_{\beta_{N}}\Big[\seg_N(q)\Big]^{\beta_{N}}_{\beta_{0}}\nonumber\\
&&\times\,\Big[\calVtilde_{k}(\Omega,\bar q)\!\Big]^{\beta_{k+1}}_{\beta_{N}}
=
\left[
\left(
\Big[\calVtilde_{k}(\Omega,\bar q)\Big]^{\mathrm{T}}
\;
\Big[\seg_N(q)\Big]^{\mathrm{T}}
\right)^{\mathrm{T}}
\;\right]^{\beta_{k+1}}_{\beta_{0}}\hspace{-2mm}.\nonumber\\
\label{eq:leftdressing}
\eea
Technically, as indicated on the rhs,
this operation can be easily implemented through a standard
right-dressing step upon transposition of the input matrices 
and back-transposition of the result.\footnote{Note that in a three-gluon vertex this transposition leads to a flip of the colour indices of the two gluons in the loop
and hence a minus sign in the colour factor. Since the colour factor has been fixed in the beginning this minus sign has to be taken into account in the colour-stripped open loop.}

As usual, before merging with unpinched open loops, 
the propagators and undressed segments  
that lie on the right side of the pinch
need to be brought back in standard form.
In case of a $\Dbar{0}$ pinch, 
all remaining $N-1$ denominators and 
segments preserve their relative position along the open 
loop. Thus, only $\Dbar{N-1}$ needs to be relabeled,
since it assumes the role of the new $\Dtilde{0}$.
Moreover, the standard form
$\Dtilde{0}=q^2-\widetilde{m}^2_0$
requires a loop momentum shift $q\to q-p_{N-1}$
for the entire open loop.
The corresponding reparametrisations for the various denominators and segments read
\bea
\begin{array}{rclcl}
\Dtilde{0}(q)&=&\Dbar{N-1}(q-p_{N-1}),   & & 
\\[3mm]
\Dtilde{i}(q)&=&\Dbar{i}(q-p_{N-1}) &
\;\mbox{for}\;\;& 1\le i\le N-2, \\[3mm]
\segtilde^{\alpha_i}_i(q) &=&\seg_{i}^{\alpha_i}(q-p_{N-1}) & 
\;\mbox{for}\;\;& k+1\le i\le N-1. 
\end{array}
\label{eq:pinchtraB}
\eea
In terms of masses and momenta this corresponds to 
$\widetilde{m}_0=m_{N-1}$,
$\widetilde{m}_i=m_i$ and
$\widetilde{p}_i =p_{i}+p_{N-1}$
for
$1\le i\le N-2$.
With these transformations the left-dressed $\Dbar{0}$-pinched open loop~\refeq{eq:OFRmergingE} 
can be merged with related unpinched objects according to 
\bea
&&\frac{\calVtilde^{\alpha_{k+1}\dots\alpha_{N-1}}_{k}(\Omegatilde^{k}_{N-1},\bar q)
\segtilde^{\alpha_{k+1}}_{k+1}(q)\cdots\segtilde_{N-1}^{\alpha_{N-1}}(q)}{\Dbar{1}\cdots \Dtilde{3} \cdots \Dtilde{N-2}\Dtilde{0}}
\equiv\nonumber\\
&&\equiv\;
\frac{\calV^{\alpha_{k+1}\dots\alpha_{N-1}}_{k}(\Omega_{N-1}^{k},\bar q)\segtilde^{\alpha_{k+1}}_{k+1}(q)\cdots\segtilde_{N-1}^{\alpha_{N-1}}(q)}
{\Dbar{1}\cdots \Dtilde{3} \cdots \Dtilde{N-2}\Dtilde{0}}
\nonumber\\[2mm]
&&\phantom{\equiv}+
\sum_{\Omega^k_N[0]}
\frac{\calVtilde_{k+1}^{\alpha_{k+1}\dots\alpha_{N-1}}(\Omega^{k+1}_N[0],\bar q)}
{\Dbar{1}\cdots \Dbar{3} \cdots \Dbar{N-1}\slashed{\Dbar{0}}}\nonumber\\ &&\phantom{\equiv}\times\,
\seg_{k+1}^{\alpha_{k+1}}(q)\cdots\seg_{N-1}^{\alpha_{N-1}}(q)\Bigg|_{q\to q-p_{N-1}}
.
\label{eq:OFRmergingD}
\eea
Apart from the loop momentum shift, $q\to q-p_{N-1}$, this formula is
entirely analogous to~\refeq{eq:OFRmergingA}.  Let us note that, since the shift
is applied to a single term, the identity~\refeq{eq:OFRmergingD} holds only
upon loop momentum integration.  Nevertheless, as far as the correctness of final results at integral
level is concerned, it can be safely applied at
the integrand level.

As demonstrated in \refse{sec:stab}, using the on-the fly reduction 
with pinch absorption
in combination with the on-the-fly techniques of \refse{sec:OFhelmerging}
results in a very fast and
numerically stable one-loop algorithm.  In particular, as compared to the
original version of {\sc OpenLoops}, we find very significant improvements, both
in terms of speed and numerical stability.
Actually, using only the new techniques of \refse{sec:OFhelmerging} without
on-the-fly reduction yields even higher CPU efficiency.  However, as we will
see, the moderate extra CPU cost that results from the on-the-fly reduction
approach is counterbalanced by a very significant gain in numerical
stability, which implies a reduced usage of quadruple precision for
exceptional phase space points.

\subsection{Cutting rule}
\label{se:cuttingrule}

As pointed out in \refse{se:dressedpinch}, 
the possibility to merge pinched $N$-point open loops and corresponding unpinched 
$(N-1)$-point open loops depends on the way they are cut.
In order to identify the relevant requirements, let us
consider the cut-open topology 
defined by the following ordered set of loop segments,
\bea
\calI_{N}=\{\calS_1,\calS_2,\dots, \calS_{N}\}.
\label{eq:unpinchedtop}
\eea
Here we have applied our standard labeling scheme, where the cut is located
between $\calS_N$ and $\calS_1$, \ie on the $\Dbar{0}$ propagator, while the
dressing recursion starts with $\calS_1$ and is directed towards $\calS_2$.  This configuration will be
referred to as a $\calS_N$/$\calS_1$ ordered cut.  Since the labeling scheme is a
consequence of the position of the cut, and not vice versa, we have to
define a cutting rule that selects $\calS_N$/$\calS_1$  out of all
possible $\calS_i$/$\calS_j$ cuts.

The cutting rule should enable the merging of pinched subtopologies
that arise from~\refeq{eq:unpinchedtop}
by pinching certain propagators $\Dbar{j}$, \ie by combining 
$\calS_j$ and $\calS_{j+1}$ in a single segment $\calS_j\oplus \calS_{j+1}$.
To this end, unless the cut propagator $\Dbar{0}$ is pinched, the cutting
rule should guarantee that the position of the cut and its direction remain
unchanged after a pinch.  More explicitly, the desired cut configurations
after a $\Dbar{j}$ pinch with $j>0$ are
\bea
\calI_{N}[1]&=&\{\calS_1\oplus \calS_2,\calS_3,\dots, \calS_{N}\},
\label{eq:pinchedtopa}\\
\calI_{N}[j]&=&\{\calS_1,\dots,\calS_{j-1}, \calS_j\oplus \calS_{j+1}, \calS_{j+2},\dots, \calS_{N}\}\nonumber\\ &&\mbox{for}\; 2\le j\le N-2,
\label{eq:pinchedtopb}\\
\calI_{N}[N-1]&=&\{\calS_1,\calS_{2},\dots,\calS_{N-1}\oplus \calS_N\}.
\label{eq:pinchedtopc}
\eea
In this way, as required for the merging operations described
in~\refses{se:dressedpinch}{se:undressedpinch}, the dressing of pinched and
unpinched objects always starts and ends with segments that contain the
external legs attached to $\calS_1$ and 
$\calS_N$, respectively.
In the case of a $\Dbar{0}$ pinch, where the original cut 
propagator disappears,
in order to enable the merging 
of left-dressed pinched subtopologies described in~\refse{se:dzeropinch}, the cut 
should be moved to the left of $\calS_N\oplus \calS_1$, so that 
\bea
\calI_{N}[0]&=&\{\calS_N\oplus \calS_1,\calS_{2},\dots, \calS_{N-1}\}.
\label{eq:pinchedtopd}
\eea
In order to ensure, at least in part, that pinched topologies are cut as in
\refeq{eq:pinchedtopa}--\refeq{eq:pinchedtopd}, we 
replace the original cutting rule~\refeq{eq:cuttingrulesel}--\refeq{eq:cuttingruledir}
by the new prescriptions
\begin{align}
F(\calS_1) &> F(\calS_k) 
\quad\forall\quad k>1,\; &\text{(selection rule)}\label{eq:newcuttingrulesel}\\
F(\calS_N) &> F(\calS_2), &\text{(direction rule)}\label{eq:newcuttingruledir}
\end{align}
where the weights $F(\calS_a)$ are defined in~\refeq{eq:cuttingruleS}.
The key property of the above cutting rule is the
pinch-invariance of the selection rule, 
which determines the first segment $\calS_1$. 
In fact, if this condition is realised for \refeq{eq:unpinchedtop}, then it is guaranteed to 
hold also for all
pinched configurations~\refeq{eq:pinchedtopa}--\refeq{eq:pinchedtopd}.
For $j=0,1$ this is an obvious consequence of
the fact that $F(\calS_1\oplus \calS_a)>F(\calS_1)$ for any $a\neq 1$.
In the other cases, the fact that $\calS_1$ remains the first subtree
in spite of the appearance of a new pinched subtree $\calS_j\oplus \calS_{j+1}$ 
with weight $F(\calS_j)+F(\calS_{j+1})$,
is guaranteed by 
\bea
F(\calS_1) > \sum_{i=2}^N F(\calS_i).
\label{eq:maxweight}
\eea
This inequality is an automatic  consequence of~\refeq{eq:newcuttingrulesel}
and of the binary nature of the weights~\refeq{eq:cuttingruleS}.
This is easily understood by observing that, 
due to $2^{\npart-1}= 1+\sum_{\alpha=1}^{\npart-1}\;2^{\alpha-1}$,
the last external particle ($\alpha=\npart$) 
outweighs the ensemble of all other particles.
Therefore the last external particle  
must belong to the leading-weight subtree $\calS_1$, 
which implies that $F(\calS_1)\ge 2^{\npart-1}$ and leads to~\refeq{eq:maxweight}.

Unfortunately, the direction rule~\refeq{eq:newcuttingruledir} is not
sufficient in order to preserve the direction of the cut.  For instance, in
case of a $\Dbar{1}$ pinch, the desired cut~\refeq{eq:pinchedtopa} requires
$F(\calS_N)>F(\calS_3)$, which does not automatically follow from $F(\calS_N)>F(\calS_2)$.  More
generally, apart from the case of a $D_3$ pinch, where the second and last
subtree do not change, there is no guarantee that the
condition~\refeq{eq:newcuttingruledir} preserves the direction of the cut.

Thus the above cutting rule does not allow one to absorb all pinched open
loops.  Nevertheless, as demonstrated in \refse {sec:stab}, it is sufficient
to obtain a very fast on-the-fly reduction algorithm.  Moreover, it should
be possible to further increase CPU efficiency, either by means of an
improved cutting rule or by inverting the dressing direction after certain
pinches.

\subsection{Rational terms of type $R_1$}
\label{sec:rationalterms}

As discussed in \refse{se:OFR}, each reduction step of type~\refeq{eq:OFRmasterformula} 
and~\refeq{eq:qmuqnuredfinalB} generates terms $\Dbar{i}-D_i=\tilq^2$
that account for the mismatch  between the $D$-dimensional and four-dimensional parts of loop denominators.
The resulting tensor integrals with $(D-4)$-dimensional
 $\tilq^{2}$ terms in the numerator
can give rise to finite terms.
As is  well known~\cite{delAguila:2004nf,Ossola:2008xq}, these so-called
rational terms of type $R_1$ can arise only in 
the presence of $1/(D-4)$ poles of ultraviolet type.
Thus, vanishing integrals of type $R_1$ can be easily identified 
by means of the simple power counting criterion
\bea
&&\int\!\mathrm{d}^D\!\momq\, \frac{q^{\mu_1}q^{\mu_2}\ldots q^{\mu_r}\,\tilq^{2s}}{\Dbar{0}\Dbar{1}\cdots\Dbar{N-1}}
=\ord(D-4)
\nonumber\\[1mm] && \qquad
\mbox{if}\quad s\ge 1
\quad\mbox{and}\quad r+2s+4 < 2N.
\label{Ronepowercounting}
\eea

In a renormalisable theory, where each loop segment increases the rank at most by one, 
$r+2s\le N$ and all integrals with $N\geq 5$ and $s\ge 1$ vanish. Thus, 
the only non-vanishing integrals of type $R_1$ that remain at the 
end of the on-the-fly reductions of \refse{se:OFR} are~\cite{delAguila:2004nf}
\bea 
\int\!\mathrm{d}^D\!\momq\, \frac{\tilde{q}^{2}}{\Dbar{0} \Dbar{1}}
&=&\frac{-\ri \pi^2}{2}\lb \mass{0}^2+\mass{1}^2-\frac{\momp{1}^2}{3}\rb \nonumber\\ &&+\mathcal{O}(D-4), \\
\int\!\mathrm{d}^D\!\momq\, \frac{q^{\mu}\,\tilde{q}^{2}}{\Dbar{0} \Dbar{1} \Dbar{2}}
&=&\frac{\ri\pi^2}{6}\lb \momp{1}+\momp{2}\rb^\mu \nonumber\\ &&+\mathcal{O}(D-4), \\
\int\!\mathrm{d}^D\!\momq\, \frac{\tilde{q}^{2}}{\Dbar{0} \Dbar{1} \Dbar{2}}
&=&\frac{-\ri \pi^2}{2}+\mathcal{O}(D-4), \\
\int\!\mathrm{d}^D\!\momq\, \frac{\tilde{q}^{4}}{\Dbar{0} \Dbar{1} \Dbar{2} \Dbar{3}}
&=&\frac{-\ri \pi^2}{6}+\mathcal{O}(D-4).
\label{eq:RoneD}
\eea

The power counting criterion~\refeq{Ronepowercounting} can 
be exploited in a way that makes it possible to discard irrelevant terms of
type $R_1$  at any intermediate step of the open-loop recursion.
To this end, for each number $k$ of dressing steps 
we anticipate the 
maximum rank $\rmaxk$
of the segments $\seg_{k+1}(q),\ldots,\seg_{N}(q)$ that remain 
to be dressed.
Given this information,
it is clear that
monomials of type
$q^{\mu_1}q^{\mu_2}\ldots q^{\mu_r}\,\tilq^{2s}$
in the dressed open loop 
cannot give rise to terms of $D$-dimensional 
rank higher than $r+2s+\rmaxk$ at the end of the recursion.
Thus, we can anticipate that,
\bea
&&\int\!\mathrm{d}^D\!\momq\, \frac{q^{\mu_1}\ldots q^{\mu_r}\,\tilq^{2s}
\seg_{k+1}(q)\dots\seg_N(q)}{\Dbar{0}\Dbar{1}\cdots\Dbar{N-1}}
=\ord(D-4)\nonumber\\ &&
\quad
\mbox{if}\quad s\ge 1
\quad\mbox{and}\quad r+2s+4+\rmaxk < 2N.
\label{RonepowercountingB}
\eea
The systematic application of this condition allows one to filter out a very large
number of $\tilq^2$ terms, thereby improving the efficiency of the algorithm.

Note also that the unpinched contributions $\calV_k(\Omega_N^k[-1])$ in the reduction identities 
\refeq{eq:OFRmasterformulaB} involve terms that reduce $r+2s-2N$, and thus the degree of ultraviolet divergence, 
by one and two.
Depending on the values of $N$ and $\rmaxk$, this can result
in vanishing $R_1$ contributions that can also be immediately discarded. For instance, in the
reduction of 
$\int\!\mathrm{d}^D\!q\, \frac{q^{\mu_1}q^{\mu_2}\tilde{q}^{2}}{\Dbar{0} \Dbar{1} \Dbar{2} \Dbar{3}}$ 
all unpinched contributions apart from those of type~\refeq{eq:RoneD}
can be neglected.

\section{Reduction identities and numerical stability}  \label{se:red}

This section deals with the reduction method of~\cite{delAguila:2004nf},
which provides the basis of the on-the-fly reduction approach of \refse{se:OFR}.  In
\refses{se:redbasis}{se:redform3} we outline the derivation of the tensor
coefficients $A_j^{\mu\nu}$ and $B^{\mu\nu}_{j,\lambda}$ in the reduction
identities \refeq{eq:qmuqnuredfinal} and \refeq{eq:qmuqnuredfinalB}.  In
doing so we set the stage for \refse{se:instabilities}, where we discuss
numerical instability problems and present a systematic approach for their
solution.

\subsection{The reduction basis} \label{se:redbasis}

The reduction identities of~\cite{delAguila:2004nf}
are based on a decomposition of the four-dimensional loop momentum,
\bea
q^\mu=\sum\limits_{i=1}^{4} c_i l_i^\mu, 
\label{eq:qdecomb}
\eea
in a basis $l_1,\dots,l_4$,
formed by 
massless momenta in two orthogonal planes, 
\bea 
l_i^2=0,\qquad
l_{1,2}\cdot l_{3,4} =0{}. 
\label{eq:redbasisprop}
\eea
This \textit{reduction basis} 
is defined in terms of the 
external momenta $p_1,p_2$, which enter the
propagators $D_1$,$D_2$. The 
basis momenta $l_{1,2}$ are chosen in the
plane spanned by  $p_1$ and  $p_2$,
\bea
l^{\mu}_{1} = p^{\mu}_{1} - \alpha_{1} p^{\mu}_{2},  \;\;\;\;\;\;\; l^{\mu}_{2} = p^{\mu}_{2} - \alpha_{2} p^{\mu}_{1}, 
\label{eq:basismom12}
\eea
while $l_{3,4}$ lie in the plane orthogonal to $p_1$,$p_2$ and are defined
as
\bea
l^{\mu}_{3} &=& \bar{v}(l_{1}) \gamma^{\mu} \left(\frac{1-\gamma^{5}}{2}\right) u(l_{2}),\nonumber\\
l^{\mu}_{4} &=& \bar{v}(l_{2}) \gamma^{\mu} \left(\frac{1-\gamma^{5}}{2}\right) u(l_{1}),
\label{eq:basismom34}
\eea
where $u$ and $\bar{v}$ are massless spinors. This definition of $l_{3,4}$ implies $l_3^{*}=e^{\ri\chi}l_4$, where 
$\chi$ is twice the phase difference between the $u$ and $v$ spinors.
The normalization of the basis is chosen such that 
\begin{equation}
\gamma = 2(l_{1} \cdot l_{2}) = -\f{1}{2}(l_{3} \cdot l_{4}){},
\label{eq:basisnorm}
\end{equation}
and the $\alpha_{1,2}$ coefficients in \refeq{eq:basismom12}
read\footnote{
The sign of the square root is chosen such that 
$\pm \sqrt{\Delta}=\mathrm{sign}(\momp{1}\cdot \momp{2})\sqrt{\Delta}$.
This guarantees that the limits 
$\lim\limits_{p_i^2\to 0}\alpha_{i}=0$ 
are approached in a smooth way.
}
\begin{equation}
\alpha_{i} = \frac{p^{2}_{i}}{(p_{1}\cdot p_{2}) \pm \sqrt{\Delta}}, 
\label{eq:al12def}
\end{equation}
where $\Delta$ is related to the 
rank-two Gram determinant $\Delta_{12}=\det(p_i\cdot p_j)$ via
\begin{equation}
\Delta = - \Delta_{12}=  (p_{1}\cdot p_{2})^2 - p^2_{1} p^2_{2}.
\label{eq:Gram12}
\end{equation}

The Gram determinant is related to the normalisation factor $\gamma$
via
\begin{equation}
\gamma = \frac{4\Delta}{(p_{1}\cdot  p_{2}) \pm \sqrt{\Delta}},
\label{eq:gammafactor}
\end{equation}
and these two parameters play a critical role for the stability of the reduction.
In fact, 
in the limit of vanishing Gram determinant,
\refeq{eq:basisnorm} and \refeq{eq:gammafactor} imply that
$(l_1\cdot l_2)\propto (l_3\cdot l_4)\propto\gamma\propto \Delta$. Thus
\bea
\lim\limits_{\Delta\to 0}\;(l_i\cdot l_j)\;=\;0\qquad\forall\; i,j,
\label{eq:parallel}
\eea
which implies that all light-like basis momenta $l_i$ become parallel to each other\footnote{
As a consequence of $l_{1,2}^\mu\in \mathbb{R}$ and
$(l_1 \cdot l_2)=E_{1} E_{2} (1-\cos(\theta_{12}))\to 0$,
the first two basis vectors become parallel to each other, \ie
$l^\mu_{1,2}\to \xi_{1,2}\, \eta^\mu$ with $\xi_{1,2}\in \mathbb{R}$
and $\eta^\mu\in \mathbb{R}$. 
As for the $\mathbb{C}$-valued basis vectors $l_{3,4}$, using 
$l_{3}^*=e^{\ri\chi}{l_{4}}$ one can define their real and imaginary parts 
as 
$l_{3(4),+}=[l_{3(4)}+e^{\ri\chi}{l_{4(3)}}]/2$,
$l_{3(4),-}=[l_{3(4)}-e^{\ri\chi}{l_{4(3)}}]/(2\ri)$,
and show that~\refeq{eq:parallel} leads to 
$(\tilde l_i\cdot \tilde l_j)\to 0$ for all real-valued vectors 
$\tilde l_i,\tilde l_j\in \{l_1,l_2,l_{3,\pm},l_{4,\pm}\}$. 
Thus, similarly as for $l_{1,2}$ we have
$\tilde l_i\to \tilde\xi_i \eta^\mu\; \forall\; i$, and we arrive at
$l^\mu_{3,4}\to z_{3,4} \eta^\mu$ with $z_3=z_4^*\in \mathbb{C}$, \ie
also the basis vectors $l_{3,4}$ become parallel to $l_{1,2}$.
}
leading to severe numerical instabilities
in the decomposition~\refeq{eq:qdecomb}.

Note that in~\cite{delAguila:2004nf} the basis momenta 
$l_i$ contain an additional normalisation factor%
\footnote{More explicitly, 
the basis momenta of~\cite{delAguila:2004nf} 
correspond to $\tilde l_i=\beta l_i$ and the various $l_i$-dependent 
quantities are related in a similar way, 
\eg $\tilde\gamma=\beta^2\gamma$, while $l_i$-independent quantities such as 
$\Delta$ and $\alpha_i$ are identical.
}
\bea
\beta=\frac{1}{1-\alpha_1\alpha_2}\;=\;\pm\,\frac{(p_1\cdot p_2)\pm\sqrt{\Delta}}{2\sqrt{\Delta}},
\label{eq:beta}
\eea
which diverges like $1/\sqrt{\Delta}$ when $\Delta\to 0$.  As a consequence,
in~\cite{delAguila:2004nf} numerical instabilities are in part visible 
as factors $\beta$ in the reduction formulas and in part hidden in the
definition of the basis vectors.  Instead, the basis
momenta defined in~\refeq{eq:basismom12}--\refeq{eq:basismom34} are stable in the
$\Delta\to 0$ limit. Thus, in the 
reduction formulas presented in \refse{se:redform4} and~\ref{se:redform3},
instabilities related to the Gram determinant~\refeq{eq:Gram12}
are fully manifest in the form of inverse powers of the parameter $\gamma\propto \Delta$.
More precisely, for $p_1^2=0$ and $p_2^2\neq 0$, Gram-determinant instabilities 
arise also from  $\alpha_2=\pm p_2^2/(2\sqrt{\Delta})$. However, the parametrisation adopted 
in~\refse{se:redform3} ensures that $\alpha_2$ is always regular.

\subsection{On-the-fly box reduction} %
\label{se:redform4}

In the following we discuss the reduction
identity~\refeq{eq:qmuqnuredfinal}, which can be rewritten in a slightly more
compact form as
\bea 
q^\mu q^\nu
 = \sum_{j=-1}^3\left[ A^{\mu\nu}_{j} +B^{\mu\nu}_{j,\lambda}\,q^{\lambda}\right]D_j,
\label{eq:qmuqnuredfinalcomp}
\eea
with $D_{-1}=1$. Since $q^\mu q^\nu$ is reconstructed in terms
of $D_0,D_1,D_2,D_3$, we denote~\refeq{eq:qmuqnuredfinalcomp}
as box reduction identity, 
although it is applicable to any integrand with $N\ge 4$ loop propagators.
The starting point for its derivation
is given by the
decomposition~\refeq{eq:qdecomb}.  Since the basis momenta $l_{1,2}$ and
$l_{3,4}$ lie in mutually orthogonal planes, it is natural to split the loop
momentum into corresponding components,
\begin{equation}\label{eq:LoopMomDec}
	q^{\mu} = \qpar^\mu+\qperp^\mu,
\end{equation}
with $\qpar^\mu=c_1 l_1^\mu+c_2 l_2^\mu$ and $\qperp^\mu=c_3 l_3^\mu+c_4
l_4^\mu$.  The respective $c_i$ coefficients can be easily related to scalar
products $(q\cdot l_i)$ using~\refeq{eq:redbasisprop}
and~\refeq{eq:basisnorm}.  This leads to,%
\footnote{This decomposition of $q$ corresponds to 
$\qpar^\mu=D^\mu/\gamma$ and $\qperp^\mu=-Q^\mu/(2\gamma)$
in~\cite{delAguila:2004nf}.}
\bea
\qpar^{\mu} &=& \frac{2}{\gamma}\big[\left(q \cdot l_{1} \right) l_{2}^{\mu} +  
\left(q \cdot l_{2} \right) l_{1}^{\mu}\big]
\quad\mbox{and} \nonumber\\
\qperp^{\mu} &=&-\frac{1}{2\gamma}\big[
 (q\cdot l_{3}) l_{4}^{\mu} + (q\cdot l_{4}) l_{3}^{\mu}\big].
\eea
The $\qpar^\mu$ component can be directly reduced to rank zero 
by reconstructing the scalar products $(q\cdot l_{1,2})$ 
in terms of $D_0,D_1,D_2$ using  
\bea
\momp{i}\cdot q &=&\frac{1}{2}\lb D_i-D_0+f_{i0} \rb,\;\;
f_{i0}=\mass{i}^2-\mass{0}^2-\momp{i}^2{}.
\label{eq:fi0}
\eea
This yields
\bea
\qpar^{\mu} &=& \frac{1}{\gamma}\big[
f_{10} r_{2}^{\mu} + f_{20} r_{1}^{\mu} + D_{1} r_{2}^{\mu} + D_{2} r_{1}^{\mu} 
\nonumber\\&&
- D_{0} \left( r_{1}^{\mu} + r_{2}^{\mu} \right)\big], 
\label{eq:qparred}
\eea
with
\begin{equation}
r^{\mu}_{1} = l^{\mu}_{1} - \alpha_{1} l^{\mu}_{2}
\qquad\mbox{and}\qquad
r^{\mu}_{2} =l^{\mu}_{2} -\alpha_{2} l^{\mu}_{1}.
\end{equation}
In order to obtain an identity that reduces also $\qperp$ to 
a linear combination of $D_0,\dots, D_3$ one has to move to rank two
by squaring~\eqref{eq:LoopMomDec} in a way that does not generate 
$\qpar^\mu\qpar^\nu$ terms. To this end one can write 
\bea
q^{\mu}q^{\nu} 
&=&
\frac{1}{2}
\Big[ \left( q^{\mu} -\qperp^{\mu} \right)
\left( q^{\nu} +\qperp^{\nu} \right)+ \qperp^{\mu}\qperp^{\nu}\Big]
 + (\mu \leftrightarrow \nu)=\nonumber\\
&=&
\frac{1}{2}
\Big[ \qpar^\mu \left( q^{\nu} +\qperp^{\nu} \right)+ \qperp^{\mu}\qperp^{\nu}\Big]
 + (\mu \leftrightarrow \nu).
\label{eq:RedEq1}
\eea
Applying~\refeq {eq:qparred} to the rhs of~\refeq{eq:RedEq1} reduces 
$\qpar^\mu q^\nu$ and $\qpar^\mu \qperp^\nu$ to rank one, such that only
\bea
\qperp^{\mu}\qperp^{\nu}=\frac{1}{4\gamma^2}\sum_{i,j=3}^4 
(q\cdot \hat l_{i})(q\cdot \hat l_{j})\,l_i^\mu l_j^\nu,
\label{eq:qperpmunu}
\eea 
with $\hat l_{3,4}=l_{4,3}$, remains to be reduced.
This is achieved by 
means of the relations~\cite{delAguila:2004nf}
\bea
(q\cdot l_{3})(q\cdot l_{4}) &=& \gamma \left( \qpar^{\lambda}\,q_{\lambda} -  D_{0} + m_{0}^2 \right), \\
(q\cdot l_{3(4)})^2 &=& \frac{\gamma}{(p_{3} \cdot l_{4(3)})} 
\nonumber \\ && \hspace{-15mm}
\times\Big[ 
\left( D_{0} + m_{0}^2 - \qpar^{\lambda}\,q_{\lambda} \right) (p_{3} \cdot l_{3(4)})
\nonumber\\
 &&\hspace{-15mm} - 
\left( D_{3} - D_{0} + f_{30} - 2 p_{3, \alpha} \qpar^{\alpha} \right) (l_{3(4),\lambda} q^{\lambda} ) \Big],\qquad
\label{eq:ql34red}
\eea
where the quadratic terms $(q\cdot l_i)(q\cdot l_j)$ with $i,j=3,4$ are
reconstructed in terms of $D_0$ and $D_3$ using also the external momentum
$\momp{3}$.

Combining \refeq{eq:qparred}--\refeq{eq:ql34red} leads to the reduction
identity~\refeq{eq:qmuqnuredfinalcomp} with
\bea
\label{eq:ABmunutensors4}
A_{-1}^{\mu \nu} &=& m_{0}^2 A_{0}^{\mu \nu},
\qquad\qquad
A_{1,2,3}^{\mu \nu} =0,
\nonumber\\
A_{0}^{\mu \nu} &=&  \frac{1}{4 \gamma} \left(\alpha L_{33}^{\mu \nu} + \frac{1}{\alpha} L^{\mu \nu}_{44} - L^{\mu \nu}_{34}\right),
\nonumber\\
B^{\mu \nu}_{\np,\lambda} &=& \sum_{i=1}^3 f_{i0} B^{\mu \nu}_{i,\lambda},
\qquad
B^{\mu \nu}_{0,\lambda} = -\sum_{i=1}^3 B^{\mu \nu}_{i,\lambda},
\nonumber\\
B^{\mu \nu}_{1,\lambda} &= & \frac{1}{4 \gamma^2} 
\Bigg[
\frac{2(p_{3}\cdot r_{2})}{(p_{3} \cdot l_{3})} 
\Big( L_{33}^{\mu \nu} l_{4,\lambda} + \frac{1}{\alpha} L_{44}^{\mu \nu} l_{3,\lambda} 
\Big) \nonumber\\ 
&-& \Big( r_{2}^{\mu} L^{\nu}_{34,\lambda} + r_{2}^{\nu} L^{\mu}_{34,\lambda} \Big)\Bigg]  
+ \frac{1}{\gamma}\Big(r_{2}^{\mu} \delta^{\nu}_{\lambda}- A_0^{\mu \nu} r_{2,\lambda} \Big)
\, ,\nonumber \\[2mm]
B^{\mu \nu}_{2,\lambda} &= & B^{\mu \nu}_{1,\lambda}\big|_{r_1\leftrightarrow r_2},
\nonumber\\
B^{\mu \nu}_{3,\lambda} &=&  - \frac{1}{4 \gamma (p_{3} \cdot l_{3})} 
\left( L_{33}^{\mu \nu} l_{4,\lambda} + \frac{1}{\alpha} L_{44}^{\mu \nu} l_{3,\lambda} \right) \, ,
\eea
where we have introduced
\bea
L^{\mu \nu}_{33}  &=& l_{3}^{\mu} l_{3}^{\nu},\qquad\qquad
L^{\mu \nu}_{44}   = l_{4}^{\mu} l_{4}^{\nu},\nonumber\\
L^{\mu \nu}_{34}  &=& l_{3}^{\mu} l_{4}^{\nu} + l_{4}^{\mu} l_{3}^{\nu}, \qquad\,
\alpha = \frac{p_3 \cdot l_{4}}{p_3 \cdot l_{3}}\, .
\label{eq:alphadef}
\eea
The relations between the
$A^{\mu\nu}_{j}$ and
$B^{\mu\nu}_{j,\lambda}$ 
tensors in the
first two lines of~\refeq{eq:ABmunutensors4} 
follow from the requirement that terms of rank different from two 
vanish on the rhs of~\refeq{eq:qmuqnuredfinalcomp}.
Note also that the tensor $L^{\mu \nu}_{34}$ can be rewritten 
in terms of $l_{1}, l_{2}$ and $g^{\mu \nu}$ as
\begin{equation}\label{eq:GmunuTrick}
L^{\mu \nu}_{34} = 4\left( l_{1}^{\mu} l_{2}^{\nu} + l_{2}^{\mu} l_{1}^{\nu} - \frac{\gamma}{2} g^{\mu \nu}\right).
\end{equation}

\subsection{On-the-fly triangle reduction}
\label{se:redform3}

The identity~\refeq{eq:qmuqnuredfinalB}, which reconstructs $q^\mu q^\nu$ in
terms of $D_0,D_1,D_2$ at the integral level, will be denoted as triangle
reduction.  Its derivation is based on the observation that 
the only terms that involve $D_3$ and $p_3$ 
in \refse{se:redform4},  \ie the squared scalar products $(q\cdot l_{3})^2$ and $(q\cdot
l_{4})^2$ in~\refeq{eq:ql34red},
do not contribute in three-point integrals of rank
$R\le 3$.  More precisely~\cite{delAguila:2004nf}, for \mbox{$i=3,4$},
\bea
\int \rd^{D}\, \bar{q}\, 
\frac{\left( q \cdot l_{i}\right)^{2}}{\bar{D_{0}} \bar{D_{1}} \bar{D_{2}}}  
\;=\;
\int \rd^{D\,} \bar{q}\, 
\frac{\left( q \cdot l_{i}\right)^{2} q^{\rho}}{\bar{D_{0}} \bar{D_{1}} \bar{D_{2}}} 
\;= 0 {}.
\label{eq:red3int}
\eea
As a consequence, the derivations of~\refse{se:redform4} are also
applicable to three-point functions at the integral level upon
replacing~\refeq{eq:qperpmunu} by
\bea
\qperp^{\mu}\qperp^{\nu} \to \frac{1}{4\gamma^2}
(q\cdot l_{3})(q\cdot l_{4})\,L^{\mu\nu}_{34}.
\label{eq:qperpmunu3}
\eea 
In this way one arrives at the reduction identities
\bea
&&\int\!\mathrm{d}^D\!\momq\,
\frac{q^\mu q^\nu S(q)}{\Dbar{0}\cdots \Dbar{2}}=\nonumber\\
&&=
\sum_{j=-1}^2
\int\!\mathrm{d}^D\!\momq\,
\frac{\left(A^{\mu\nu}_{j}+B^{\mu\nu}_{j,\lambda}\,q^\lambda\right) S(q)}{\Dbar{0}\cdots 
\slashed{\Dbar{j}}\cdots\Dbar{2}},
\label{eq:3pointintredA}
\eea
where $S(q)=S+S_\rho q^\rho$ is an arbitrary rank-one polynomial, and the
tensors $A_j^{\mu\nu}$ and  $B_{j,\lambda}^{\mu\nu}$ are obtained
from~\refeq{eq:ABmunutensors4} through the trivial replacements
\bea
L^{\mu \nu}_{33} \to 0, \quad
L^{\mu \nu}_{44} \to 0.
\label{eq:triangleOFRrepl}
\eea

\subsection{Treatment of Gram-determinant instabilities}
\label{se:instabilities}

As pointed out in \refse{se:redbasis}, when the rank-two Gram determinant
$\Delta_{12}$ tends to zero the reduction
basis~\refeq{eq:basismom12}--\refeq{eq:basismom34} becomes degenerate.  This
leads to spurious singularities that manifest themselves as factors
$\gamma^{-k}\propto \Delta_{12}^{-k}$ in the reduction identities.
In practice, the residues of $\Delta_{12}^{-k}$ poles are suppressed at
$\mathcal{O}(\Delta_{12}^{k})$ as a result of subtle numerical cancellations
between various contributions.  Thus, for $\Delta_{12}\to 0$ the results
of the reduction are finite but suffer from severe numerical instabilities.
As can be seen from~\refeq{eq:ABmunutensors4}, spurious singularities reach
the maximum power $k=2$, \ie each reduction step results in a numerical
instability that scales quadratically in the inverse Gram determinant
$\Delta_{12}$.

\begin{figure*}[t!]
\begin{center}
 \includegraphics[width=0.3\textwidth]{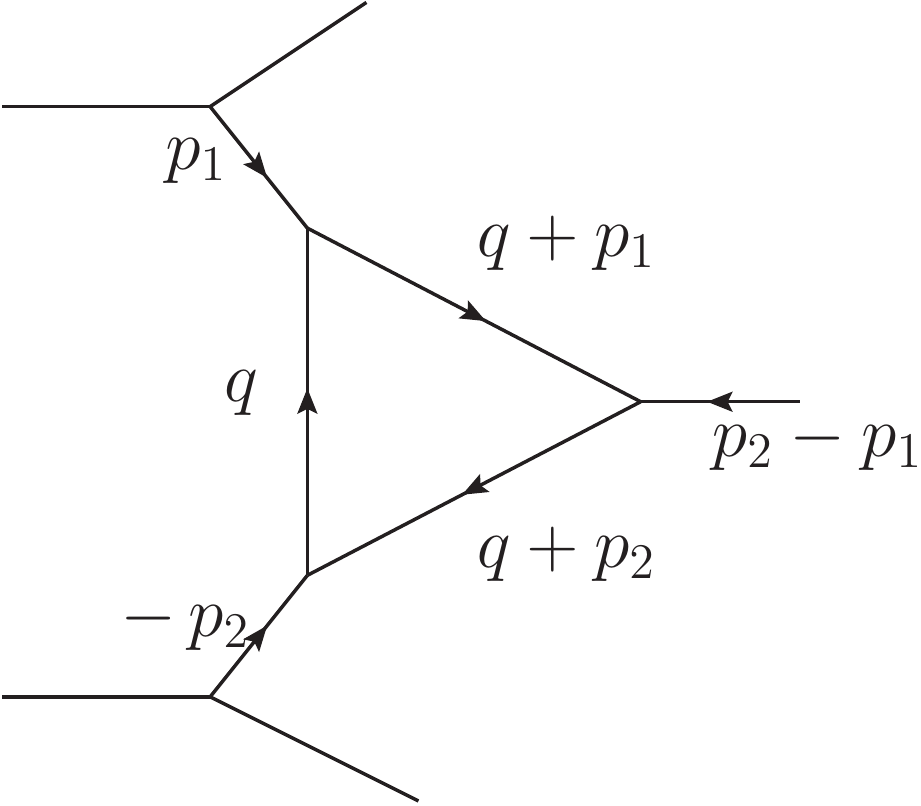}
 \end{center}
 \caption{Triangle $t$-channel  (sub)topology that gives rise to 
$\Delta_{12}\to 0$ numerical instabilities when
$(\momp{2}-\momp{1})^2=0$ and $p_1^2\to p_2^2$.
\label{fig:tchanneltriangle}}
\end{figure*}

The reduction~\refeq{eq:ABmunutensors4} involves also spurious singularities
related to the rank-three Gram determinant, $\Delta_{123}$, which enter
through the terms $(l_3\cdot p_3)^{-1}\propto
|\Delta_{123}|^{-1/2}$~\cite{delAguila:2004nf}.  
However, as compared to the $\Delta_{123}\to 0$ case, $\Delta_{12}\to 0$
instabilities are statistically more likely and are also more enhanced due to their 
$\Delta_{12}^{-2}$ scaling behaviour.
In fact, studying high statistics samples for various representative
processes, we have found that the numerical instabilities of the on-the-fly 
reductions of \refses{se:redform4}{se:redform3} are very strongly correlated to the
parameter $\gamma\propto \Delta_{12}$. 
Therefore, as we will see in the following, avoiding $\Delta_{12}\to 0$ spurious singularities in a
systematic way makes it possible to reach
excellent numerical stability.

\subsubsection{Box reduction}
\label{eq:boxstability}

In the case of the on-the-fly box reduction~\refeq{eq:qmuqnuredfinalcomp},
poles in $\Delta_{12}$ arise only through the factors $1/\gamma^2$ and
$1/\gamma$ in~\refeq{eq:ABmunutensors4}, which are a direct consequence of 
the choice of the
momenta $p_1,p_2$ for the construction of the reduction
basis~\refeq{eq:basismom12}--\refeq{eq:basismom34}.
Since rank-two Gram-determinant instabilities depend on only two of the
three available momenta, they can be easily avoided by constructing the
basis with $p_1, p_3$ or $p_2, p_3$ instead of $p_1, p_2$, depending on the
values of the respective Gram determinants
$\Delta_{13},\Delta_{23},\Delta_{12}$.

In practice, in order to avoid small rank-two Gram determinants 
we perform the box reduction upon a permutation,
\bea
\{D_1, D_2, D_3\} \;\longrightarrow\; \{D_{i_1}, D_{i_2}, D_{i_3}\},
\label{eq:propperm}
\eea
which orders loop denominators with related momenta and masses
in such a way that 
\bea
\frac{|\Delta_{i_1i_2}|}{Q^4_{i_1i_2}}\;=\; \max\left\{
\frac{|\Delta_{12}|}{Q^4_{12}},\,
\frac{|\Delta_{13}|}{Q^4_{13}},\,
\frac{|\Delta_{23}|}{Q^4_{23}}\right\}.
\label{eq:maxgramdet}
\eea
The scales $Q^2_{ij}$, which render the above ratios dimensionless,
are defined as the largest element of the respective Gram matrices, \ie
\bea
Q^2_{ij}=\max\{|p_i\cdot p_i|, |p_i^2|, |p_j^2|\}.
\label{eq:q2ijdef}
\eea
 Note that the
permutation~\refeq{eq:propperm} can be applied without changing the order of
the corresponding segments $S_i(q)$, \ie without any modification of the
open-loop dressing recursion.  Moreover, the choice of the optimal
permutation~\refeq{eq:propperm}--\refeq{eq:maxgramdet} can be done in a
fully flexible way at runtime and locally in individual reduction steps,
depending on the kinematics of the actual phase space point.
In practice~\refeq{eq:propperm} is applied only 
to compute the reduction basis and 
the coefficients~\refeq{eq:ABmunutensors4},
which are then converted back to the original ordering.

Avoiding a spurious $\Delta_{12}\to 0$ singularity
with~\refeq{eq:propperm}--\refeq{eq:maxgramdet} does not guarantee its disappearance
in future reduction steps.
In fact, all reduced contributions where $D_1$ and  $D_2$
remain unpinched will still involve the same small Gram determinant.
However, the permutation trick~\refeq{eq:propperm} can be iterated 
as long as $N\ge 4$ loop denominators are available.
In this way, rank-two Gram-determinant instabilities can be
isolated in triangle contributions, which 
arise only at later steps of the open-loop recursion
for loop diagrams with $N> 3$.

\subsubsection{Triangle reduction}
\label{eq:trianglestability}

For the case of triangle topologies one can show that, excluding regions where
the external particles become soft or 
collinear,
vanishing $\Delta_{12}$ Gram determinants can 
arise only from the $t$-channel topology depicted
in \reffi{fig:tchanneltriangle}, where the triangle
exhibits two space-like external momenta, 
$p_1$ and $p_2$, 
and a time-like external momentum, $p_2-p_1$. Since the Gram determinant vanishes when 
$p_1^2\to p_2^2$, we adopt the parametrisation
\bea
\momp{1}^2 = -p^2<0,\;\;
\momp{2}^2 = -p^2 (1+\delta),\;\;
(\momp{2}-\momp{1})^2 = 0,
\label{eq:triangleparam}
\eea
where $p_1$ and $p_2$ can be ordered such that  $\delta>0$. The parameters 
$\Delta$ and $\gamma$ are related to $\delta$ via
\bea 
\sqrt{\Delta} = \frac{p^2}{2} \delta 
\qquad\mbox{and}\qquad
\gamma = -p^2 \delta^2 {},
\eea
\ie the $\Delta\to 0$ limit corresponds to $\delta\to 0$.

In kinematic regions with small $\delta$, the numerical reduction of rank-$r$ triangles via
iterated on-the-fly reductions~\refeq{eq:ABmunutensors4} 
and
subsequent rank-one reductions \refeq{eq:PVtrianglered}--\refeq{eq:Bmunudef}
can lead to spurious singularities up to order $1/\delta^{4r-2}$.  
In order to avoid numerical instabilities, we first perform a full {\it
algebraic} reduction of 3-point tensor integrals, 
\bea
C^{\mu_1\dots\mu_r}
&=& 
\frac{(2\pi\mu)^{2\eps}}{\ri\pi^2}
\int\!\mathrm{d}^D\!\momq\,\frac{q^\mu_1\cdots q^{\mu_r}}{\Dbar{0}\Dbar{1}\Dbar{2}},
\label{eq:Cmudefrank}
\eea
to scalar integrals.  This leads to a cancellation of the leading spurious
singularites, and the resulting analytic expressions for rank-$r$ triangles
involve only poles up to order $1/\delta^{r+1}$.  For instance, for the case
of triangles with massless internal lines, $m_0=m_1=m_2=0$, reducing tensor
integrals of rank $r=1,2,3$ to scalar integrals we obtain
\bea
\label{eq:exactriangA}
&&C^\mu = \frac{2}{\delta ^2 p^2} \left\{
B_0(-p^2,0,0) \left[-\momp{1}^{\mu}\lb 1+\f{\delta}{2} \rb+\momp{2}^{\mu}\right]       \right. \nonumber \\ & & \left.
+B_0\left(-p^2 (1+\delta),0,0\right) \left[\momp{1}^{\mu}  (1+\delta)  
- \momp{2}^{\mu} \lb 1+\f{\delta}{2}\rb \right]   \right\}        \nonumber \\ & & 
+\frac{1}{\delta } C_0\left(-p^2,-p^2 (1+\delta),0,0,0\right) \nonumber 
\Big[-\momp{1}^{\mu}(1+\delta)+\momp{2}^{\mu}\Big]\\ & &+\frac{1}{\delta p^2} B_0(0,0,0) \left[\momp{2}^{\mu}-\momp{1}^{\mu}\right] ,
\\[2mm]
\label{eq:exactriangB}
&&C^{\mu\nu}=  
B_0(-p^2,0,0) \left[-\frac{g^{\mu\nu}}{4 \delta }
+\f{\momp{11}^{\mu\nu}}{p^2} \left(\frac{3}{\delta ^3}+\frac{5}{\delta ^2}+\frac{3}{2 \delta}\right)\right. \nonumber \\ & & \left.
-\f{\momp{12}^{\mu\nu}}{p^2} \left(\frac{3}{\delta ^3}+\frac{5}{2 \delta ^2}\right)+\f{ \momp{22}^{\mu\nu}}{p^2}\frac{3}{\delta ^3}\right]
+B_0\left(-p^2 (1+\delta),0,0\right) \nonumber \\ & &\times
\left[\frac{g^{\mu\nu}}{4 \delta }+\frac{1}{4} g^{\mu\nu}
-\f{\momp{11}^{\mu\nu} }{p^2}\left(\frac{3}{\delta ^3 }+\frac{6}{\delta ^2   }   +\frac{3}{\delta  }\right)                              
   +\f{\momp{12}^{\mu\nu} }{p^2} \left(\frac{3}{\delta ^3 }\nonumber \right.\right.\\ 
   & &\left.\left.+\frac{7}{2 \delta ^2 }+\frac{1}{2 \delta  }\right)                
   +\f{\momp{22}^{\mu\nu} }{p^2} \left(-\frac{3}{\delta ^3   }-\frac{1}{\delta ^2 }+\frac{1}{2 \delta  }\right)\right]             
+B_0(0,0,0)  \nonumber \\ & &\times\left[\f{\momp{11}^{\mu\nu}}{p^2} \left(\frac{1}{\delta ^2 }+\frac{3}{2 \delta }\right)
-\f{\momp{12}^{\mu\nu}}{p^2} \left(\frac{1}{\delta ^2 }+\frac{1}{2 \delta  }\right)                     
   +\f{\momp{22}^{\mu\nu}}{p^2} \left(\frac{1}{\delta ^2 }   \right.\right. \nonumber \\ & & \left.\left. -\frac{1}{2 \delta    }\right)\right]    
+C_0\left(-p^2,-p^2 (1+\delta),0,0,0\right) \Big[
\momp{11}^{\mu\nu}\left(\frac{1}{\delta ^2} \right. \nonumber \\ & & \left.+\frac{2}{\delta }+1\right)                     
   -\momp{12}^{\mu\nu}\left(\frac{1}{\delta ^2}+\frac{1}{\delta }\right)  +\frac{\momp{22}^{\mu\nu}}{\delta ^2}\Big]  
-\f{1}{2} \left[-\frac{1}{2} g^{\mu\nu} \right.\nonumber \\ & &\left.
+\f{\momp{11}^{\mu\nu} }{p^2}  \left(\frac{2}{\delta ^2 }+\frac{2}{\delta  }\right)                     
   -\f{\momp{12}^{\mu\nu}}{p^2} \left(\frac{2}{\delta ^2 }+\frac{1}{\delta  }\right)+\frac{2 \momp{22}^{\mu\nu}}{\delta ^2 p^2}\right],
\\[2mm]
\label{eq:exactriangC}
&&C^{\mu\nu\rho}=
B_0(-p^2,0,0) 
\left[
\momp{111}^{\mu\nu\rho} \left(-\frac{11}{3 \delta ^4 p^2}-\frac{10}{\delta ^3 p^2} \right.\right. \nonumber \\ & & \left.\left.
-\frac{17}{2 \delta ^2 p^2}-\frac{11}{6 \delta  p^2}\right)
+\momp{112}^{\mu\nu\rho} \left(\frac{11}{3 \delta ^4 p^2}+\frac{20}{3\delta ^3 p^2} 
+\frac{17}{6 \delta ^2 p^2}\right)\right. \nonumber \\ & & \left.
+\momp{122}^{\mu\nu\rho} \left(-\frac{11}{3 \delta ^4 p^2}-\frac{10}{3 \delta ^3 p^2}\right)
+\momp{222}^{\mu\nu\rho} \frac{11}{3 \delta ^4 p^2}                                                                                         
+\momp{1}^{\{\mu}g^{\nu\rho\}}\right. \nonumber \\ & & \left.\times\left(\frac{1}{12 \delta ^2}+\frac{1}{6 \delta}\right)
-\frac{\momp{2}^{\{\mu}g^{\nu\rho\}}}{12 \delta ^2}\right]  
+B_0\left(-p^2 (1+\delta),0,0\right)\nonumber \\ & & \times\left.\Big[  
\momp{111}^{\mu\nu\rho} \left(\frac{11}{3 \delta ^4 p^2}+\frac{11}{\delta ^3 p^2}
+\frac{11}{\delta ^2 p^2}+\frac{11}{3 \delta  p^2}\right)
+\momp{112}^{\mu\nu\rho} \right. \nonumber \\ & & \times\left.\left(-\frac{11}{3 \delta ^4 p^2}-\frac{23}{3 \delta ^3 p^2}
-\frac{13}{3 \delta ^2 p^2}-\frac{1}{3 \delta  p^2}\right)
+\momp{122}^{\mu\nu\rho} \left(\frac{11}{3 \delta ^4 p^2}\right.\right. \nonumber \\ & & \left.\left.
+\frac{13}{3 \delta ^3 p^2}
+\frac{1}{2 \delta ^2 p^2}-\frac{1}{6 \delta  p^2}\right)
+\momp{222}^{\mu\nu\rho} \left(-\frac{11}{3 \delta ^4 p^2}-\frac{1}{\delta ^3 p^2}\right.\right. \nonumber \\ & & \left.\left.
+\frac{1}{2 \delta ^2 p^2}-\frac{1}{3 \delta p^2}\right)
+\momp{1}^{\{\mu}g^{\nu\rho\}}\left(-\frac{1}{12 \delta ^2}-\frac{1}{6 \delta }-\frac{1}{12}\right) \right. \nonumber \\ & & \left.
+\momp{2}^{\{\mu}g^{\nu\rho\}}\left(\frac{1}{12 \delta ^2}-\frac{1}{12}\right) \right] 
+B_0(0,0,0) 
\left[\momp{111}^{\mu\nu\rho} \left(-\frac{1}{\delta ^3 p^2}\right.\right. \nonumber \\ & & \left. \left.
-\frac{5}{2 \delta ^2 p^2}-\frac{11}{6 \delta  p^2}\right) 
+\momp{112}^{\mu\nu\rho} \left(\frac{1}{\delta ^3 p^2}+\frac{3}{2 \delta ^2 p^2}+\frac{1}{3 \delta  p^2}\right)\right. \nonumber \\ & & \left.
+\momp{122}^{\mu\nu\rho} \left(-\frac{1}{\delta ^3 p^2}-\frac{1}{2 \delta ^2 p^2}+\frac{1}{6 \delta  p^2}\right)
+\momp{222}^{\mu\nu\rho} \left(\frac{1}{\delta ^3 p^2}\right.\right. \nonumber \\ & & \left. \left.
-\frac{1}{2 \delta ^2 p^2}+\frac{1}{3 \delta  p^2}\right)\right]
+C_0\left(-p^2,-p^2 (1+\delta),0,0,0\right) \nonumber\\ &&\times
\Big[ 
\momp{111}^{\mu\nu\rho}\left(-\frac{1}{\delta ^3}
-\frac{3}{\delta ^2}-\frac{3}{\delta }-1\right)
+\momp{112}^{\mu\nu\rho}\left(\frac{1}{\delta ^3}+\frac{2}{\delta ^2}+\frac{1}{\delta }\right)  \nonumber \\ & & 
+\momp{122}^{\mu\nu\rho}\left(-\frac{1}{\delta ^3}-\frac{1}{\delta^2}\right)
+\frac{\momp{222}^{\mu\nu\rho}}{\delta ^3}\Big] 
-\f{1}{2}
\left[\momp{111}^{\mu\nu\rho} \left(-\frac{10}{3 \delta ^3 p^2}\right.\right. \nonumber \\ & & \left.\left.
-\frac{22}{3 \delta ^2 p^2}-\frac{37}{9 \delta  p^2}\right)
+\momp{112}^{\mu\nu\rho} \left(\frac{10}{3 \delta^3 p^2}+\frac{14}{3 \delta ^2 p^2}+\frac{10}{9 \delta  p^2}\right)\right. \nonumber \\ & & \left.
+\momp{122}^{\mu\nu\rho} \left(-\frac{10}{3 \delta ^3 p^2}-\frac{2}{\delta ^2 p^2}+\frac{2}{9 \delta  p^2}\right)
+\momp{222}^{\mu\nu\rho} \left(\frac{10}{3 \delta ^3 p^2}\right.\right. \nonumber \\ & & \left.\left.
-\frac{2}{3 \delta ^2 p^2}+\frac{1}{9 \delta  p^2}\right)
+\momp{1}^{\{\mu}g^{\nu\rho\}}\left(\frac{1}{6 \delta }+\frac{5}{18}\right)
+\momp{2}^{\{\mu}g^{\nu\rho\}} \right. \nonumber \\ & & \times\left.\left(\frac{1}{9}-\frac{1}{6 \delta }\right) \right],
\eea
with the tensors
\bea
\momp{ij}^{\mu\nu}&=&\sum_{\pi(i,j)}\momp{i}^\mu \momp{j}^\nu,\qquad
\momp{ijk}^{\mu\nu\rho}=\sum_{\pi(i,j,k)}\momp{i}^\mu \momp{j}^\nu\momp{k}^\rho,\nonumber\\
\momp{i}^{\{\mu}g^{\nu\rho\}}&=&
\momp{i}^{\mu}g^{\nu\rho} +\momp{i}^{\nu}g^{\mu\rho} +\momp{i}^{\rho}g^{\mu\nu},
\eea
where $i,j,k=1,2$, and the sums are restricted to inequivalent permutations, \eg
\mbox{$\momp{11}^{\mu\nu}=\momp{1}^\mu \momp{1}^\nu$,} \mbox{
$\momp{112}^{\mu\nu}=
\momp{1}^\mu \momp{1}^\nu\momp{2}^\rho+
\momp{1}^\mu \momp{2}^\nu\momp{1}^\rho+
\momp{2}^\mu \momp{1}^\nu\momp{1}^\rho$,} 
etc.

Analytic expression of type~\refeq{eq:exactriangA}--\refeq{eq:exactriangC}
guarantee a reduced sensitivity to Gram-determinant instabilities.
Thus they are used as 
default for the reduction of triangles configurations
of type \refeq{eq:triangleparam} with $\delta>\deltathr$.
The freely adjustable threshold parameter $\deltathr$
is set to $\deltathr=10^{-3}$.
To avoid numerical instabilities in regions with
$\delta< \deltathr$ we perform 
systematic expansions in $\delta$.  In particular, for a 
complete cancellation of the $1/\delta$ poles also 
the $\delta$-dependent $C_0$ and $B_0$ scalar integrals have to be
expanded in $\delta$.
To this end we use {\sc LiteRed}~\cite{Lee:2013mka}, and expanding the
residues of $\delta^{-k}$ poles up to order $\delta^{k+m}$ we obtain 
regular Taylor series including terms up to order $\delta^m$.

For the case $m_1=m_2=m_3=0$, 
expanding up to order $\delta^2$
yields 
\bea
\label{eq:expandtriangA}
 C^\mu&=& \f{\momp{1}^{\mu}+\momp{2}^{\mu}}{2 p^2} \left(1-\dbzero\right)
+ \delta\; \f{\momp{1}^{\mu}+2\momp{2}^{\mu}}{6 p^2} \dbzero
\nonumber\\ &&
- \delta^2\; \f{\momp{1}^{\mu}+3\momp{2}^{\mu}}{12 p^2}\; \left(\dbzero +\f{1}{2}\right)
   +\mathcal{O}(\delta^3) {},
 \\[2mm]
\label{eq:expandtriangB}
C^{\mu\nu}&=& \f{2 \momp{11}^{\mu\nu}+\momp{12}^{\mu\nu}+2 \momp{22}^{\mu\nu}}{6 p^2}
\left(\dbzero -\f{1}{2}  \right)+  
\frac{1}{4} g^{\mu\nu} \bzero
\nonumber\\ &&
-\delta\left[\f{\momp{11}^{\mu\nu}+\momp{12}^{\mu\nu}+3 \momp{22}^{\mu\nu}}{12 p^2}  
\left(\dbzero +\f{1}{2}\right)  
 -\frac{1}{8} g^{\mu\nu}\right]\nonumber   \\ & &
+ \delta^2\left[\f{2 \momp{11}^{\mu\nu}+3 \momp{12}^{\mu\nu}+12 \momp{22}^{\mu\nu}}{60 p^2}
\left(\dbzero +1 \right) \right.\nonumber\\ &&\left.
+\frac{1}{24} g^{\mu\nu}\right]
+\mathcal{O}(\delta^3) {},
 \\[2mm]
\label{eq:expandtriangC}
C^{\mu\nu\rho}&=&
\f{3\momp{111}^{\mu\nu\rho}+ \momp{112}^{\mu\nu\rho}+ \momp{122}^{\mu\nu\rho}+3 \momp{222}^{\mu\nu\rho}}{12 p^2}
\left(\f{1}{6}-\dbzero\right) \nonumber\\ &&
- \f{ \momp{1}^{\{\mu}g^{\nu\rho\}} + p^2 \momp{2}^{\{\mu}g^{\nu\rho\}}}{72} 
\left(6 \bzero +1 \right)\nonumber\\ & &
+\delta\left[\f{3 \momp{111}^{\mu\nu\rho}+2 \momp{112}^{\mu\nu\rho}+3 \momp{122}^{\mu\nu\rho}+12 \momp{222}^{\mu\nu\rho}}{60 p^2}  \right.\nonumber\\ &&\left.
\times\left(\dbzero+\f{5}{6}\right)
+\f{1}{36}\momp{1}^{\{\mu}g^{\nu\rho\}}+\f{1}{18}\momp{2}^{\{\mu}g^{\nu\rho\}}
\right]\nonumber\\ & &
-\delta^2\left[
\f{\momp{111}^{\mu\nu\rho}+\momp{112}^{\mu\nu\rho}+2 \momp{122}^{\mu\nu\rho}+10 \momp{222}^{\mu\nu\rho}}{60 p^2} \right.\nonumber\\ &&\left.
\times\left(\dbzero+\f{4}{3}\right)
-\f{1}{144}  \momp{1}^{\{\mu}g^{\nu\rho\}}-\f{1}{48} \momp{2}^{\{\mu}g^{\nu\rho\}}
\right]\nonumber\\ &&
+\mathcal{O}(\delta^3) {},
\eea
where 
\bea
\bzero &=& B_0(-p^2,0,0),\nonumber\\
\dbzero &=& B_0(-p^2,0,0) - B_0(0,0,0).
\eea
Similar results have been
obtained for the case of massive internal propagators.
More precisely, we have implemented all needed mass
configurations for NLO QCD calculations, \ie $(\mass{0}, \mass{1},
\mass{2})=(0,0,0)$, $(m,m,m)$, $(0,m,m)$ and $(m,0,0)$, with $m>0$,
including terms up to order $\delta^2$ in the expansions.  The extra cases
needed for NLO EW calculations, \ie $(0,m_1,m_2)$, $(m_1,m_2,m_2)$ and $(m_1,m_2,m_3)$, with
$m_i>0$, will be implemented soon.

\begin{figure*}[t!]
\bce
\includegraphics[width=0.46\textwidth]{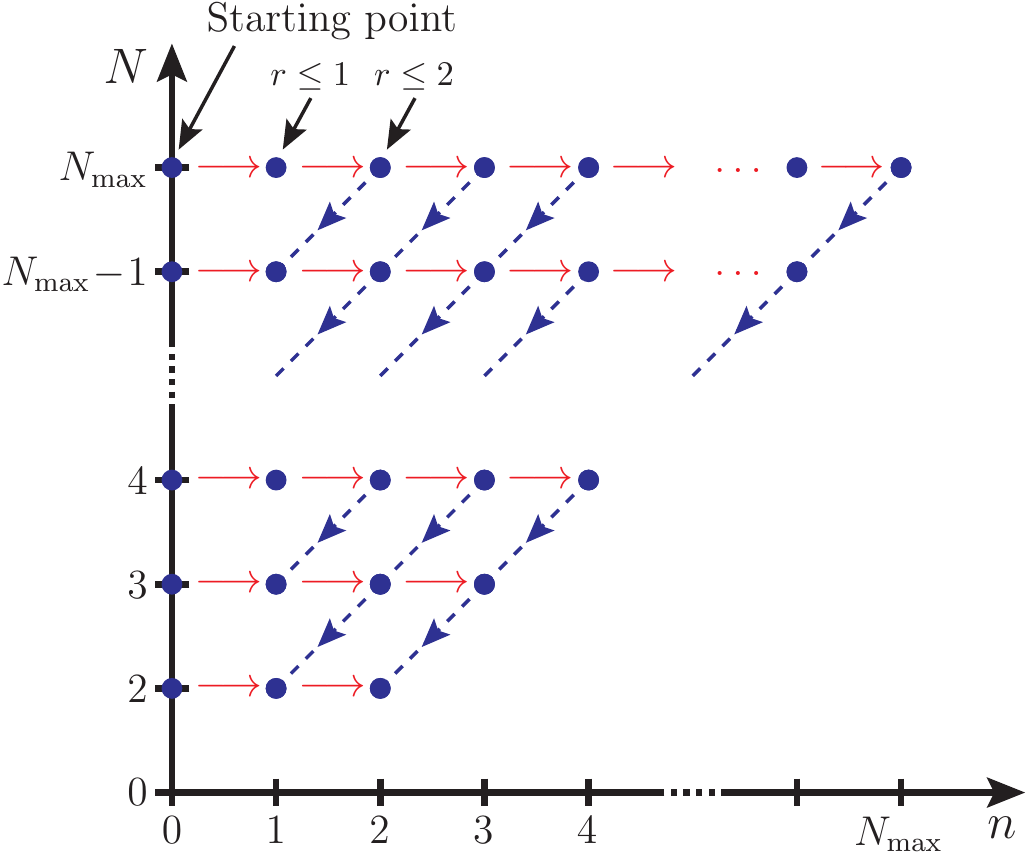}
\ece
\caption{Evolution of the total number $N$ of loop segments and 
the number $n$ of dressed segments during the 
open-loop recursion.
Horizontal and diagonal arrows describe, respectively,
dressing steps and the generation of pinched subtopologies 
in the on-the-fly reduction.
Corresponding unpinched contributions, where the rank is reduced but $(N,n)$
remains unchanged are not shown.
The algorithm starts at $(\nmax,0)$ and 
first proceeds towards the highest possible
$n$ before moving one step lower in $N$.}
\label{fig:OL2alg}
\end{figure*}

Since the analytic expressions~\refeq{eq:exactriangA}--\refeq{eq:exactriangC}
and their expansions~\refeq{eq:expandtriangA}--\refeq{eq:expandtriangC}
reduce triangles of rank $r$ to rank zero in a single step, 
in order to be able to apply them in a fully flexible way,
the on-the-fly reduction of triangles~\refeq{eq:qmuqnuredfinalB}
has to be postponed at the end of the open-loop recursion,
where triangle contributions have reached the maximum rank.
This means that, in addition to the contributions listed in \refeq{list:TIs},
we also generate $N=3$ terms with $R=3$.  After taking the
trace~\refeq{eq:mergingtrace}, depending on the actual value of $\delta$, 
the reduction of triangles is done either using the expansion
 formulas or the on-the-fly reduction~\refeq{eq:qmuqnuredfinalB}
followed by Passarino-Veltman reduction steps~\refeq{eq:PVtrianglered}--\refeq{eq:Bmunudef}.

\section{Implementation and performance} \label{sec:algorithm}

This section summarises the key structure of the 
new algorithm, outlines some aspects of its implementation,
and presents technical performance studies.

\subsection{Structure and implementation of the new algorithm}

Similarly as for the original open-loop method, given a certain scattering
process the algorithm starts with the generation of all tree and one-loop
diagrams.  This is done in symbolic form, including only topological
information and particle content.  
One-loop diagrams are colour stripped and cut-open, and the interference
of their colour structure with the Born amplitude is taken as initial
condition for the open-loop recursion.

The recursion is organised by grouping open loops according to the total
number $N$ of segments and the number $n$ of dressed segments. 
Dressing steps increase $n$ and reduction steps reduce $N$ or keep it constant. Thus, 
as illustrated  in \reffi{fig:OL2alg},
the $(N,n)$ groups are dressed through a series of 
iterations with $N=\nmax,\nmax-1,\nmax-2,\dots$,
where $N$ is kept fixed while all segments $n=0,1,2,\dots,N$ 
are processed.
A step of the open-loop recursion, to be applied to all objects in 
an $(N,n)$-group, consists of the following operations:
\begin{itemize}

 \item[1.] merge all open loops with the same one-loop topology, the same
cut and the same undressed segments into a single object.  Note that one
can merge open loops of different rank.

 \item[2.] For each open loop with $n<N$ and rank $R$ determine
the rank $\Rnext$ that would be reached by performing the subsequent dressing step.

\item[2.a] If $\Rnext=3$ or if $n=N$  and $R=2$ avoid the dressing step and
perform an on-the-fly reduction, which generates
unpinched and pinched terms with $R=1$.
The unpinched terms remain in the $(N,n)$ group.
As for the pinched terms, 
if the adjacent segments
of a pinched propagator are dressed they can be reduced to a single effective segment,
thereby turning $(N,n)\to (N-1,n-1)$. Otherwise, contributions 
with an undressed pinch stay in the $(N,n)$ group.

 \item[2.b] If $n<N$ and $\Rnext\le 2$ perform a dressing step, which turns $(N,n)\to
(N,n+1)$.  If this step dresses a pinched propagator that was previously
undressed, the corresponding segments can be reduced to a single effective
segment turning $(N,n+1)\to (N-1,n)$.

 \item[3.] Sort all open loops into the proper group and repeat steps 1--3
for all open loops in the $(N,n)$-group until the group is empty or $n=N$
and all open loops in the group have rank $R\le\Rlast$ (see below).
\end{itemize}

{
\renewcommand{\arraystretch}{1.25}
\begin{table*}[t]
\begin{center}
\begin{tabular}{c|cccc}
method                & \parbox{22mm}{\bce tensor  \\  reduction \ece} 
& \parbox{22mm}{\bce scalar  \\  integrals \ece} 
& \parbox{22mm}{\bce diagram  \\   merging  \ece} 
& \parbox{22mm}{\bce helicity  \\  summation  \ece}\\[-1.5ex]
\hline
{\sc OpenLoops\,2}  & on-the-fly       & {\sc Collier}    & on-the-fly      & on-the-fly   \\
{\sc OpenLoops\,2+Collier}     & {\sc Collier}    & {\sc Collier}    & on-the-fly      & on-the-fly   \\
{\sc OpenLoops\,1+Collier}     & {\sc Collier}    & {\sc Collier}    & parent--child   & standard   \\ 
{\sc OpenLoops\,1+Cuttools}    & {\sc Cuttools}   & {\sc OneLOop}    & parent--child   & standard   \\
Quad Precision  & {\sc Cuttools}   & {\sc OneLOop}          & parent--child   & standard   
\end{tabular}
\end{center}
\caption{List of the different variants of the {\sc OpenLoops} program 
that are compared in \refses{se:speedbench}{se:stabbench}. 
As third-party tools we use {\sc Collier1.2}~\cite{Denner:2016kdg}, {\sc Cuttools1.9.5}~\cite{Ossola:2007ax} and {\sc OneLOop3.6.1}~\cite{vanHameren:2010cp}. 
}
\label{tab:methods}
\end{table*}
}

Topologies with $N=2$ are dressed without reduction, and, in order to enable
Gram-determinant expansions (see~\refse{eq:trianglestability}), also $N=3$
contributions are dressed without on-the-fly reduction.
Thus $\Rlast=N$ for $N=2,3$, while
open loops with $N\ge 4$ are reduced on-the-fly down to $\Rlast=1$.
The algorithm starts at $(N,n)=(\nmax,0)$ and terminates with the
dressing of two-point contributions at $(N,n)=(2,2)$.
At this point all open loops are closed with the trace
operation~\refeq{eq:mergingtrace}, and the last reduction steps described
in~\ref{app:intred} are applied.

The above on-the-fly algorithm has been implemented in the framework of the
original {\sc OpenLoops} program~\cite{hepforge}, which consists of a computer-algebraic code generator
written in {\sc Mathematica} and a numerical part written in {\sc
Fortran\,90}. 
Given an arbitrary Standard Model process, the {\sc Mathematica} 
generator simulates the full chain of recursion steps in
symbolic form and translates it into {\sc Fortran\,90} code for the
calculation of the actual scattering amplitude.  The only external tools
that need to be interfaced to the new {\sc OpenLoops} program are
{\sc Feynarts}~\cite{Hahn:2000kx}, for the generation 
of tree and one-loop diagrams, and {\sc Collier}~\cite{Denner:2016kdg},
for the calculation of scalar integrals.
All other aspects of the open-loop method are directly implemented as
process-independent {\sc Mathematica} and {\sc Fortran} routines.  This
includes the management of colour algebra, the kernels of the dressing
recursion at tree and one-loop level, the on-the-fly and integral
reductions, the helicity bookkeeping system, $R_2$ rational terms, UV
counterterms, and several other aspects.

The entire program is fully automated, the new on-the-fly methods are
implemented and widely tested at NLO QCD and they will soon be extended to NLO EW.
These methods will be made publicly available with
the upcoming release of {\sc OpenLoops\,2}.
Similarly as for {\sc OpenLoops\,1},
numerical routines generated with the new on-the-fly techniques 
will be accessible through an automated download and installation 
system and the standard {\sc OpenLoops} interfaces 
to a variety of public Monte Carlo programs.
In addition to the on-the-fly approach, {\sc OpenLoops\,2} will support also
the original open-loop method, which requires additional third-party tools
such as {\sc Collier} or {\sc Cuttools}~\cite{Ossola:2007ax} and {\sc
OneLOop}~\cite{vanHameren:2010cp} for the reduction to scalar
integrals.

\subsection{Technical performance} \label{sec:stab}

In this section we study the technical performance of the new algorithm.
Similarly as in~\cite{Cascioli:2011va}, we present speed and stability
benchmarks for the one-loop QCD corrections to four families of partonic
processes,
\begin{itemize}
\item[(a)] $\mathrm{u \bar{u}} \to \mathrm{t \bar{t}} + n\, \mathrm{g}$,
\item[(b)] $\mathrm{g g} \to \mathrm{t \bar{t}} + n\, \mathrm{g}$, 
\item[(c)] $\mathrm{u \bar{d}} \to \mathrm{W^{+} g} + n\, \mathrm{g}$,
\item[(d)] $\mathrm{u \bar{u}} \to \mathrm{W^{+} W^{-}} + n\, \mathrm{g}$,
\vspace{-3ex}
\bea
\label{eq:proclist}
\eea
\end{itemize}
with $n=0,1,2,3$ additional gluons, i.e.~including processes with up to 
7 scattering particles. Top quarks and $W$ bosons are not decayed,
and sums over the colour and helicity degrees of freedom of the external
particles are included throughout.

Benchmarks obtained with the new open-loop algorithm, denoted as
{\sc OpenLoops\,2}, are based on {\sc Collier} for the calculation of 
scalar integrals.
In order to highlight the effect of the new on-the-fly methods of
\refses{sec:OFhelmerging}{sec:red}, we also consider variants of the {\sc
OpenLoops} program where these new methods are not used.  
Specifically, as detailed in~\refta{tab:methods}, we restrict the on-the-fly
approach to helicity sums and diagram merging, using {\sc Collier} for tensor
reduction. This approach is denoted as {\sc OpenLoops\,2+\Collier}.  
Alternatively, we apply the original open-loop method in combination with
tensor integrals (denoted as {\sc OpenLoops\,1+Collier}) or OPP reduction (denoted as {\sc
OpenLoops\,1+Cuttools}). The {\sc OpenLoops\,1+Cuttools} mode relies on 
{\sc OneLOop} for the scalar integrals and is used also 
to generate benchmarks in quadruple precision. 

By default, {\sc OpenLoops} calculations are monitored through a built-in
stability system that estimates the level of instability of one-loop results and
automatically triggers re-evaluations in double or quadruple precision for
critical phase space points.  
In the following, in order to avoid any bias in the comparisons, the stability
system is switched off.  In this way, one-loop amplitudes are computed only
once per phase space point in double precision.  
Unless stated otherwise, Gram-determinant expansions are always kept active,
both in the  on-the-fly reduction of {\sc OpenLoops2} and in {\sc Collier}.

\begin{figure*}[t!]
\centering
\includegraphics[width=0.7\textwidth]{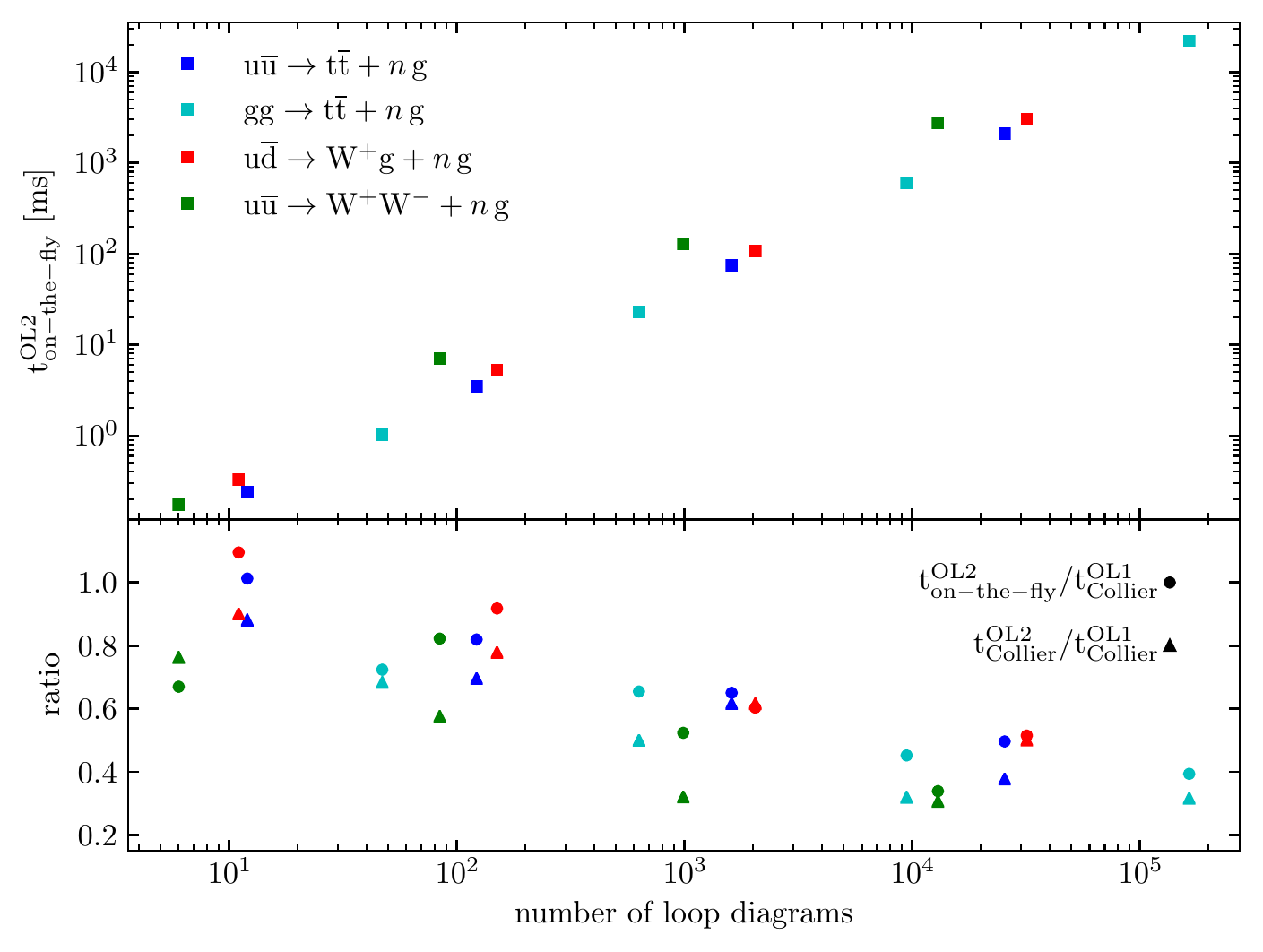}
\caption{Runtimes per phase space point for the calculation of the one-loop
scattering probability~\eqref{eq:W01} on a single $\text{Intel}$ i7-4790K
core with gfortran-4.8.5.  
Results for the processes in~\refeq{eq:proclist} are plotted versus the
number of one-loop diagrams.  Timings of {\sc OpenLoops\,2} with on-the-fly
reduction ($t^{\OLtwo}_{\onthefly}$) are shown in the upper frame.
The lower frame presents a comparison to {\sc OpenLoops\,2+Collier} and {\sc
OpenLoops\,1+Collier} (see~\refta{tab:methods}).}
\label{fig:ol2speed}
\end{figure*}

\subsubsection{Speed benchmarks}
\label{se:speedbench}

To illustrate the speed of the new algorithm, in~\reffi{fig:ol2speed} we
plot runtimes per phase space point for the calculation of the one-loop
scattering probability~\refeq{eq:W01}.  
The processes listed in~\refeq{eq:proclist} involve a number of one-loop
Feynman diagrams that ranges from the order of $1$ to $10^5$.
The corresponding runtimes, measured on a
single
$\text{Intel}$ i7-4790K core with gfortran-4.8.5, 
vary from the order of $10^{-1}$ to $10^4$\,ms.
In this range, we confirm that 
runtimes tend to grow linearly with the 
number of one-loop diagrams up to $2\to 4$ processes~\cite{Cascioli:2011va},
and we find that this scaling behaviour persists 
up to $2\to 5$ processes.
%
%
As compared to {\sc OpenLoops\,1+Collier}, the new algorithm with on-the-fly
reduction is up to a factor 2 to 3
faster for multi-particle processes.
Depending on the process, using {\sc OpenLoops\,2+Collier}, \ie restricting
the on-the-fly approach to helicity sums plus diagram merging and reducing
tensor integrals with {\sc Collier}, can result in a further significant
speed-up.  
However, the moderate slowdown caused by the on-the-fly reduction can be
counterbalanced by the improved numerical stability, which implies a reduced
need of re-evaluations in quadruple precision (see \refse{se:stabbench}).

\subsubsection{Stability benchmarks}
\label{se:stabbench}

In this section we study the numerical stability  of the new open-loop
algorithm.  To this end, one-loop scattering probability densities computed
in double precision ($\calW^{\ssst{DP}}_{\oneloop}$) are compared against
benchmarks in quadruple precision ($\calW^{\ssst{QP}}_{\oneloop}$).
More precisely, defining the relative difference between two results as
\bea
\label{eq:relativedeviation} 
\calA(\calW_a,\calW_b) &=& \log_{10}
\left|\frac{\calW_a-\calW_b}{\calWmin}\right| \quad\mbox{with} \nonumber\\[2mm]
\calWmin &=& \min\left\{|\calW_a|,|\calW_b|\right\}, 
\eea
we estimate the instability of double-precision results as\footnote{Note
that, in order to avoid possible sources of bias, the quad-precision benchmark $\calW^{\ssst{QP,R}}_{\oneloop}$ is
computed with rescaled kinematics as detailed below.}
\begin{equation}
\mathcal{A}_{\ssst{DP}} = \calA\left(\calW^{\ssst{DP}}_{\oneloop}, \calW^{\ssst{QP,R}}_{\oneloop} \right).
\label{eq:dpacc}
\end{equation}
This quantity can be regarded, up to a minus sign, as the number of correct
digits of the double-precision evaluation.

\begin{figure*}[t!]
\centering
\includegraphics[width=0.7\textwidth]{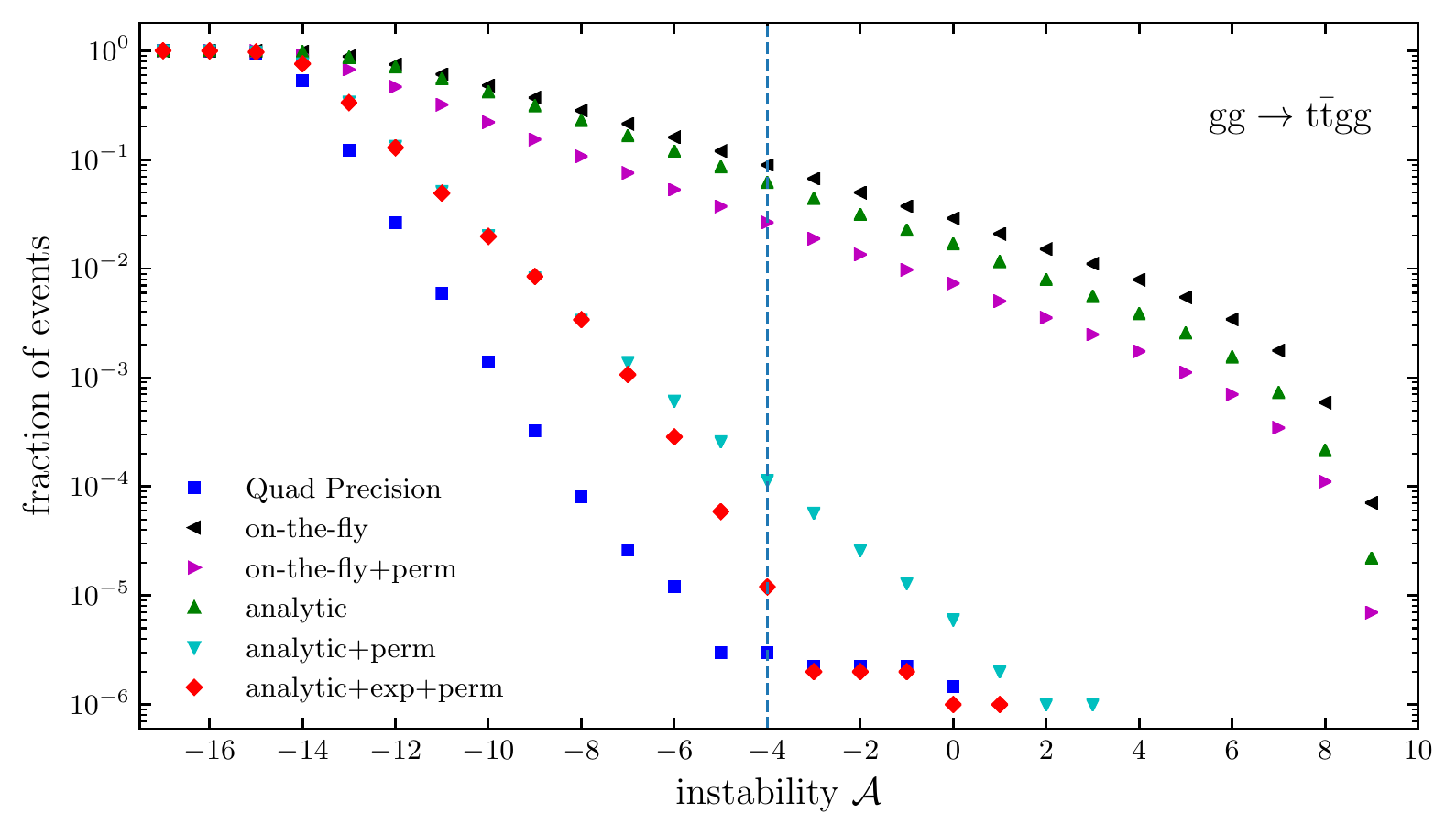}
\caption{Probability of finding events with instability
$\calA>\calA_{\mathrm{min}}$ as a function of $\calA_{\mathrm{min}}$ in a
sample of $10^6$ $\mathrm{g g} \to \mathrm{t\bar{t}gg}$ events.  
The stability of quad-precision benchmarks (blue) is compared to different variants
of the  {\sc
OpenLoops\,2} algorithm in 
double precision.
Unstable results
without special treatment of Gram determinants (``on-the-fly'')  are stabilised
using the permutation trick~\refeq{eq:propperm}  for boxes (``perm''),
exact analytic formulas for triangles (``analytic'')
and Gram-determinant expansions for $\delta<\deltathr$ (``exp'').
See \refse{eq:trianglestability}.
}
\label{fig:ggttggexp1}
\end{figure*}

To estimate the intrinsic accuracy of quad-precision benchmarks, computed
using {\sc OpenLoops\,1+Cuttools} and {\sc OneLoop}, we use a so-called
rescaling test~\cite{Badger:2010nx,Cascioli:2011va}, where scattering amplitudes are
computed with rescaled masses and momenta and scaled back according to their
mass dimensionality.
Thus for a given phase space point the accuracy of the quadruple
precision benchmarks is assessed as
\begin{equation}
\mathcal{A}_{\ssst{QP}} = \calA\left(\calW^{\ssst{QP}}_{\oneloop}, \calW^{\ssst{QP,R}}_{\oneloop} \right),
\label{eq:qpacc}
\end{equation}
where $\calW^{\ssst{QP}}_{\oneloop}$ and $\calW^{\ssst{QP,R}}_{\oneloop}$
are the original and rescaled quad-precision evaluations.
This quantity represents the finite resolution of the instability estimate~\refeq{eq:dpacc}.
As we will see, 
quad-precision benchmarks can become more unstable than double-precision
results obtained with {\sc OpenLoops\,2}.
In this case, the instability estimate~\refeq{eq:dpacc}
yields $\mathcal{A}_{\ssst{DP}}\sim \mathcal{A}_{\ssst{QP}}$
but should be interpreted
as $\mathcal{A}_{\ssst{DP}} < \mathcal{A}_{\ssst{QP}}$.

To assess the stability of {\sc OpenLoops\,2}, for each process
in~\refeq{eq:proclist} we have studied a sample of $10^6$ homogeneously distributed
phase space points at $\sqrt{s}=1$\,TeV.
To exclude soft and collinear regions we have required $p_{i,\mathrm{T}} >
50$\,GeV and $\Delta R_{ij} > 0.5$ for all massless final-state QCD partons.


Fig.~\ref{fig:ggttggexp1} illustrates 
the effect of Gram-determinant instabilities and the goodness of the
solutions introduced in~\refse{se:instabilities} in the case
of  $\mathrm{g g} \to \mathrm{t \bar{t}gg}$.
For this challenging multi-particle process,
using the {\sc OpenLoops\,2} 
on-the-fly reductions 
without any special treatment
of Gram determinants  we observe an extremely
high level of numerical instability in double precision.
The probabilities to obtain one-loop results with less than four or zero
correct digits are around $10^{-1}$ and $10^{-2}$, respectively,  and 
the tail of the stability distribution extends up to 
a level of instability of ten orders of magnitude and more.
Applying the permutation trick~\refeq{eq:propperm} (``perm'') and using
analytic expressions for three-point integrals (``analytic'') result in a 
dramatic stability
improvement for the box and triangle reductions, respectively.
Combining these two improvements (``analytic+perm'')
reduces the the probability of finding points with only few correct digits
by three orders of magnitude, and yields a maximum level of instability 
around $10^2$.
Finally, switching on the Gram-determinant expansions 
for $\delta<\deltathr$
leads
to a further very drastic reduction of the probability of finding results
with less than 3--4 correct digits.
In this range, we observe an overlap with the tail of the quad-precision
distribution.  As discussed above, this indicates that {\sc OpenLoops 2} in
double precision is more stable than the quad-precision benchmarks, and its
estimated instability represents only an upper bound.  Most likely, the tail
of the true {\sc OpenLoops\,2} stability distribution ends at $10^{-3}$.

%
\begin{figure*}[ht]
\begin{minipage}{0.33\textwidth}
\bce
{\scriptsize on-the-fly}
\includegraphics[height=50mm,trim=0 0 48 0 ,clip]{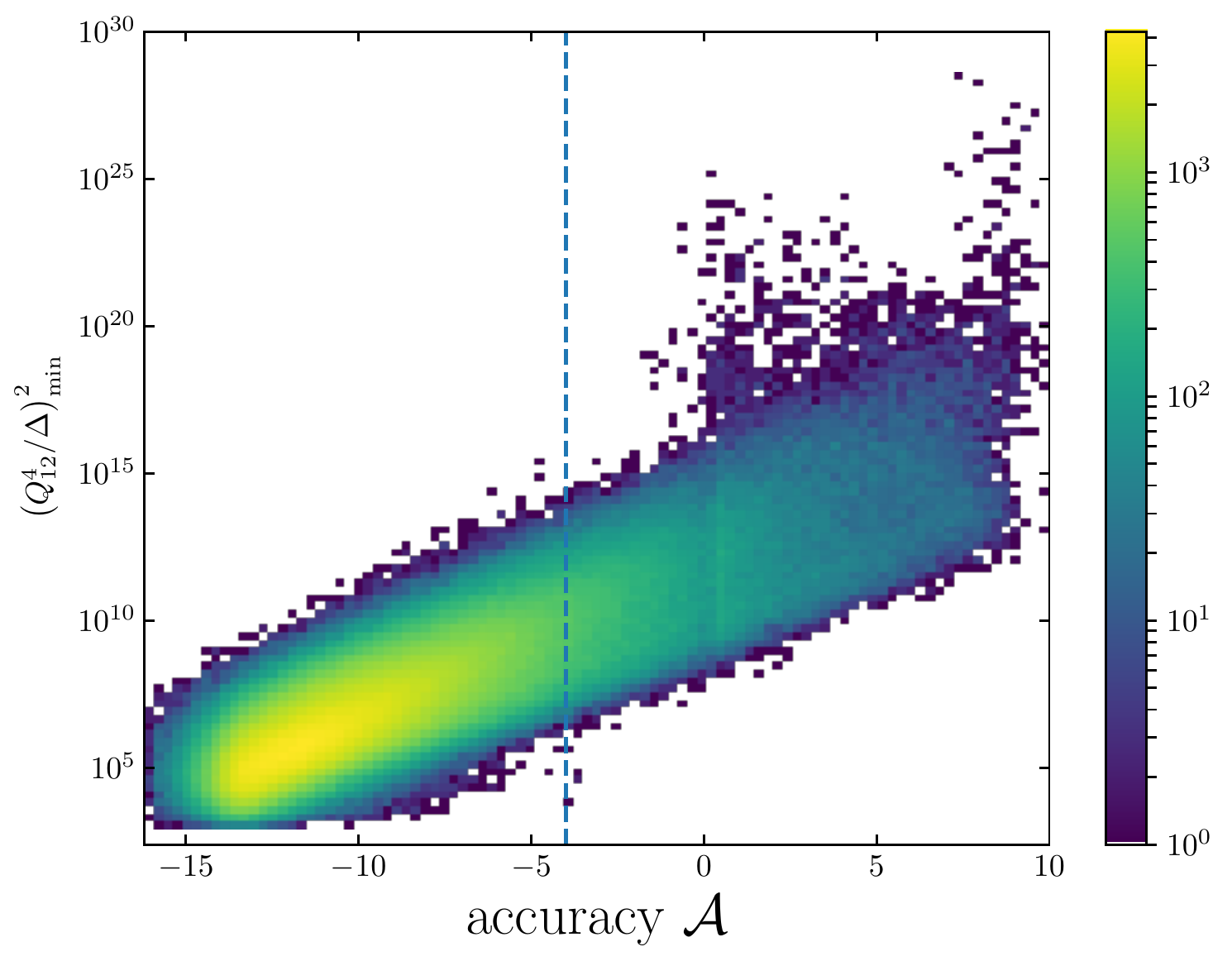} 
\ece
\end{minipage}
\begin{minipage}{0.31\textwidth}
\bce
{\scriptsize analytic + perm}
\includegraphics[height=50mm,trim=23 0 48 0 ,clip]{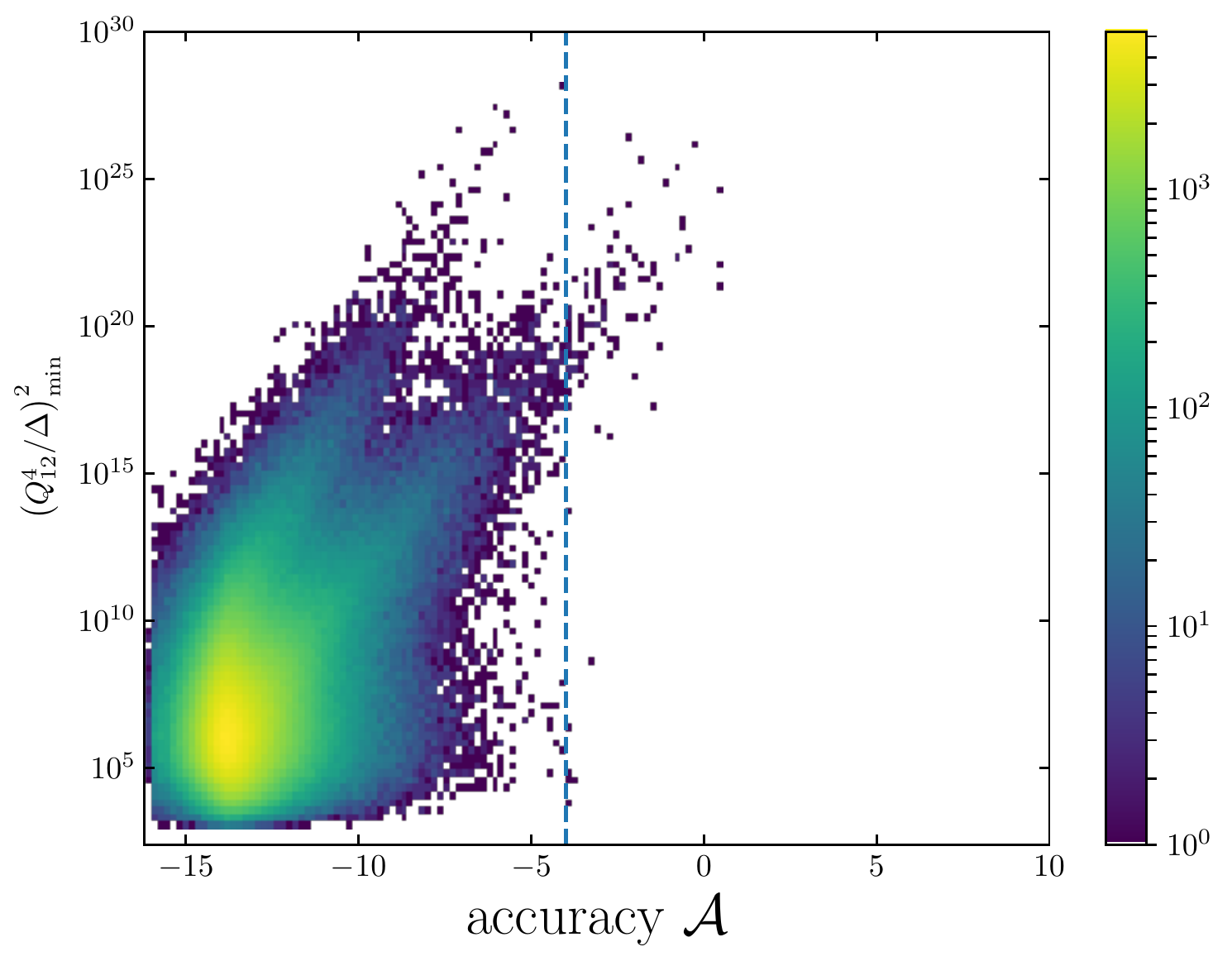} 
\ece
\end{minipage}
\begin{minipage}{0.33\textwidth}
\bce
{\scriptsize analytic + perm + exp}
\includegraphics[height=50mm,trim=23 0 0 0 ,clip]{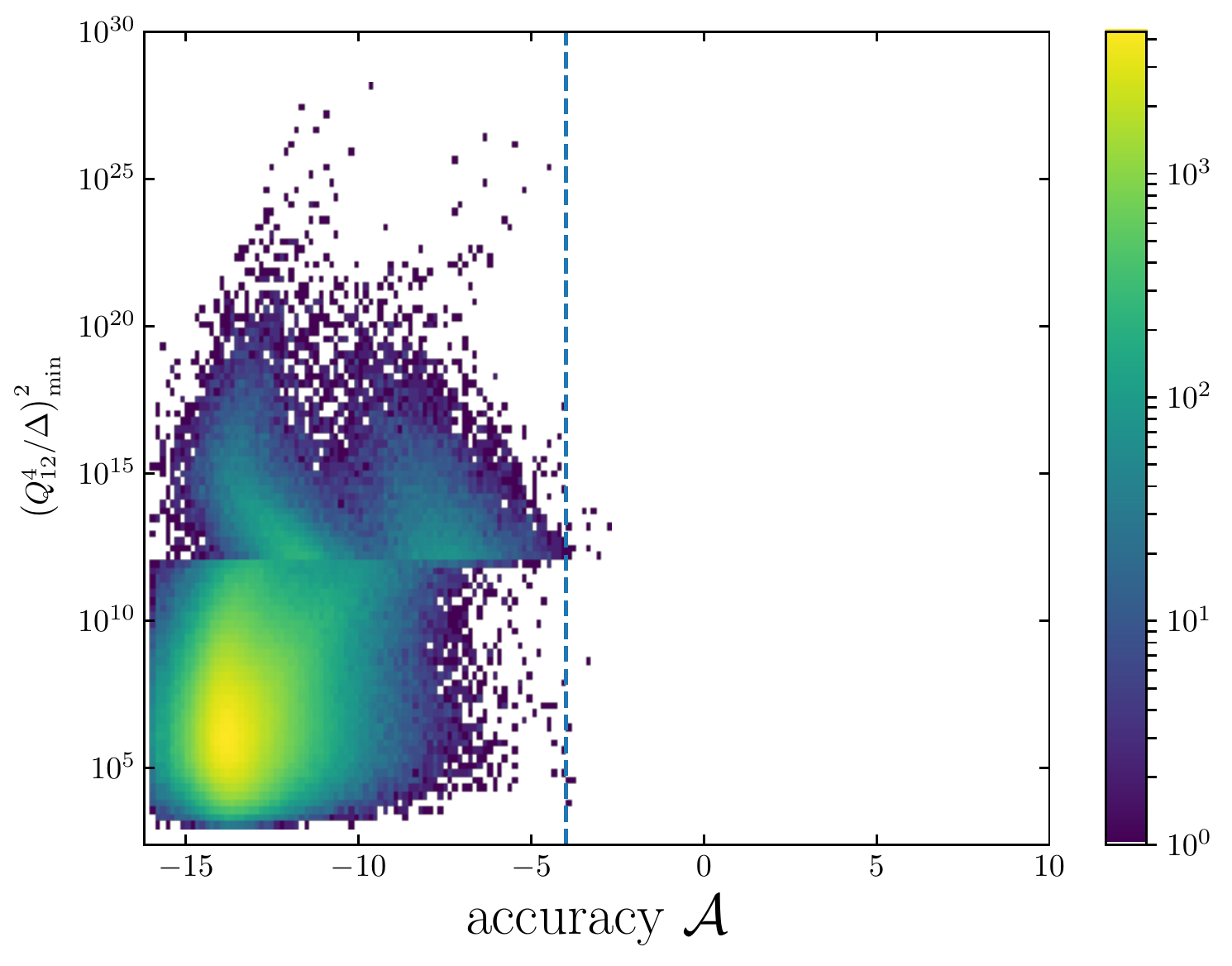}
\ece
\end{minipage}
\caption{Correlation between the instability $\calA$ of {\sc OpenLoops\,2}
in double precision and the largest
$(Q^4/\Delta)^2$ in the event, where $\Delta$ is any rank-two Gram
determinant and $Q^2$ is the maximum scale in the corresponding 
Gram matrix. See \refeq{eq:q2ijdef}.
Probability densities correspond to $10^6$ events.  Unstable results without special treatment of
Gram determinants (left) are stabilised using the permutation
trick~\refeq{eq:propperm} for box reduction 
and analytic expressions for triangle reduction
(middle) plus Gram-determinant expansions for $\delta<\deltathr$ 
(right).}
\label{fig:ggttggexp2}
\end{figure*}


To gain more insights into the origin of numerical instabilities
in the on-the-fly reduction of {\sc OpenLoops\,2},
let us investigate 
the correlation between the
instability~\refeq{eq:dpacc} and rank-two Gram determinants $\Delta$
in the $\mathrm{g g} \to \mathrm{t\bar{t}gg}$ sample
of \reffi{fig:ggttggexp1}.
More precisely, in~\reffi{fig:ggttggexp2} 
we consider the minimal value of the dimensionless parameter
$\Delta_{ij}/Q_{ij}^4$ in the event, where $Q_{ij}^2$ is the largest
$|p_i\cdot p_j|$ in the corresponding Gram matrix (see \refse{eq:boxstability}).
As demonstrated by the left plot in~\reffi{fig:ggttggexp2}, the instability
of the entire scattering amplitude features a remarkably strong correlation
with rank-two Gram determinants over twenty orders of magnitude.
Moreover we observe a  quadratic or faster scaling in
$Q^4/\Delta$, consistent with the form of the $\gamma^{-2}\sim \Delta^{-2}$
poles in \refeq{eq:ABmunutensors4}.
The middle plot shows the combined effect of the permutation
trick~\refeq{eq:propperm}, which avoids the smallest Gram determinant of the
event in all reductions with $N\ge 4$ loop denominators, 
and the attenuation of spurious singularites
through analytic expressions for three-point configurations
of type \refeq{eq:triangleparam}.
In this way the
probability of having less than four correct digits is reduced to 0.1 permil.
Finally, in the right plot we see that points with less than 3--4 correct digits
disappear completely when Gram-determinant expansions are switched
on.\footnote{Points with unreliable quad-precision benchmarks
($\mathcal{A}_{\ssst{QP}}\sim \mathcal{A}_{\ssst{DP}}$) are 
not considered in~\reffi{fig:ggttggexp2}.}
As one can clearly recognise in the right plot, the threshold for the
activation of Gram-determinant expansions corresponds to $(Q^4/\Delta)^2\sim
\deltathr^{-4}=10^{12}$.

\begin{figure*}[ht]
\centering
\includegraphics[width=0.55\textwidth]{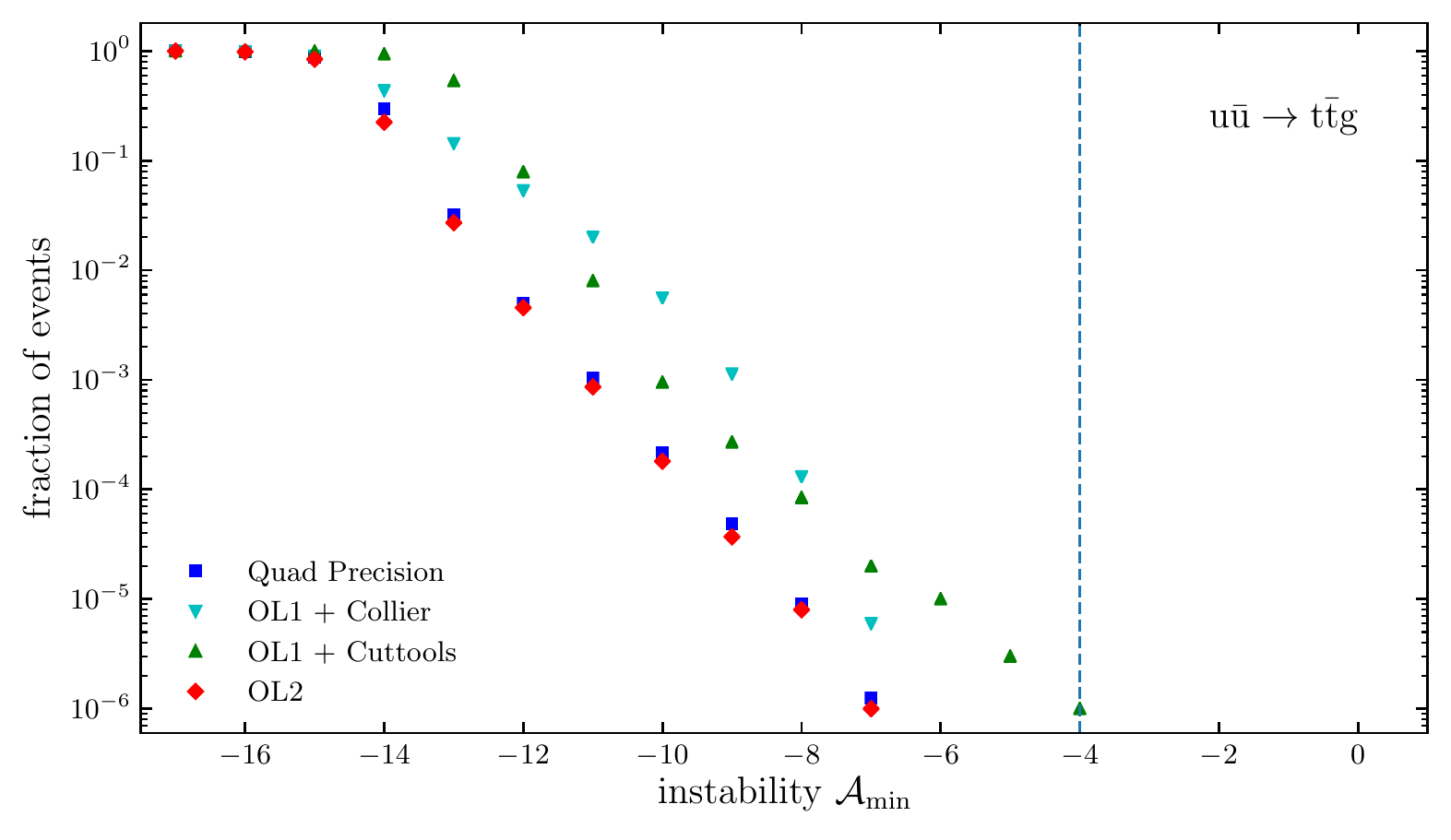} \\ [-5mm]
\includegraphics[width=0.55\textwidth]{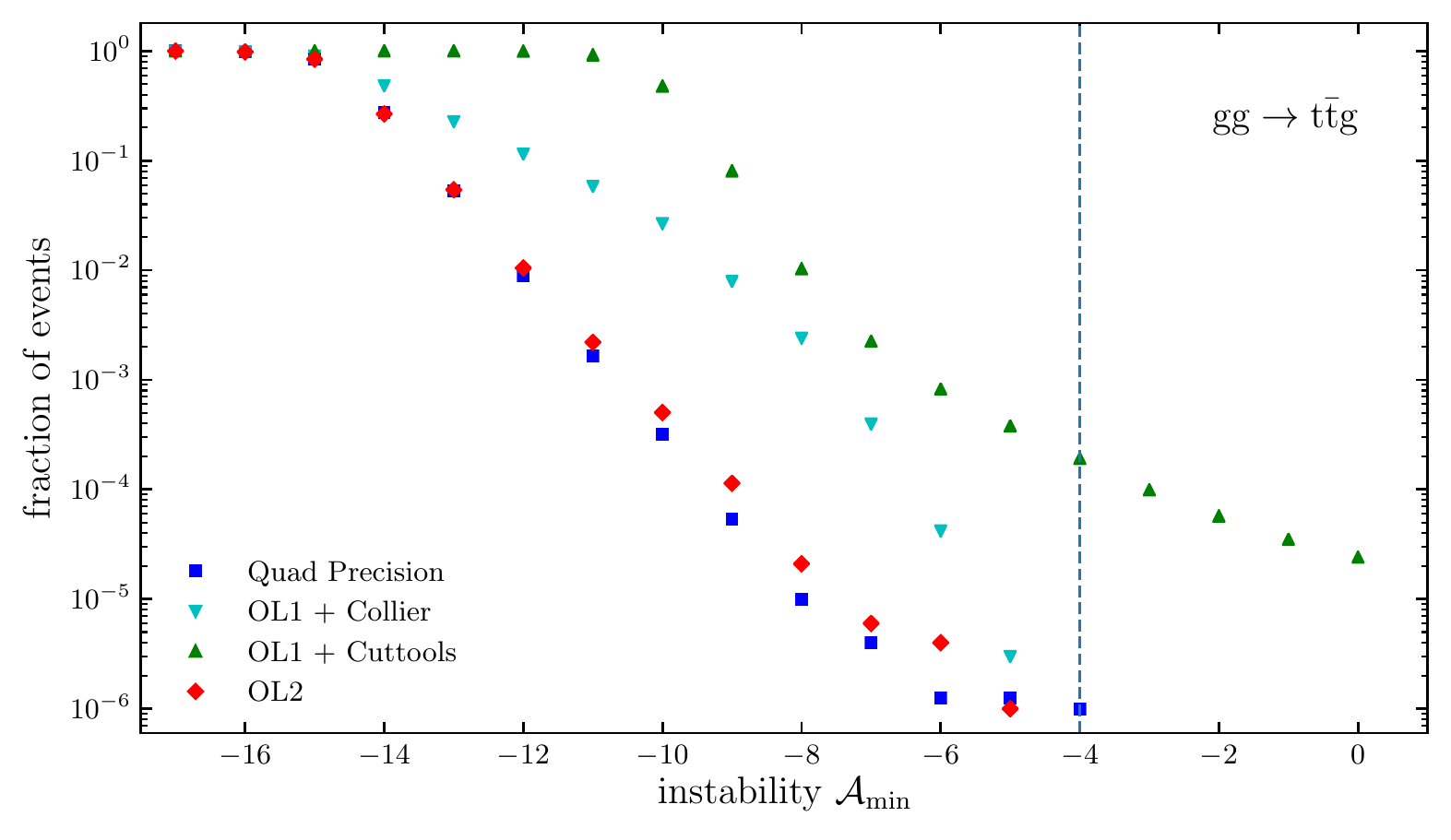} \\ [-5mm]
\includegraphics[width=0.55\textwidth]{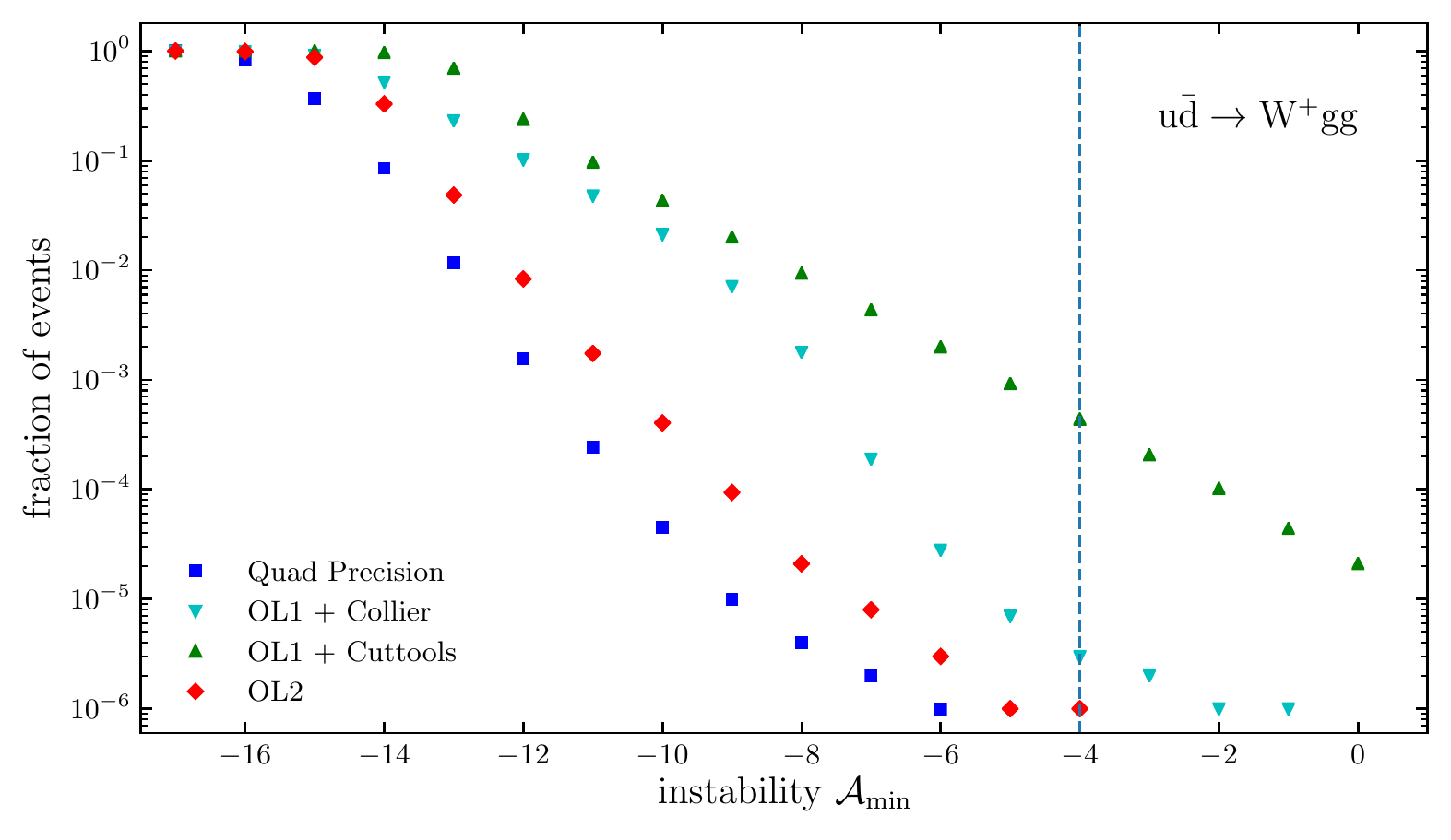} \\ [-5mm]
\includegraphics[width=0.55\textwidth]{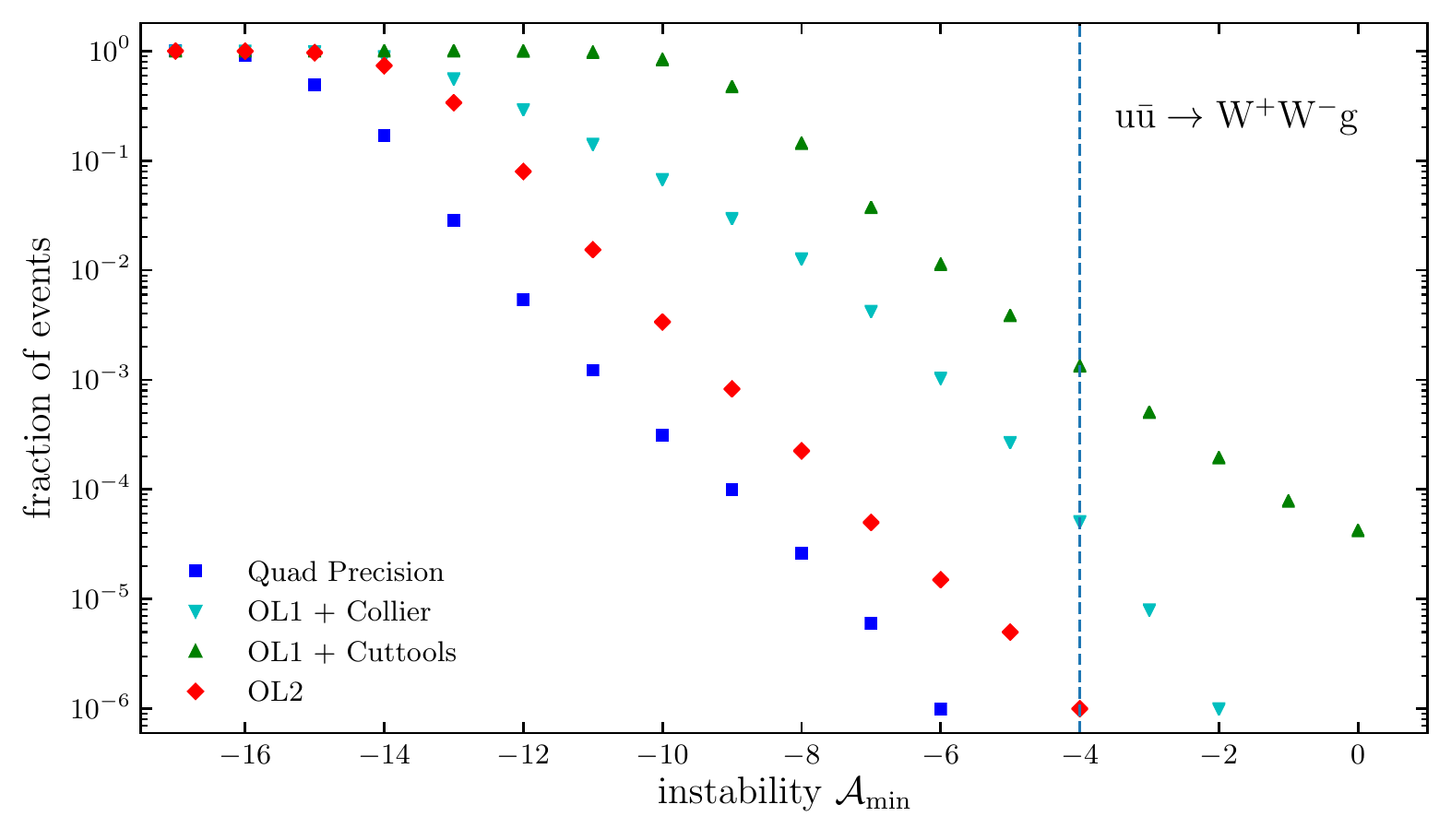} 
\caption{Stability distributions for 
$2 \to 3$  processes defined as in \reffi{fig:ggttggexp1}. The stability of 
{\sc OpenLoops\,2} with on-the-fly reduction is
compared to {\sc OpenLoops\,1} with {\sc Collier} or {\sc Cuttools}.
The instability of the employed quad precision benchmarks is also shown.}
\label{fig:2to3}
\end{figure*}

Finally, in \reffis{fig:2to3}{fig:2to4} we compare the stability 
of {\sc OpenLoops\,2} against {\sc OpenLoops\,1+Collier} and {\sc
OpenLoops\,1+Cuttools} for the $2\to 3$ and $2\to 4$ processes
in~\refeq{eq:proclist}.
In {\sc OpenLoops\,2} the stability improvements of~\refse{se:instabilities}
are applied throughout. 
The results of {\sc OpenLoops\,1+Cuttools} feature the highest instability tails for
all considered processes.  The probability of finding less than four correct
digits can exceed $10^{-3}$ in $2\to 3$ and $10^{-2}$ in $2\to 4$ processes, while
the fraction of fully unstable points with $\calA\ge 0$ 
can reach $10^{-3}$ in $2\to 4$ processes.
Switching to {\sc OpenLoops\,1+Collier} we find that, depending on the
process, the probability of finding only a few correct digits goes down by
one to three orders of magnitude, while in eight samples of $10^6$ points we
do not find a single result with $\calA>0$.%
\footnote{As discussed above, due to the insufficient quality of
quad-precision benchmarks, instability estimates in the tail of $\mathrm{g
g} \to \mathrm{t\bar{t}gg}$ are not significant.}
\begin{figure*}[ht]
\centering
\includegraphics[width=0.55\textwidth]{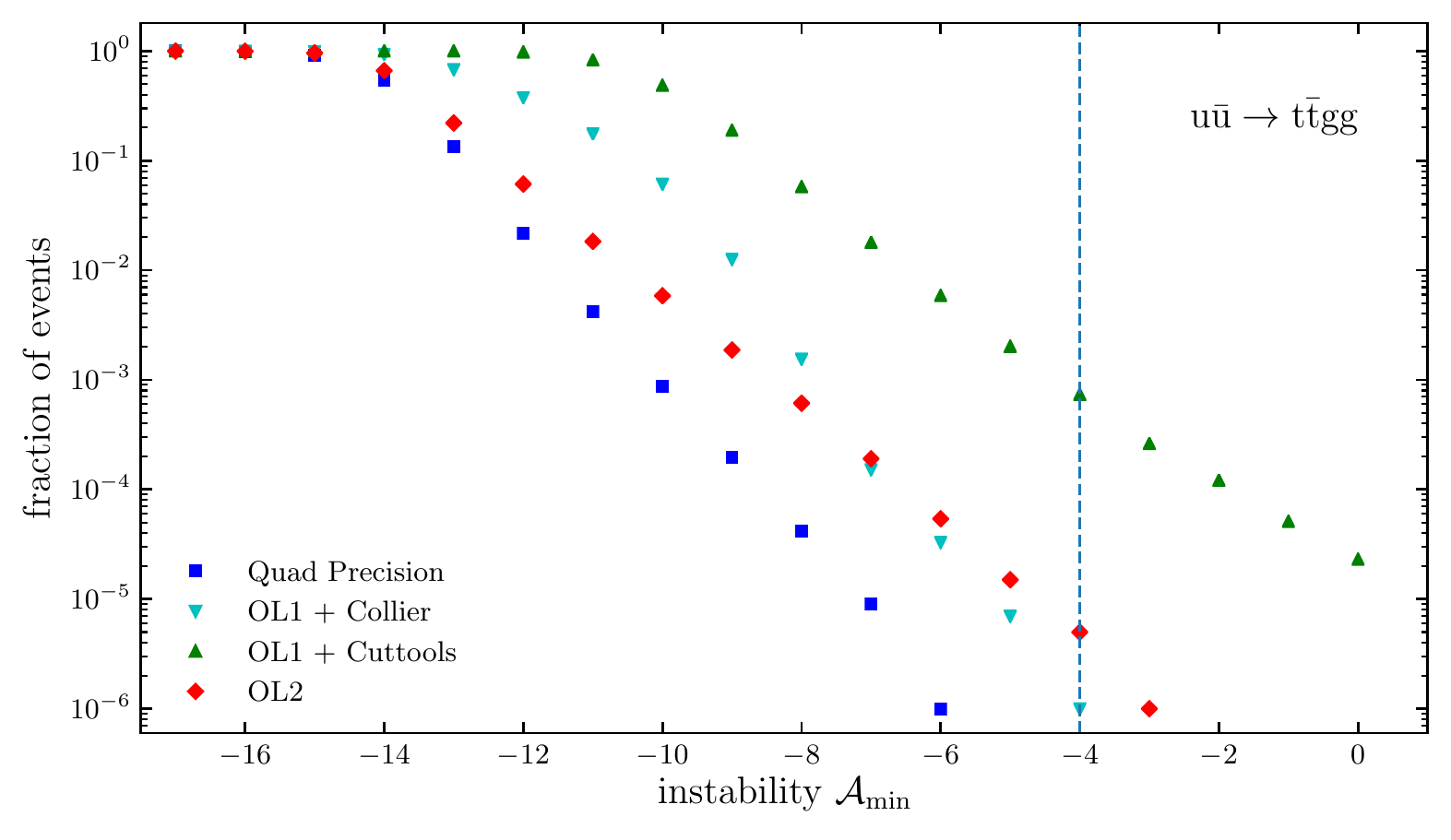} \\ [-5mm]
\includegraphics[width=0.55\textwidth]{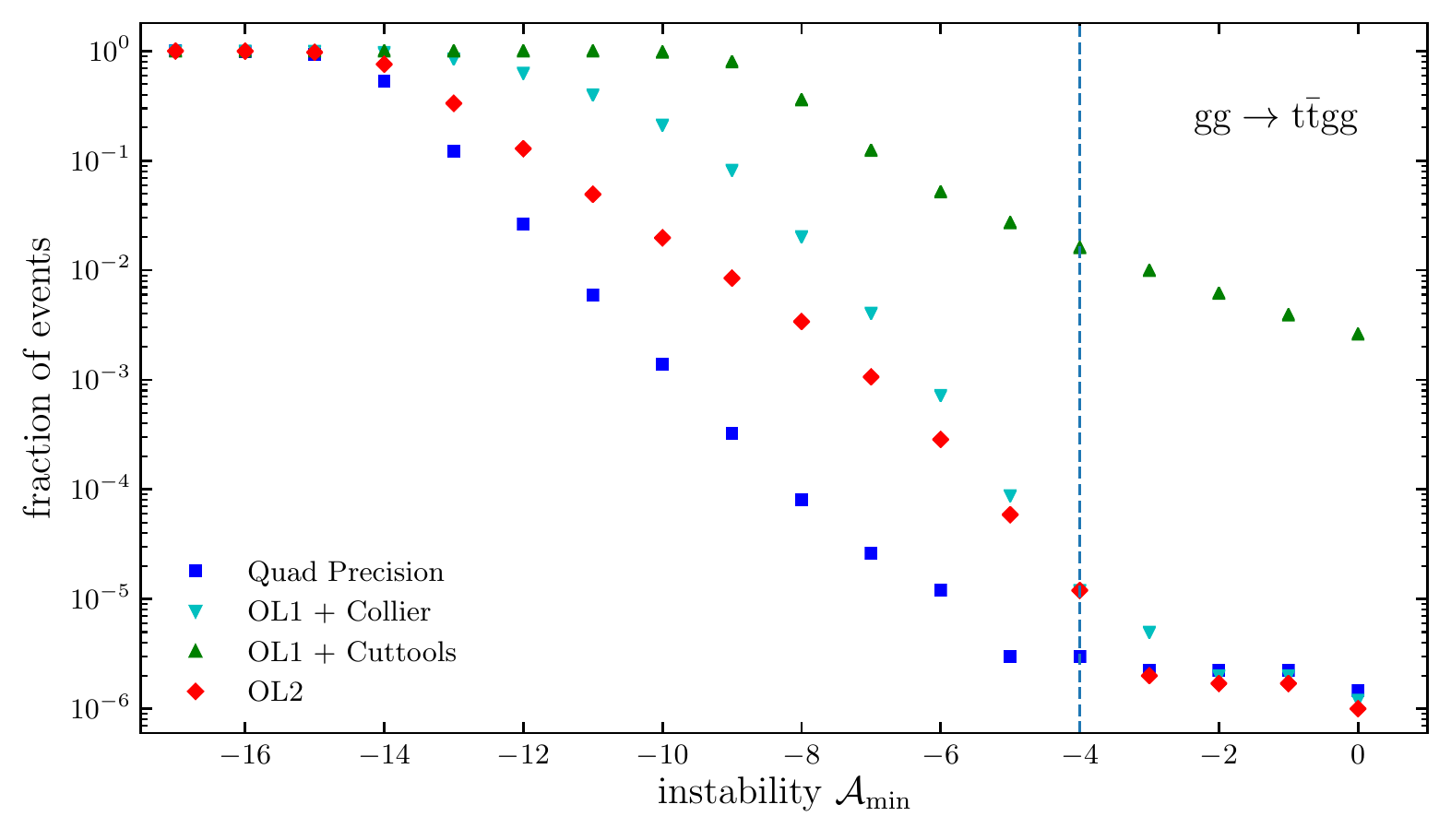} \\ [-5mm]
\includegraphics[width=0.55\textwidth]{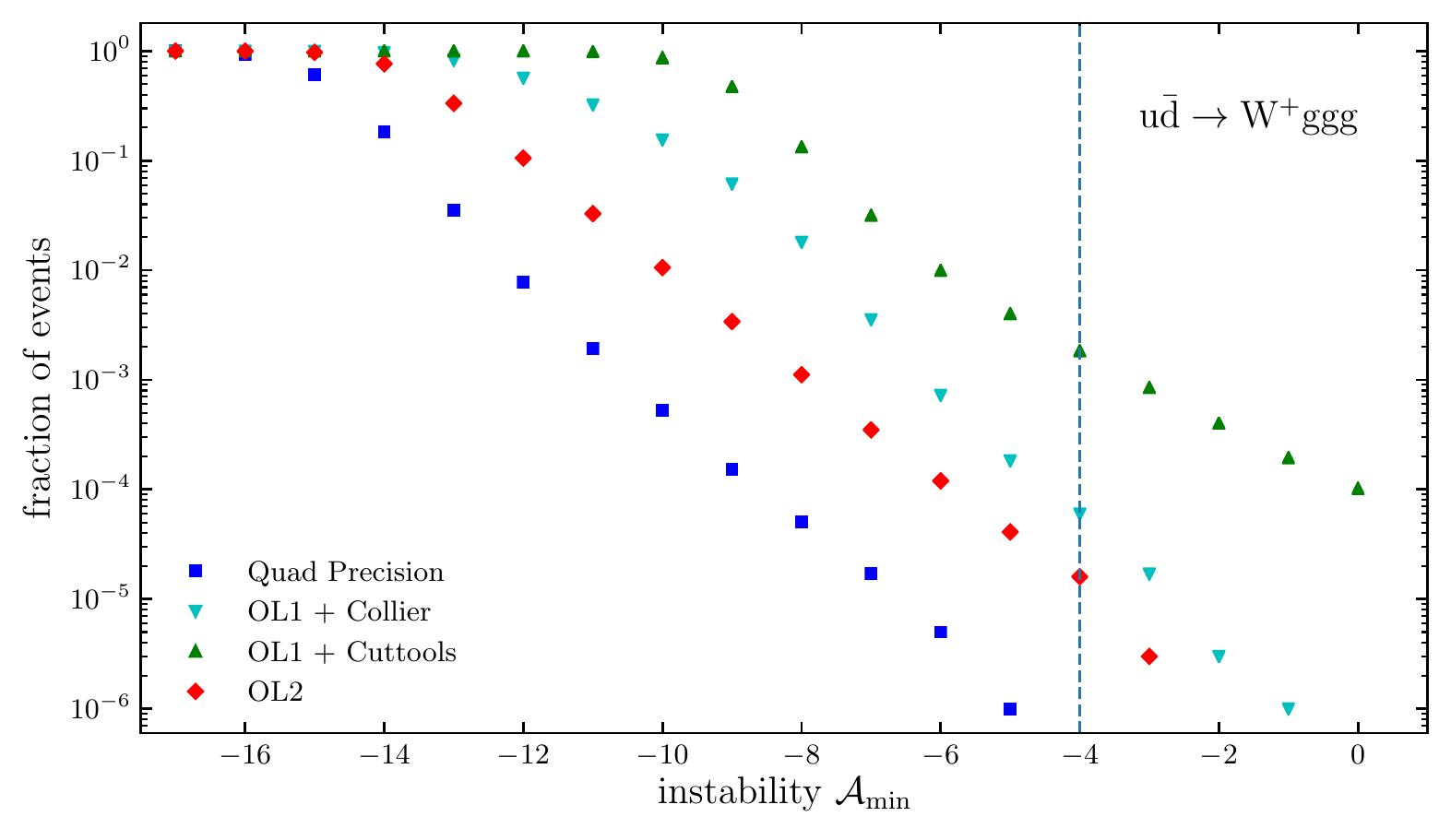} \\ [-5mm]
\includegraphics[width=0.55\textwidth]{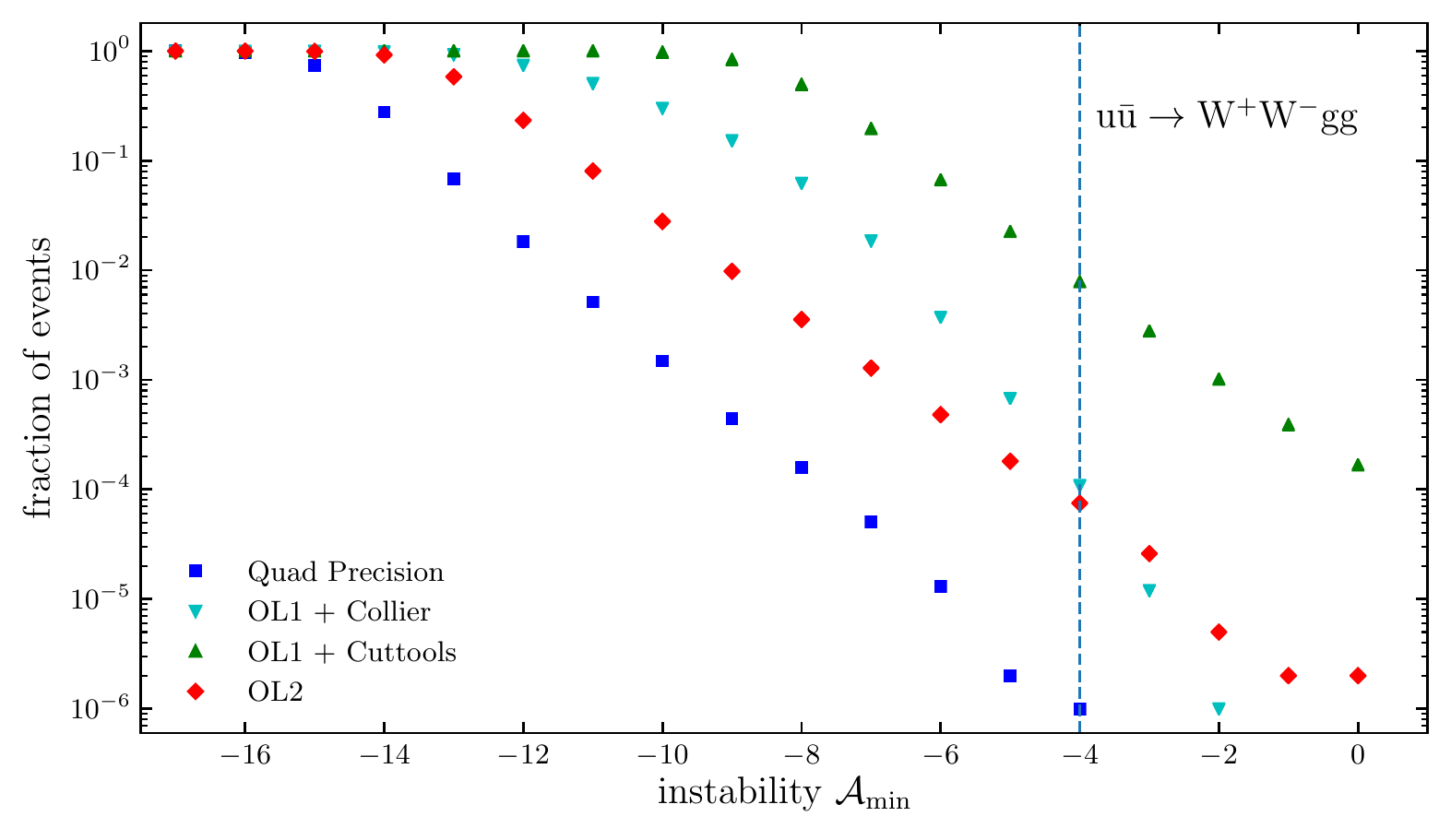} \\ 
\caption{Stability distributions for 
$2 \to 4$  processes defined as in \reffi{fig:2to3}.} 
\label{fig:2to4}
\end{figure*}

Using {\sc OpenLoops\,2} can lead to a further significant stability
improvement.  This is especially evident for $2\to 3$ processes, 
where the stability of the on-the-fly reduction in {\sc OpenLoops\,2} is
remarkably close to the quad-precision benchmarks and even superior
than quad precision for the
case of $\mathrm{t\bar{t}g}$ production.  
When quad precision is sufficiently accurate to resolve the instabilities  of {\sc
OpenLoops\,2} we observe improvements of one--two orders of magnitude with
respect to {\sc OpenLoops\,1+Collier}.  
In the case of $2\to 4$ processes, depending on the process and the
considered number of digits, {\sc OpenLoops\,2} can perform somewhat better
or slightly worse than {\sc OpenLoops\,1+Collier}, like in the case of
$\mathrm{u \bar{d}} \to \mathrm{W^{+}gg g}$ or $\mathrm{u \bar{u}} \to
\mathrm{W^{+}W^-g g}$, respectively.  However, both approaches guarantee
excellent numerical stability.

\section{Conclusions and Outlook} \label{se:conclusions}

We have presented a new approach for the automated calculation of scattering
amplitudes at one loop.  This new technique is based on the open-loop
approach, where cut-open loop integrands are factored into 
a product of loop-momentum dependent segments
that are combined through recursive tensorial multiplications.

The key idea behind the new method is that various operations, which are
typically done at the level of full Feynman diagrams or amplitudes, can be
performed {\it on-the-fly} during the open-loop recursion, \ie after the
multiplication of each loop segment.
Since it exploits the factorised structure of open loops in a systematic way,
this on-the-fly approach can reduce the complexity of certain operations in a 
very significant way.

We have first applied the on-the-fly method to helicity summations and to
the merging of topologically equivalent open loops, finding 
speed-up factors of up to two or three
as compared to the original 
open-loop algorithm.
Moreover, using the integrand reduction method by del\,Aguila and Pittau,
we have introduced an on-the-fly technique for the reduction of open loops. 
With this approach, the construction of loop amplitudes
and their reduction are interleaved step by step within a 
single numerical recursion.
In this way, objects with tensor rank higher than two are avoided
throughout, and the complexity of the calculations is reduced in 
a very drastic way.
The proliferation of pinched subtopologies that emerge from the
reduction is avoided by absorbing them on-the-fly into topologically
equivalent open loops.

The employed integrand reduction method suffers from severe numerical
instabilities that are dominated by kinematic regions with small rank-two
Gram determinants $\Delta$ and scale like $1/\Delta^{2}$.  
In the reduction of $N$-point objects with $N\ge
4$, we have shown that $\Delta$-instabilities can easily be avoided
through appropriate permutations of the loop denominators.
In this way we were able to isolate $\Delta$-instabilities 
in triangle topologies with a particular kinematic
configuration and to cure them 
by means of analytic expansions in $\Delta$.
This approach is the first example of an integrand reduction algorithm that is
essentially free from Gram-determinant instabilities.  The level of
stability that is achieved in double precision 
is competitive with public implementations of OPP
reduction in quadruple precision.

The new algorithm is fully automated and validated at NLO QCD and can be extended to electroweak 
interactions.
It will become publicly available in the upcoming release of {\sc OpenLoops\,2}.
Its technical features can be  especially beneficial in NLO calculations for
challenging multi-particle processes.  Moreover, in view of its excellent
numerical stability, the new algorithm is very attractive for the calculation
of real--virtual contributions at NNLO.
Finally, the idea of simplifying the construction of loop amplitudes 
through the factorisation of loop integrands and their on-the-fly reduction
may open new interesting perspectives for the
automation of two-loop calculations.

\section*{Acknowledgments}
We are grateful to P.~Maierh\"ofer and J.~Lindert for discussions and
extensive technical support.  
This research was supported in part by the Swiss National Science Foundation
(SNF) under contracts PP00P2-128552 and BSCGI0-157722 as well as by the
Research Executive Agency (REA) of the European Union under the Grant
Agreement PITN--GA--2012--316704 ({\it HiggsTools}).

\appendix

\section{Reduction of the remaining tensor integrals} \label{app:intred}

After the on-the-fly reduction described in \refse{se:OFR} we are left with
a few types of low-rank integrals~\refeq{list:TIs} that need to be
reduced to the standard basis of one-loop scalar integrals.
The relevant reductions are performed upon taking the trace~\refeq{eq:mergingtrace}
as described in the following. 
In this appendix we restrict ourselves to tensor integrals 
with four-dimensional $q^\mu$ terms in the numerator, while 
contributions with additional $\tilde{q}^{2}$ terms in the numerator are
described in \refse{sec:rationalterms}.

\subsection{Five- and higher-point
integrals with $R=0,1$} 
\label{sec:intred5}

Integrals with $N\ge 5$ loop propagators and a numerator of rank $R\le 1$,
\bea
\calN(q)=\calN+\calN_\mu q^\mu,
\eea
can be reduced to a linear combination of scalar boxes,
\bea
\label{eq:OPPformula_rank01}
&&\int\!\mathrm{d}^D\!\momq\,
\f{\calN(q)}{\Dbar{0}\Dbar{1}\cdots \Dbar{N-1}}= \nonumber\\ &&=
\sum_{i_0 < i_1 < i_2 < i_3}^{N-1}
\int\!\mathrm{d}^D\!\momq\,
\f{d_{i_0 i_1 i_2 i_3}} 
{\Dbar{i_0}\Dbar{i_1}\Dbar{i_2}\Dbar{i_3}}.
\eea
To determine the box coefficients 
we use the OPP reduction formula~\cite{Ossola:2006us} 
\bea
d_{i_0 i_1 i_2 i_3}&=& \f{1}{2}\left[R_{i_0 i_1 i_2 i_3}(q_0^{+})+R_{i_0 i_1 i_2 i_3}(q_0^{-})\right],\quad
\eea
where $q_0^{\pm}$ are the solutions of the quadruple-cut conditions
$\Dbar{i_0}=\Dbar{i_1}=\Dbar{i_2}=\Dbar{i_3}=0$, and 
\bea
R_{i_0 i_1 i_2 i_3}(q)=\f{\calN(q)}{\prod\limits_{i \ne i_0, i_1, i_2, i_3}^{N-1} \Dbar{i}}
\label{eq:oppres}
\eea
are the corresponding residues.

For the special case $d_{0123}$, where $p_{i_0}=p_0=0$,
the explicit solutions of the quadruple cut equations 
read
\begin{equation}
q_0^{\pm} = x_1\,l_1 + x_2\,l_2 + x_3^{\pm} \,l_4 + x_4^{\pm}\,l_4 ,
\label{eq:quadcutsol}
\end{equation}
where $l_i$ are the basis momenta defined in~\refeq{eq:basismom12}--\refeq{eq:basismom34}.
The $x_{1,2}$ coefficients are given by 
\bea
x_{1,2} = \frac{1}{\gamma}\left(d_{2,1} - d_0 \left(1-\alpha_{2,1} \right) 
-d_{1,2}\, \alpha_{2,1}\right),\qquad
\eea
with $d_i = m_i^2 - p_i^2$.
The terms $\alpha_{1,2}$ and $\gamma$ are defined in \refse{se:red}, 
and the remaining coefficients in~\refeq{eq:quadcutsol} are given by
\bea
x_4^{\pm} =  a x_3^{\pm} =-\frac{\left( b \pm c_0 \right)}{2},
\eea
with $a = (p_3\cdot l_3)/(p_3\cdot l_4)$ and
\bea
b &=& \frac{1}{p_3\cdot l_4} \left(\frac{d_0 - d_3}{2} 
+ x_1 \left(p_3 \cdot l_1\right) + x_2 \left(p_3 \cdot l_2 \right) \right),
\nonumber\\
c_0 &=& \sqrt{b^2 - a\left(x_1\,x_2 - \frac{d_0}{\gamma}\right)}.
\label{eq:opplast}
\eea
 
The other coefficients $d_{i_0i_1i_2i_3}$ can be obtained from the above
formulas through an obvious reparametrisation of masses and momenta
($i_0i_1i_2i_3\to 0123$) and a subsequent momentum shift, $p_i^\mu\to
p_i^\mu-p_0^\mu$ for $i=0,1,2,3$, which has to be applied throughout
in~\refeq{eq:oppres}--\refeq{eq:opplast}.

\subsection{Four-point functions with $R=1$} \label{sec:intred4}

The reduction of rank-one boxes,
\bea
\int\!\mathrm{d}^D\!\momq\,
\frac{q^\mu}{\Dbar{0}\cdots \Dbar{3}}
&=&
\sum_{j=-1}^3
\int\!\mathrm{d}^D\!\momq\,
\frac{A_j^\mu}{\Dbar{0}\cdots 
\slashed{\Dbar{j}}\cdots
\Dbar{3}},
\label{eq:rankoneboxred}\qquad\;
\eea
results into a scalar box ($j=-1$) and four scalar triangles ($j=0,1,2,3$).
For the corresponding coefficients we use the reduction formulas~\cite{delAguila:2004nf},
\bea
A^{\mu}_{1,2}  &=& \frac{1}{2 \gamma}\left[r^{\mu}_{2,1} - \frac{p_{3} \cdot r_{2,1}}{p_{3} \cdot l_{3}} \left( l^{\mu}_{3}
+ \frac{1}{\alpha} l^{\mu}_{4} \right) \right], \nonumber\\
A^{\mu}_{3}  &=& \frac{1}{4} \frac{1}{p_{3} \cdot l_{3}} \left( l^{\mu}_{3} + \frac{1}{\alpha} l^{\mu}_{4} \right),
\eea
and
\bea
A^{\mu}_{-1} = \sum_{i=1}^3 f_{i0} A^{\mu}_{i}, \qquad
A^{\mu}_{0} = -\sum_{i=1}^3 A^{\mu}_{i}.
\eea
The relevant ingredients, $\gamma, \alpha, l^\mu_i, r^\mu_i, f_{i0}$, 
are defined in \refse{se:red}.

\subsection{Three-point integrals with $R=1$} \label{sec:intred3}

For rank-one triangles we use the covariant decomposition
\bea
&& C^\mu(\momp{1},\momp{2},\mass{0},\mass{1},\mass{2}) = \ceps
\int\!\mathrm{d}^D\!\momq\,\frac{q^\mu}{\Dbar{0}\Dbar{1}\Dbar{2}}=\label{eq:Cmudef}\nonumber\\
&&= \sum_{i=1}^2\momp{i}^\mu C_i(\momp{1}^2,\momp{2}^2,\mass{0},\mass{1},\mass{2}),
\label{eq:PVtrianglered}
\eea
where $\ceps={(2\pi\mu)^{2\eps}}/{(\ri\pi^2)}$,
and we compute the $C_i$ coefficients via 
Passarino--Veltman reduction~\cite{Passarino:1978jh,Denner:1991kt}.

\subsection{Two-point integrals with $R=1,2$} \label{sec:intred2}

Also for two-point integrals with rank $R=1,2$ we perform a Passarino--Veltman
reduction~\cite{Passarino:1978jh,Denner:1991kt} based on the covariant decomposition
\bea
 B^\mu(\momp{1},\mass{0},\mass{1}) &=& \ceps\int\!\mathrm{d}^D\!\momq\, \frac{q^\mu}{\Dbar{0} \Dbar{1}}=\nonumber\\
&=& \momp{1}^\mu B_1(\momp{1}^2,\mass{0},\mass{1}) \label{eq:Bmudef},\\
 B^{\mu\nu}(\momp{1},\mass{0},\mass{1}) &=& \ceps\int\!\mathrm{d}^D\!\momq\, \frac{q^\mu q^\nu}{\Dbar{0} \Dbar{1}}=\nonumber\\
&=&\momp{1}^\mu\momp{1}^\nu B_{11}(\momp{1}^2,\mass{0},\mass{1})\nonumber\\&+& g^{\mu\nu} B_{00}(\momp{1}^2,\mass{0},\mass{1}).
\label{eq:Bmunudef}
\eea
Besides the standard reduction formulas for $B_1$, $B_{11}$, and $B_{00}$, for
$p^2=0$ we have implemented the special cases
\bea
B_1(0, \mass{0}, \mass{0}) &=& -\f{1}{2} B_0 (0, \mass{0}, \mass{0}),\\
B_1(0, \mass{0}, \mass{1}) &=& \f{1}{  4 (\mass{0}^2 - \mass{1}^2)} \Big(2 A_0(\mass{1}) \nonumber\\ &&
- 2\mass{0}^2 B_0(0, \mass{0}, \mass{1})-1\Big),\\
B_{11}(0,0,0) &=& \f{1}{3} B_{0}(0,0,0)=0,\\
B_{11}(0,\mass{0},\mass{0}) &=& -\f{1}{3} + \f{A_0(\mass{0})}{3 \mass{0}^2},\\
B_{11}(0,\mass{0},\mass{1}) &=& \f{1}{18 (\mass{0} - \mass{1})^3 (\mass{0} + \mass{1})^3}
\nonumber\\ && \hspace{-15mm}
\times\,\Big(5 \mass{0}^6 - 9 \mass{0}^2 \mass{1}^4 + 4 \mass{1}^6 
+ 6 \mass{0}^4 A_0(\mass{0}) \nonumber\\ && \hspace{-15mm}\phantom{\times\,}
- 6 (3 \mass{0}^4 - 3 \mass{0}^2 \mass{1}^2 + \mass{1}^4) A_0(\mass{1})\Big)
.
\nonumber\\
\eea
with the scalar master integrals
\bea
B_0(\momp{1}^2,\mass{0},\mass{1})&=&\ceps\int\!\mathrm{d}^D\!\momq\, \frac{1}{\Dbar{0} \Dbar{1}}, \nonumber\\
A_0(\mass{0}^2)&=&\ceps\int\!\mathrm{d}^D\!\momq\, \frac{1}{\Dbar{0}}. \label{eq:A0B0def}
\eea

\bibliographystyle{JHEP}

\providecommand{\href}[2]{#2}\begingroup\raggedright\endgroup

\end{document}